\documentclass[english]{article}
\usepackage{lmodern}

\usepackage[T1]{fontenc}
\usepackage[latin9]{inputenc}
\usepackage[a4paper]{geometry}
\usepackage{dsfont}
\usepackage{color}
\usepackage{babel}
\usepackage{amsmath}
\usepackage{amsthm}
\usepackage{amssymb}
\usepackage{stmaryrd}
\usepackage{graphicx}
\usepackage{setspace}
\usepackage{esint}
\usepackage{caption}
\usepackage{natbib}
\usepackage[unicode=true,bookmarks=true,bookmarksnumbered=false,bookmarksopen=false,breaklinks=false,pdfborder={0 0 0},pdfborderstyle={},backref=page,colorlinks=true]{hyperref}
\hypersetup{pdftitle={Unbiased MCMC},pdfauthor={Jacob OLeary Atchade},linkcolor=RoyalBlue,citecolor=RoyalBlue}

\usepackage{float}
\usepackage{subfig}
\usepackage[dvipsnames,svgnames,x11names,hyperref]{xcolor}
\usepackage{nicefrac}
\usepackage{algorithm}

\makeatletter

\newfloat{algorithm}{tbp}{loa}
\providecommand{\algorithmname}{Algorithm}
\floatname{algorithm}{\protect\algorithmname}

\theoremstyle{plain}

\theoremstyle{plain}
\newtheorem{prop}{\protect\propositionname}[section]

\newcommand{\R}{\mathbb{R}}

\numberwithin{equation}{section}

\def\Xset{\mathsf{X}} 
\def\F{\mathcal{F}} 

\def\D{\mathcal{D}}

\def\Cset{\mathcal{C}} 

\theoremstyle{plain}
\newtheorem{assumption}{Assumption}[section]

\makeatother

\providecommand{\propositionname}{Proposition}
\providecommand{\theoremname}{Theorem}

\begin{document}

\title{Unbiased Markov chain Monte Carlo with couplings}

\author{
    Pierre E. Jacob\thanks{Department of Statistics, Harvard University, Cambridge, USA. Email:
pjacob@fas.harvard.edu},
    John O'Leary\thanks{Department of Statistics, Harvard University, Cambridge, USA. Email:
joleary@g.harvard.edu},
Yves F. Atchad\'e\thanks{Department of Mathematics \& Statistics, Boston University, Boston, USA.
Email: atchade@bu.edu}
}

\maketitle

\begin{abstract}
    Markov chain Monte Carlo (MCMC) methods provide consistent approximations
    of integrals as the number of iterations goes to infinity. MCMC estimators
    are generally biased after any fixed number of iterations.  We propose to remove this bias by using couplings of Markov
    chains together with a telescopic sum argument of Glynn and Rhee (2014).
    The resulting unbiased estimators can be computed independently 
    in parallel.  We discuss practical couplings
    for popular MCMC algorithms.  We establish the theoretical validity of
    the proposed estimators and study their efficiency relative to the
    underlying MCMC algorithms.  Finally, we illustrate the performance and limitations of the
    method on toy examples, on an Ising model around its critical temperature, on a high-dimensional variable selection problem,
    and on an approximation of the cut distribution arising in Bayesian inference for models made of multiple modules.
\end{abstract}

\section{Introduction \label{sec:introduction}}

Markov chain Monte Carlo (MCMC) methods constitute a popular class of
algorithms to approximate high-dimensional integrals arising in statistics and
other fields
\citep{liu2008monte,robert:casella:2004,brooks2011handbook,green2015bayesian}.
These iterative methods provide estimators that are consistent as the number of
iterations grows large but potentially biased for any fixed number of
iterations, which discourages the parallel execution of many short chains
\citep{rosenthal2000parallel}.  Consequently, efforts have focused on exploiting
parallel processors within each iteration
\citep{tjelmeland2004using,brockwell2006parallel,lee2010utility,jacob2011using,calderhead2014general,goudie2017massively,yang2017parallelizable} and
on the design of parallel chains targeting different distributions
\citep{altekar2004parallel,wang2015parallelizing,srivastava2015wasp}.
Still, MCMC estimators are ultimately justified by asymptotics in the
number of iterations, which is discordant with current trends in computing hardware, 
characterized by increasing parallelism but stagnating clock speeds.

In this paper we propose a general construction to produce unbiased estimators of integrals with respect to a target probability distribution from MCMC kernels.
The lack of bias means that these estimators can be implemented on parallel processors in the
framework of \citet{Glynn1991}, without communication between processors.
Confidence intervals can be constructed with asymptotic guarantees in the number of processors, 
in contrast with standard MCMC confidence intervals that are justified
asymptotically in the number of iterations \citep[e.g.][]{flegal2008markov,gongflegal,atchade2016markov,vats2018strong}.
The lack of bias has additional benefits, as discussed in Section \ref{subsec:cut-distribution} in which we make use of its interplay with the law of iterated expectations to perform modular inference;
see also the discussion in Section \ref{sec:discussion}.

Our contribution follows the path-breaking work of \citet{glynn2014exact},
which uses couplings to construct unbiased estimators of integrals with respect to an invariant distribution.
They illustrate their construction on Markov chains represented by iterated random functions, leveraging the contraction properties of such functions. \citet{glynn2014exact} also consider 
Harris recurrent chains for which an explicit minorization condition holds.
Previously, \citet{McLeish:2011} employed similar debiasing techniques to obtain ``nearly unbiased''
estimators from a single MCMC chain.
More recently \citet{jacob2017smoothing} remove the bias from 
conditional particle filters \citep{andrieu:doucet:holenstein:2010} by coupling chains so that they meet in finite time.
The present article brings this type of ``Rhee--Glynn'' construction to 
generic MCMC algorithms, with a novel analysis of estimator efficiency and a variety of examples.
Our proposed construction involves couplings of MCMC algorithms, which we discuss for 
generic Metropolis--Hastings and Gibbs samplers. 

Couplings have been used to study the convergence properties of MCMC algorithms from both theoretical and practical points of view
\citep[e.g.][]{reutter1995general,johnson1996studying,rosenthal1997faithful,johnson1998coupling,neal1999circularly,roberts2004general,johnson2004,johndrow2017coupling}.
Couplings also underpin perfect samplers
\citep{propp:wilson:1996,murdoch1998exact,casella:lavine:robert:2001,flegal2012exact,leedoucetperfectsimulation,huber2016perfect}.
A notable aspect of the approach of \citet{glynn2014exact} preserved in our method is that only two chains have to be coupled 
for the proposed estimator to be unbiased, without further assumptions on the state space or target distribution.
Thus the approach applies more broadly than 
perfect samplers \citep[see][]{glynn2016exact} while yielding unbiased estimators
rather than exact samples. Coupling pairs of Markov chains also forms the basis of the approach of
\citet{neal1999circularly}, with a similar motivation for parallel computation. The proposed estimation technique
also shares aims with regeneration methods \citep[e.g.][]{mykland1995regeneration,brockwell:kadane:05}, 
and we propose a numerical comparison in Section \ref{subsec:Gibbs-pump-failures}.

In Section \ref{sec:Two-coupled-chains} we introduce our estimators
and present a coupling of random walk Metropolis--Hastings chains as an illustration.
In Section \ref{sec:theory} we establish the efficiency properties of these estimators,
discuss the verification of key assumptions, and describe the use of the proposed estimators
on parallel processors in light of results from e.g. \citet{Glynn1991}. In Section \ref{sec:Practical-couplings} we describe how to couple some important MCMC algorithms and illustrate the effect of dimension on algorithm performance with a multivariate Normal target.
Section \ref{sec:illustrations} contains more challenging examples including a multimodal target, a comparison with regeneration methods, 
sampling problems in large-dimensional discrete spaces arising in Bayesian variable selection and Ising models, and an application to modular inference.
We discuss our findings in Section \ref{sec:discussion}.
Scripts in \texttt{R} \citep{RCRAN} are available at \url{https://github.com/pierrejacob/unbiasedmcmc} and supplementary materials
are available online.

\section{Unbiased estimation from coupled chains\label{sec:Two-coupled-chains}}

\subsection{Rhee--Glynn estimator \label{subsec:Basic-construction}}

Given a target probability distribution $\pi$ on a Polish space $\mathcal{X}$ and a measurable real-valued test function $h$ integrable with
respect to $\pi$, we want to estimate the expectation $\mathbb{E}_\pi[h(X)]=\int h(x)\pi(dx)$.
Let $P$ denote a Markov transition kernel on $\mathcal{X}$ that leaves $\pi$ invariant,
and let $\pi_0$ be some initial probability distribution on $\mathcal{X}$. Our estimators are based on a coupled pair of Markov chains $(X_{t})_{t\geq0}$
and $(Y_{t})_{t\geq0}$, which marginally start from $\pi_0$ and evolve according 
to $P$. In particular, we suppose that $\bar P$ is a transition kernel on the joint space $\mathcal{X}\times\mathcal{X}$ such that $\bar{P}((x,y),A\times \mathcal{X}) = P(x,A)$
and $\bar{P}((x,y),\mathcal{X}\times A) = P(y,A)$
for any $x,y\in\mathcal{X}$ and any measurable set $A$. We then construct the coupled Markov chain $(X_t,Y_t)_{t\geq 0}$ as follows. We draw $(X_0,Y_0)$ such that $X_0\sim \pi_0$ and $Y_0\sim \pi_0$. Given $(X_0,Y_0)$ we draw $X_1\sim P(X_0, \cdot)$, and then for any $t\geq 1$, given $X_0, (X_1,Y_0),\ldots,(X_t,Y_{t-1})$, we draw $(X_{t+1},Y_{t}) \sim \bar{P}((X_{t},Y_{t-1}),\cdot)$. 
We consider the following assumptions.

\begin{assumption} \label{assumption:marginaldistributions}
As $t\to\infty$, $\mathbb{E}[h(X_{t})]\to\mathbb{E}_\pi[h(X)]$. Furthermore, there
exists an $\eta>0$ and $D<\infty$ such that $\mathbb{E}[|h(X_{t})|^{2+\eta}]\leq D$ for all $t\geq0$.
\end{assumption}

\begin{assumption} \label{assumption:meetingtime} The chains are such that the meeting time  $\tau:=\inf\{t\geq1:\ X_{t}=Y_{t-1}\}$ satisfies $\mathbb{P}(\tau>t)\leq C\;\delta^{t}$ for all $t\geq 0$, for some constants $C<\infty$ and $\delta\in(0,1)$.
\end{assumption}

\begin{assumption} \label{assumption:sticktogether}The chains stay
together after meeting, i.e. $X_{t}=Y_{t-1}$ for all $t\geq\tau$.
\end{assumption}

By construction, each of the marginal chains $(X_t)_{t\geq 0}$ and
$(Y_t)_{t\geq 0}$ has initial distribution $\pi_0$ and transition kernel $P$.
Assumption \ref{assumption:marginaldistributions} requires these 
chains to result in a uniformly bounded $(2+\eta)$-moment of $h$;
more discussion on moments of Markov chains can be found in \citet{tweedie1983existence}.
Since $X_0$ and $Y_0$ may be drawn from any coupling of $\pi_0$ with itself,
it is possible to set $X_0 = Y_0$. However, $X_1$ is then generated from $P(X_0,\cdot)$, so that
$X_1 \neq Y_0$ in general. Thus one cannot force the meeting time to be small by setting $X_0 = Y_0$.
Assumption \ref{assumption:meetingtime} puts a condition on the coupling operated by $\bar{P}$,
and would not in general be satisfied for an independent coupling. 
Coupled kernels must be carefully designed, using e.g. common random numbers
and maximal couplings,
for Assumption \ref{assumption:meetingtime} to be satisfied. We present a simple case in Section \ref{subsec:exampleMH}
and further examples in Section \ref{sec:Practical-couplings}. 
We stress that the state space is not assumed to be discrete, and that the constants $D$ and $\eta$ of Assumption \ref{assumption:marginaldistributions}
and $C$ and $\delta$ of Assumption \ref{assumption:meetingtime} do not need to be known
to implement the proposed approach.
Assumption \ref{assumption:sticktogether} typically holds 
by design; coupled chains that stay identical after meeting 
are termed ``faithful'' in \citet{rosenthal1997faithful}.

Under these assumptions we introduce the following motivation for an unbiased estimator of $\mathbb{E}_\pi[h(X)]$, following \citet{glynn2014exact}.
We begin by writing $\mathbb{E}_\pi[h(X)]$ as $\lim_{t\to\infty}\mathbb{E}[h(X_{t})]$.
Then for any fixed $k\geq 0$,
\begin{align*}
\mathbb{E}_\pi[h(X)] & =\mathbb{E}[h(X_{k})]+\sum_{t=k+1}^{\infty}(\mathbb{E}[h(X_{t})]-\mathbb{E}[h(X_{t-1})]) & \text{expanding the limit as a telescoping sum,}\\
 & =\mathbb{E}[h(X_{k})]+\sum_{t=k+1}^{\infty}(\mathbb{E}[h(X_{t})]-\mathbb{E}[h(Y_{t-1})]) & \text{since the chains have the same marginals,}\\
 & =\mathbb{E}[h(X_{k})+\sum_{t=k+1}^{\infty}(h(X_{t})-h(Y_{t-1}))\text{]} & \text{swapping the expectations and limit,}\\
 & =\mathbb{E}[h(X_{k})+\sum_{t=k+1}^{\tau-1}(h(X_{t})-h(Y_{t-1}))] & \text{by Assumption \ref{assumption:sticktogether}.}
\end{align*}
We note that the sum in the last equation is zero if $k+1>\tau-1$. The heuristic argument above suggests that the estimator 
${H_{k}(X,Y)=h(X_{k})+\sum_{t=k+1}^{\tau-1}(h(X_{t})-h(Y_{t-1}))}$
should have expectation $\mathbb{E}_\pi[h(X)]$.
We observe that this estimator requires $\tau-1$ calls to $\bar{P}$
and $\max(1,k+1-\tau)$ calls to $P$; thus under Assumption \ref{assumption:meetingtime} its cost has a finite expectation.

In Section \ref{sec:theory} we establish the validity of the estimator under
the three conditions above; this formally justifies the swap of expectation and limit. 
The estimator can be viewed as a debiased version of $h(X_k)$.
Unbiasedness is guaranteed for any choice of $k\geq 0$, but both the cost and variance of $H_k(X,Y)$ are sensitive to $k$;
see Section \ref{subsec:variance}.
Thanks to this unbiasedness property, we can sample
$R\in \mathbb{N}$ independent copies of $H_k(X,Y)$ in parallel and average the results to estimate $\mathbb{E}_\pi[h(X)]$
consistently as $R\to \infty$; we defer further considerations on the use of unbiased estimators
on parallel processors to Section \ref{sec:glynnheidelberger}.

Before presenting examples and enhancements to the estimator above, 
we discuss the relationship between our approach and existing work.
There is a rich literature applying forward couplings to study Markov
chains convergence \citep{johnson1996studying,johnson1998coupling,thorisson2000coupling,
lindvall2002lectures, rosenthal2002quantitative,johnson2004,douc:moulines:rosenthal:04,nikooienejad2016bayesian},
and to obtain new algorithms such as perfect samplers \citep{huber2016perfect} 
and the methods of \citet{neal1999circularly} and \citet{neal2001improving}.
Our approach is closely related to \citet{glynn2014exact},
who employ pairs of Markov chains to obtain unbiased estimators. 
The present work combines similar arguments with couplings of MCMC 
algorithms and proposes further improvements to remove bias at a reduced
loss of efficiency.

Indeed \citet{glynn2014exact} did not apply their methodology 
to the MCMC setting. They consider chains associated with contractive iterated random functions 
\citep[see also][]{diaconis1999iterated}, and Harris recurrent chains
with an explicit minorization condition. A minorization condition refers to a small set $\mathcal{C}$, $\lambda>0$, an integer $m\geq 1$,
and a probability measure $\nu$ such that for all $x\in \mathcal{C}$ and some measurable set $A$,
$P^m(x,A) \geq \lambda \nu(A)$. 
Such a condition is said to be explicit if the set, constant and probability measure are known by the user.
Finding explicit small sets that are useful in practice can present a technical challenge, even for MCMC experts
\citep[see discussion and references in][]{cowles1998simulation}.
When available, explicit minorization conditions can also be employed 
to identify regeneration times, yielding estimators 
amenable to parallel computation in the framework of \citet{mykland1995regeneration} and \citet{brockwell:kadane:05}.
By contrast \citet{johnson1996studying,johnson1998coupling} and \citet{neal1999circularly}
address the question of coupling MCMC algorithms
so that pairs of chains meet exactly,
without analytical knowledge on the target distribution. The present article
focuses on the use of couplings of this type in the framework of \citet{glynn2014exact}.

\subsection{Coupled Metropolis--Hastings example\label{subsec:exampleMH}}

Before further examination of our estimator and its properties, 
we present a coupling of Metropolis--Hastings (MH) chains 
that will typically satisfy Assumptions \ref{assumption:marginaldistributions}-\ref{assumption:sticktogether} in
realistic settings; this coupling was proposed in \citet{johnson1998coupling} as part of a method to diagnose convergence. We postpone discussion of other couplings of MCMC algorithms to 
Section \ref{sec:Practical-couplings}.
We recall that each iteration $t$ of the MH algorithm \citep{hastings:1970} begins by drawing a proposal $X^{\star}$
from a Markov kernel $q(X_{t},\cdot)$, where $X_t$ is the current state.
The next state is set to $X_{t+1}=X^{\star}$ if 
${U \leq \pi(X^{\star})q(X^{\star},X_{t})}/({\pi(X_{t})q(X_{t},X^{\star})})$,
where $U$ denotes a uniform random variable on $[0,1]$, and $X_{t+1}=X_{t}$ otherwise.

We define a pair of chains so that each proceeds marginally according to the MH algorithm and jointly so that 
the chains will meet exactly after a random number of steps. We suppose that the pair of chains 
are in states $X_t$ and $Y_{t-1}$, and consider how to generate $X_{t+1}$ and $Y_t$
so that $\{X_{t+1}=Y_{t}\}$ might occur.

If $X_t\neq Y_{t-1}$, the event  $\{X_{t+1}=Y_{t}\}$ cannot occur if both chains reject their respective proposals, $X^\star$ and $Y^\star$. 
Meeting will occur if these proposals are identical and if both are accepted.
Marginally, the proposals follow ${X^{\star}|X_{t}\sim q(X_{t},\cdot)}$
and $Y^{\star}|Y_{t-1}\sim q(Y_{t-1},\cdot)$. If $q(x,x^\star)$ can be evaluated for
all $x,x^\star$, then one can sample from a maximal 
coupling between the two proposal distributions, which is a coupling 
of $q(X_{t},\cdot)$ and $q(Y_{t-1},\cdot)$ maximizing the probability
of the event $\{X^\star = Y^\star\}$. How to sample from maximal couplings of continuous distributions
is described in \citet{thorisson2000coupling}
and in Section \ref{subsec:maximalcoupling}.
One can accept or reject the two proposals using a common uniform
random variable $U$. The chains will stay together 
after they meet: at each step after meeting, the proposals will be identical with probability one,
and jointly accepted or rejected with a common uniform variable. 
This coupling requires neither explicit minorization conditions nor contractive properties of a random function
representation of the chain.

\subsection{Time-averaged estimator \label{subsec:Improvements}}

To motivate our next estimator, we note that we can compute
$H_{k}(X,Y)$ for several values of $k$ from the same realization of the coupled chains, and that the average of these is unbiased as well.
For any fixed integer $m$ with $m\geq k$, we can run coupled chains for $\max(m,\tau)$ iterations,
compute the estimator $H_{\ell}(X,Y)$ for each $\ell\in\{k,\ldots,m\}$,
and take the average ${H_{k:m}(X,Y)=(m-k+1)^{-1}\sum_{\ell=k}^{m}H_{\ell}(X,Y)}$,
as we summarize in Algorithm \ref{alg:Unbiased-estimator-avg}.
We refer to $H_{k:m}(X,Y)$ as the \emph{time-averaged estimator};
the estimator $H_k(X,Y)$ is retrieved when $m=k$.
Alternatively we could average the estimators $H_{\ell}(X,Y)$
using weights $w_\ell\in \mathbb{R}$ for $\ell\in\{k,\ldots,m\}$, 
to obtain $\sum_{\ell = k}^m w_\ell H_\ell(X,Y)$. 
This will be unbiased if $\sum_{\ell = k}^m w_\ell = 1$. 

\begin{algorithm}
\begin{enumerate}
\item Draw $X_{0}$,$Y_{0}$ from an initial distribution $\pi_{0}$
and draw $X_{1}\sim P(X_{0},\cdot)$.
\item Set $t = 1$. While $t < \max(m,\tau)$, where $\tau=\inf\{t\geq1:\;X_{t}=Y_{t-1}\}$,
    \begin{itemize}
        \item draw $(X_{t+1},Y_{t})\sim\bar{P}((X_{t},Y_{t-1}),\cdot)$,
        \item set $t \leftarrow t + 1$.
    \end{itemize}
\item For each $\ell\in\left\{ k,...,m\right\} $, compute $H_{\ell}(X,Y) = h(X_{\ell})+\sum_{t=\ell+1}^{\tau-1}(h(X_{t})-h(Y_{t-1}))$.
\item[] Return $H_{k:m}(X,Y) = (m-k+1)^{-1}\sum_{\ell=k}^{m}H_{\ell}(X,Y)$; or compute $H_{k:m}(X,Y)$ with \eqref{eq:averageestimator}.
\end{enumerate}
\caption{Unbiased ``time-averaged'' estimator $H_{k:m}(X,Y)$ of $\mathbb{E}_\pi[h(X)]$.\label{alg:Unbiased-estimator-avg}}
\end{algorithm}

Rearranging terms in $(m-k+1)^{-1}\sum_{\ell=k}^{m}H_{\ell}(X,Y)$,
we can write the time-averaged estimator as
\begin{align}\label{eq:averageestimator}
H_{k:m}(X,Y) & =\frac{1}{m-k+1}\sum_{\ell=k}^{m}h(X_{\ell}) + 
\sum_{\ell=k+1}^{\tau-1}\min\left(1, \frac{\ell-k}{m-k+1}\right)(h(X_{\ell}) - h(Y_{\ell-1})).
\end{align}
The term $(m-k+1)^{-1}\sum_{\ell=k}^{m}h(X_{\ell})$ corresponds to
a standard MCMC average with $m$ total iterations and a burn-in period of $k-1$ iterations.
We can interpret the other term as a bias correction. If $\tau\leq k+1$, then the correction term equals zero. 
This provides some intuition for the choice
of $k$ and $m$: large $k$ values lead to the bias correction being equal to zero with large probability,
and large values of $m$ result in $H_{k:m}(X,Y)$ being similar to an estimator obtained from a long MCMC run.
Thus we expect the variance of $H_{k:m}(X,Y)$ to be similar to that of MCMC estimators for 
appropriate choices of $k$ and $m$.

The estimator $H_{k:m}(X,Y)$ requires $\tau-1$ calls to $\bar{P}$
and $\max(1,m+1-\tau)$ calls to $P$, which is overall comparable to $m$ calls to $P$ when $m$ is large.
Indeed, for the proposed couplings, calls to $\bar{P}$ are approximately twice as expensive as calls to $P$.
Therefore, the cost of $H_{k:m}(X,Y)$ is comparable to $2(\tau-1) + \max(1,m+1-\tau)$
iterations of the underlying MCMC algorithm.
Thus both the variance and the cost of $H_{k:m}(X,Y)$ will approach those of MCMC estimators for
large values of $k$ and $m$.
This motivates the use of the estimator $H_{k:m}(X,Y)$ with $m>k$, which allows us to control the loss of efficiency associated with the removal of burn-in bias in contrast with the basic
estimator $H_k(X,Y)$ of Section \ref{subsec:Basic-construction}.
We discuss the choice of $k$ and $m$ in further detail in Section \ref{sec:theory} and in the subsequent experiments.
A variant of \eqref{eq:averageestimator} can be obtained by considering a time lag greater than one between
the two chains $(X_t)_{t \geq 0}$ and $(Y_t)_{t \geq 0}$, with the meeting time defined as the first time $t$ for which $\{X_t = Y_{t-\text{lag}}\}$ occurs. This introduces another tuning parameter but is found to be fruitful in \citet{biswas2019estimating}.

We conclude this section with a few remarks on practical implementations.
First, the test function $h$ does not have to be specified at run-time in
Algorithm \ref{alg:Unbiased-estimator-avg}.  One can store the coupled chains and choose the test function later.
Also, one typically resorts to thinning the output of an MCMC sampler if the memory cost of
storing chains is prohibitive, or if the cost of evaluating the test function
of interest is significant compared to the cost of each MCMC iteration
\citep[e.g.][]{owen2017statistically}.  This is feasible in the proposed
framework: one could consider a variation of Algorithm
\ref{alg:Unbiased-estimator-avg} where each call to the Markov kernels $P$ and
$\bar{P}$ would be replaced by multiple calls to them.  We also observe that
the proposed estimators can take values outside of the range of the test
function $h$; for instance they can take negative values even if the
range of the test function contains only non-negative values. 

Finally, we stress the difficulty inherent in choosing an initial distribution
$\pi_0$. The estimators are unbiased for any choice of $\pi_0$, including point
masses, but this choice has an impact on both
the computing cost and the variance.  There is also a choice about whether to
draw $X_0$ and $Y_0$ independently from $\pi_0$ or not; in
our experiments we use independent draws.  We will see in Section
\ref{subsec:Random-Walk-Metropolis-bimodal} that unfortunate choices of initial
distributions can severely affect the performance of the proposed estimators. This
suggests trying more than one choice of initialization, especially in
the setting of multimodal targets.  Overall the choice of $\pi_0$ and its relative importance compared to standard MCMC
are open questions.

\subsection{Signed measure estimator \label{sec:signedmeasure}}

We can formulate the proposed estimation procedure in terms of a signed measure $\hat{\pi}$ defined by
\begin{align}\label{eq:piestimator}
    \hat{\pi} & = \frac{1}{m-k+1}\sum_{\ell=k}^{m} \delta_{X_{\ell}} + 
    \sum_{\ell=k+1}^{\tau - 1}\min\left(1, \frac{\ell-k}{m-k+1}\right)( \delta_{X_{\ell}} - \delta_{Y_{\ell-1}}),
\end{align}
obtained by replacing test function evaluations by delta masses in \eqref{eq:averageestimator},
as in Section 4 of \citet{glynn2014exact}. 
The measure $\hat{\pi}$ is of the form
    $\hat{\pi} = \sum_{\ell = 1}^N \omega_\ell \delta_{Z_\ell}$
    where the weights satisfy $\sum_{\ell=1}^N  \omega_\ell = 1$
and where the atoms $(Z_\ell)$ are values 
among the history of the coupled chains. Some of the weights $(\omega_\ell)$
may be negative, making $\hat{\pi}$ a signed empirical measure. In this view the unbiasedness property
states $\mathbb{E}[\sum_{\ell = 1}^N \omega_\ell h(Z_\ell)]= \mathbb{E}_\pi[h(X)]$ for a test function $h$.

One can consider the convergence behavior of $\hat{\pi}^R = R^{-1} \sum_{r=1}^R \hat{\pi}^{(r)}$ 
towards $\pi$, where $(\hat{\pi}^{(r)})$ for $r\in \{1,\ldots,R\}$ are  independent replications of $\hat{\pi}$.
\citet{glynn2014exact} obtain a Glivenko--Cantelli result for a similar measure 
related to their estimator. In the current setting, assume for simplicity that $\pi$ is univariate or else
consider only one of its marginals. 
To emphasize the importance of the number of replications $R$, we rewrite the weights
and atoms as $\hat{\pi}^R =
\sum_{\ell = 1}^{N_R} \omega_\ell \delta_{Z_\ell}$.  Introduce the function $s\mapsto
\hat{F}^R(s) = \sum_{\ell = 1}^{N_R} \omega_\ell \mathds{1}(Z_{\ell} \leq
s)$ on $\mathbb{R}$.
Proposition \ref{prop:glivenko} below states that $\hat{F}^R$ converges to $F$ as $R\to\infty$ uniformly with probability one, 
where $F$  is the cumulative distribution function of $\pi$.

The function $s\mapsto \hat{F}^R(s)$ is not monotonically increasing because of negative weights among $(\omega_\ell)$,
which motivates the following comments regarding the estimation of quantiles of $\pi$.
Assume from now on that the pairs $(\omega_\ell, Z_\ell)$ are ordered such that $Z_\ell \leq Z_{\ell+1}$.
For any $q\in(0,1)$ there might more than one index $\ell$
such that $\sum_{i=1}^{\ell-1} \omega_i \leq q$ and $\sum_{i=1}^{\ell} \omega_i > q$; the quantile estimate
might be defined as $Z_\ell$ for any such $\ell$.
The convergence of $\hat{F}^R$ to $F$ indicates that all such estimates are expected to converge to the $q$-th quantile of
$\pi$. Therefore the signed measure representation leads to a way of estimating quantiles of the target distribution
in a consistent way as $R\to \infty$. The construction of confidence intervals for these quantiles, perhaps 
by bootstrapping the $R$ independent copies, stands as an interesting area for future research.
Another route to estimate quantiles of $\pi$ would be to project marginals of $\hat{\pi}^R$ onto the space of
probability measures, for instance using a generalization of the Wasserstein metric to signed measures \citep{mainini2012description}. 
One could also estimate $F$ using isotonic regression \citep{chatterjee2015risk},
considering $\hat{F}^R(s)$ for various values $s$ as noisy measurements of $F(s)$.

\section{Properties and parallel implementation \label{sec:theory}}

The proofs of the results of this section are in the supplementary materials.
Our first result establishes the basic validity of the proposed estimators. 

\begin{prop}
\label{prop:unbiased}Under Assumptions \ref{assumption:marginaldistributions}-\ref{assumption:sticktogether},
for all $k\geq 0$ and $m\geq k$, the estimator $H_{k:m}(X,Y)$ has expectation $\mathbb{E}_\pi[h(X)]$, a finite variance,
and a finite expected computing time.
\end{prop}

A direct consequence of Proposition \ref{prop:unbiased}
is that an average of $R$ independent copies of $H_{k:m}(X,Y)$ converges to $\mathbb{E}_\pi[h(X)]$ as $R\to \infty$.
We discuss more sophisticated results on unbiased estimators and parallel processing in Section \ref{sec:glynnheidelberger} and other uses of such estimators in Sections \ref{subsec:cut-distribution} and \ref{sec:discussion}.
Following \citet{glynn2014exact}, we provide Proposition \ref{prop:glivenko} on the signed measure estimator of \eqref{eq:piestimator}. We recall that such estimators apply to univariate target distributions or to the marginal distributions of a multivariate target. 

\begin{prop}
\label{prop:glivenko} Under Assumptions \ref{assumption:meetingtime}-\ref{assumption:sticktogether},
for all $m\geq k \geq 0$,
and assuming that $(X_t)_{t\geq 0}$ converges to $\pi$ in total variation, 
introduce the function $s\mapsto \hat{F}^R(s) = \sum_{\ell = 1}^{N_R} \omega_\ell \mathds{1}(Z_{\ell} \leq s)$,
where $(\omega_\ell, Z_{\ell})_{\ell=1}^{N_R}$ are weighted atoms obtained from $R$ independent copies of $\hat{\pi}$ in \eqref{eq:piestimator}.
Denote by $F$ the cumulative distribution function of $\pi$. Then 
$\sup_{s\in\mathbb{R}} |\hat{F}^R(s) - F(s)| \xrightarrow[R\to\infty]{} 0 $ almost surely.
\end{prop}

Section \ref{subsec:variance} studies the variance and efficiency of $H_{k:m}(X,Y)$,
Section \ref{subsec:Bounding-coupling-probabilities} concerns 
the verification of Assumption \ref{assumption:meetingtime} using drift conditions,
and Section \ref{sec:glynnheidelberger} discusses estimation on parallel processors in the presence of a budget constraint.

\subsection{Variance and efficiency \label{subsec:variance}}

We consider the impact of $k$ and $m$ on the efficiency of the proposed estimators, which will then suggest guidelines for the choice of these tuning parameters.
Estimators $H^{(r)}_{k:m}(X,Y)$, for $r=1,\ldots,R$, can be generated independently 
and averaged. More estimators can be produced in a given computing budget if each estimator is cheaper to produce. 
The trade-off can be understood in the framework of \citet{glynn1992asymptotic},
see also \citet{Rhee:Glynn:2012,glynn2014exact},
by defining the asymptotic inefficiency as the product of the variance and expected cost of the estimator.
That product is the asymptotic variance of $R^{-1}\sum_{r=1}^R H^{(r)}_{k:m}(X,Y)$
as the computational budget, as opposed to the number of estimators $R$, goes to infinity \citep{glynn1992asymptotic}.
Of primary interest is the comparison of this asymptotic inefficiency 
with the asymptotic variance of standard MCMC estimators. 
We start by writing the time-averaged estimator of \eqref{eq:averageestimator} 
as 
\begin{align*}
    H_{k:m}(X,Y) &=  \text{MCMC}_{k:m} + \text{BC}_{k:m},
\end{align*}
where $\text{MCMC}_{k:m}$ is the MCMC average $(m-k+1)^{-1} \sum_{\ell = k}^m h(X_\ell)$ and $\text{BC}_{k:m}$ is the bias correction term.
The variance of $H_{k:m}(X,Y)$ can be written
\begin{align*}
    \mathbb{V}[H_{k:m}(X,Y)] &= \mathbb{E}\left[(\text{MCMC}_{k:m} - \mathbb{E}_\pi[h(X)])^2 \right]
    + 2 \mathbb{E}\left[(\text{MCMC}_{k:m} - \mathbb{E}_\pi[h(X)]) \text{BC}_{k:m} \right]
    + \mathbb{E}\left[\text{BC}_{k:m}^2 \right].
\end{align*}
Defining the mean squared error of the MCMC estimator as $\text{MSE}_{k:m}=\mathbb{E}\left[(\text{MCMC}_{k:m} - \mathbb{E}_\pi[h(X)])^2 \right]$,
the Cauchy--Schwarz inequality yields
\begin{equation} \label{eq:varianceHkmbound}
    \mathbb{V}[H_{k:m}(X,Y)] \leq \text{MSE}_{k:m} 
    + 2 \sqrt{\text{MSE}_{k:m}} \sqrt{\mathbb{E}\left[\text{BC}_{k:m}^2 \right]}
    + \mathbb{E}\left[\text{BC}_{k:m}^2 \right].
\end{equation}
To bound $\mathbb{E}[\text{BC}_{k:m}^2]$, we introduce a geometric drift condition on the Markov kernel $P$.

\begin{assumption}\label{assumption:drift}
    The Markov kernel $P$ is $\pi$-invariant, $\varphi$-irreducible and aperiodic, and there exists a measurable function 
    $V:\;\mathcal{X}\to [1,\infty)$, $\lambda\in (0,1)$, $b<\infty$ and a small set $\mathcal{C}$ such that
        for all $x\in \mathcal{X}$,
    \[\int P(x,dy) V(y) \leq \lambda V(x) + b\mathds{1}(x \in \mathcal{C}).\]
\end{assumption}

We refer the reader to \cite{meyn:tweedie:1993} for the definitions and core theoretical tools for working with Markov chains on a general state space, in particular Chapter 5 for aperiodicity,
$\varphi$-irreducibility and small sets, and Chapter 15 for geometric drift
conditions; see also \citet{roberts2004general}. 
Geometric drift conditions are known to hold for various MCMC
algorithms \citep[e.g.][]{roberts1996exponential,roberts1996geometric,jarner:hansen00,atchade05,khare2013,choi2013polya,pal2014}.
Assumption \ref{assumption:drift} often plays a central role in establishing geometric ergodicity  \citep[e.g. Theorem 9 in][]{roberts2004general}.
We show next that this assumption enables an informative bound 
on $\mathbb{E}[\text{BC}_{k:m}^2]$.

\begin{prop}\label{prop1}
    Suppose that Assumptions \ref{assumption:meetingtime}-\ref{assumption:sticktogether} and \ref{assumption:drift} hold,
    with a function $V$ for which the integral $\int V(x) \pi_0(dx)$ is finite. If the
    function $h$ is such that $\sup_{x\in\mathcal{X}} |h(x)|/V(x)^\beta<\infty$
    for some $\beta\in [0,1/2)$, then for all $m \geq k\geq 0$ we have
        \[\mathbb{E}[\text{BC}_{k:m}^2] \leq \frac{C_{\delta,\beta} \delta_\beta^k}{(m-k+1)^2}, \]
    for some constants $C_{\delta,\beta} < +\infty$, and $\delta_\beta = \delta ^{1-2\beta} \in (0,1)$, with $\delta\in(0,1)$ as in Assumption \ref{assumption:meetingtime}.
\end{prop}

Using Proposition \ref{prop1}, equation \eqref{eq:varianceHkmbound} becomes 
\begin{equation} \label{eq:varianceHkmbound2}
    \mathbb{V}[H_{k:m}(X,Y)] \leq \text{MSE}_{k:m} 
    + 2 \sqrt{\text{MSE}_{k:m}} \frac{\sqrt{C_{\delta,\beta} \delta_\beta^{k}}}{m-k+1}
    +\frac{C_{\delta,\beta} \delta_\beta^k}{(m-k+1)^2}.
\end{equation}
The variance of $H_{k:m}(X,Y)$ is thus
bounded by the mean squared error of an MCMC estimator plus
additive terms that vanish geometrically in $k$ and polynomially in $m-k$. 

In order to facilitate the comparison between the efficiency of $H_{k:m}(X,Y)$
and that of MCMC estimators, we add simplifying assumptions.
First, the right-most terms of \eqref{eq:varianceHkmbound2} decrease geometrically with $k$, 
at a rate driven by $\delta_\beta = \delta^{1-2 \beta}$ where $\delta$ is as in Assumption \ref{assumption:meetingtime}.
This motivates a choice of $k$ depending on the distribution of the meeting time $\tau$. In practice, we can 
sample independent realizations of the meeting time and choose $k$ 
such that $\mathbb{P}(\tau > k)$ is small; i.e. we choose $k$ as a large quantile of the meeting times.

Dropping the third term on the right-hand side of \eqref{eq:varianceHkmbound2},
which is of smaller magnitude than the second term, assuming that $\text{MSE}_{k:m}>0$ and that $m> \tau$ with large probability, we obtain the approximate inequality 
\begin{multline*}
    \mathbb{E}[2(\tau-1) + \max(1, m+1-\tau)] \mathbb{V}[H_{k:m}(X,Y)] \lessapprox \left(m + \mathbb{E}(\tau)\right)\mathbb{V}[H_{k:m}(X,Y)]\\
    \lessapprox (m-k+1) \text{MSE}_{k:m} \left(1 + \frac{k+\mathbb{E}(\tau)}{m-k+1}\right)\left(1 + \frac{2}{\sqrt{m-k+1}} \sqrt{\frac{C_{\delta,\beta}\delta_\beta^k}{(m-k+1)\text{MSE}_{k:m}}}\right).
\end{multline*}

As $k$ increases we expect $(m-k+1) \text{MSE}_{k:m}$ 
to converge to $\mathbb{V}[(m-k+1)^{-1/2}\sum_{t=k}^m h(X_t)]$,
where $X_k$ would be distributed according to $\pi$. Denote this variance by $V_{k,m}$.
The limit of $V_{k,m}$ as $m\to \infty$ 
is the asymptotic variance of the MCMC estimator, denoted by $V_\infty$. Hence, for $k$ and $m-k$ 
both large, the loss of efficiency of the method compared to standard MCMC is 
approximately $1 + (k+\mathbb{E}(\tau))/(m-k)$.

This informal series of approximations 
suggests that we can retrieve an asymptotic efficiency comparable to the underlying MCMC 
estimators with appropriate choices of $k$ and $m$ that depend on the distribution of the meeting time $\tau$.
These choices are thus sensitive to the coupling of the chains, and not only to the performance of the underlying MCMC algorithm.
Choosing $m$ as a multiple of $k$, such as $5k$ or $10k$, makes intuitive sense when considering that $k/m$ is the proportion
of iterations that are simply discarded in the event that $\tau<k$.
In other words, the bias of MCMC can be removed at the cost of an increased variance,
which can in turn be reduced by choosing large enough values of $k$ and $m$.
This results in a tradeoff with the desired level of parallelism:
one might prefer to keep $k$ and $m$ small, yielding a suboptimal efficiency for $H_{k:m}(X,Y)$,
but enabling more independent copies to be generated in a given computing time.

\subsection{Verifying Assumption \ref{assumption:meetingtime} \label{subsec:Bounding-coupling-probabilities}}

We discuss how Assumption \ref{assumption:drift} on the Markov kernel $P$ can be used to verify Assumption \ref{assumption:meetingtime}, on the shape of the meeting time distribution. Informally, Assumption
\ref{assumption:drift} guarantees that the bivariate chain
$\{(X_t,Y_{t-1}),\;t\geq 1\}$ visits $\Cset\times\Cset$ infinitely often, where $\Cset$ is a small set.
If there is a positive probability of the event $\{X_{t+1} =Y_t\}$
for every $t$ such that 
$(X_t,Y_{t-1})\in\Cset\times\Cset$, then we expect Assumption
\ref{assumption:meetingtime} to hold. The next result formalizes that
intuition. The proof is based on a modification of an argument by
\cite{douc:moulines:rosenthal:04}. We introduce $\D=
\{(x,y)\in\mathcal{X}\times \mathcal{X}:\; x=y\}$. Then Assumption
\ref{assumption:sticktogether} reads $\bar P((x,x),\D) =1$ for all $x\in
\mathcal{X}$.

\begin{prop}\label{prop:vgeometric}
    Suppose that $P$ satisfies Assumption \ref{assumption:drift} with a small set $\mathcal{C}$
    of the form $\mathcal{C} = \{x: V(x)\leq L\}$ where $\lambda + b/(1+L) < 1$. 
    Suppose also that there exists $\epsilon\in (0,1)$ such that
\begin{equation}\label{mino}
\inf_{(x,y)\in\mathcal{C}\times\mathcal{C}} \bar P((x,y),\D) \geq \epsilon.
\end{equation}
Then there exists a finite constant $C'$ and a $\kappa \in (0,1)$, such that for all $n\geq 1$,
\[\mathbb{P}(\tau>n)\leq C' \pi_0(V) \kappa^n,\]
where $\pi_0(V) = \int V(x) \pi_0(dx)$.
Hence Assumption \ref{assumption:meetingtime} holds as long as $\pi_0(V)<\infty$.
\end{prop}

Note that if Assumption \ref{assumption:drift} holds with a small set of the
form $\Cset = \{x:\;V(x)\leq L\}$ for some $L>0$, then it also holds for $\Cset
=\{x:\;V(x)\leq L'\}$ for all $L'\geq L$. In that case one can always choose
$L$ large enough so that  $\lambda + b/(1+L) < 1$.  Hence the main
restriction in Proposition \ref{prop:vgeometric} is the assumption that the
small sets in Assumption \ref{assumption:drift} are of the form $ \{x: V(x)\leq
L\}$, i.e. level sets of $V$.  This is known to be true in some cases. For
instance it is known from Theorem 2.2 of \citet{roberts1996geometric} that for a
large class of Metropolis-Hastings algorithms, any non-empty compact set is a
small set, and therefore for these algorithms it suffices to check that the
level sets of the drift function $V$ are compact.  Common examples of drift
functions include $V(x) = c/\sqrt{\pi(x)}$
\citep{roberts1996geometric,jarner:hansen00,atchade05},  $V(x) = ce^{b|x|}$
\citep{roberts1996exponential} or the example in \citet{pal2014}, which all have
compact level sets under mild regularity conditions.

The work of \citet{middleton2018unbiased} contains results 
that generalize Propositions \ref{prop1} and \ref{prop:vgeometric}
to Markov chains satisfying polynomial drift
conditions \citep[e.g][]{andrieu2015convergence}, leading to polynomial tails
for the associated meeting times.

\subsection{Parallel implementation under budget constraints\label{sec:glynnheidelberger}}

Our main motivation for unbiased estimators comes from parallel processing;
see Sections \ref{subsec:cut-distribution} and \ref{sec:discussion} for other motivations.
Independent unbiased estimators with finite variance can be generated on separate machines,
and combined into consistent and asymptotically Normal estimators.
If the number of estimators is pre-specified, this follows from the central limit theorem for i.i.d. variables.
We might prefer to specify a time budget,
and generate as many estimators as possible within the budget.
The lack of bias allows the application of a variety of results on
budget-constrained parallel simulations, which we briefly review here, following
\citet{Glynn1990,Glynn1991}. 

We denote the proposed estimator by $H$ and its expectation, which is the object of interest here, by $\pi(h)$. Generating $H$ takes a random time $C$.
We write $N(t)$ for
 the number of independent copies of $H$ that can be produced by time $t$.
The sequence $(H_n,C_n)_{n\in \mathbb{N}}$ refers to i.i.d. copies of 
$(H,C)$, so that we can write $N(t) = \sup \{n\geq 0: C_{1}+\ldots+C_{n} \leq t\}$, with $N(t) = 0$ if $t<C_1$.
We add the subscript $\mathrm{p}$ to refer to objects associated with processor $\mathrm{p}\in \{1,\ldots,\mathrm{P}\}$. 

A first result is that the estimator $\bar{H}_{\mathrm{p}}(t)$, defined for all $1\leq \mathrm{p} \leq \mathrm{P}$ as $0$ if $N_\mathrm{p}(t) = 0$,
and by the sample average of $H_{\mathrm{p}1},\ldots,H_{\mathrm{p}N_\mathrm{p}(t)}$ otherwise, is biased:
$\mathbb{E}[\bar{H}_\mathrm{p}(t)] = \mathbb{E}[H] - \mathbb{E}[H\mathds{1}(C > t)]$.
Corollary 6 of \citet{Glynn1990} states that 
if $\mathbb{E}[|H \exp(\alpha C)|]<\infty$ 
for some $\alpha>0$, then the bias is negligible compared to $\exp(-\alpha t)$ as $t \to \infty$.
By Cauchy--Schwarz,
$\mathbb{E}[|H \exp(\alpha C)|]^2$ is less than the product of 
$\mathbb{E}[H^2]$ and $\mathbb{E}[\exp(2 \alpha C)|]$. 
In our context, 
$\mathbb{E}[H^2]$ is finite under Proposition \ref{prop:unbiased}, and 
$\mathbb{E}[\exp(2 \alpha C)|]$ is finite for a range of values of $\alpha$ 
that depends on the value of $\delta$ in Assumption \ref{assumption:meetingtime}.

We can define an unbiased estimator of $\pi(h)$ with a slight
modification of $\bar{H}_{\mathrm{p}}(t)$.  For all ${\mathrm{p} \in \{1,\dots,\mathrm{P}\}}$, set
$\tilde{H}_\mathrm{p}(t) = \bar{H}_\mathrm{p}(t)$ if $N_\mathrm{p}(t)>0$ and $\tilde{H}_\mathrm{p}(t) = H_{\mathrm{p}1}$ if
$N_\mathrm{p}(t)=0$.  
With $\tilde{N}_\mathrm{p}(t) =  \max(1,N_\mathrm{p}(t))$, then $\tilde{H}_\mathrm{p}(t)$ is the sample average of $H_{\mathrm{p}1},\ldots,H_{\mathrm{p}\tilde{N}_\mathrm{p}(t)}$.
The computation of $\tilde{N}_\mathrm{p}(t)$ requires the completion of $H_{\mathrm{p}1}$,
and thus we cannot necessarily return $\tilde{H}_\mathrm{p}(t)$ at time $t$, in contrast with $\bar{H}_\mathrm{p}(t)$.
On the other hand, we have $\mathbb{E}[\tilde{H}_{\mathrm{p}}(t)] =
\mathbb{E}[H]=\pi(h)$, i.e. the estimator is unbiased, provided that
$\mathbb{E}[|H|]<\infty$ (Corollary 7 of \citet{Glynn1990}).  We denote the
average of $\tilde{H}_\mathrm{p}(t)$ over $\mathrm{P}$ processors by
$\tilde{H}(\mathrm{P},t) = \mathrm{P}^{-1}\sum_{\mathrm{p}=1}^\mathrm{P} \tilde{H}_{\mathrm{p}}(t)$, which is unbiased for $\pi(h)$.

Asymptotic results on $\tilde{H}(\mathrm{P},t)$ can be found in \citet{Glynn1991} and are summarized below. We
first have the consistency results:
$\lim_{t\to \infty} \tilde{H}(\mathrm{P},t) = \lim_{\mathrm{P}\to \infty} \tilde{H}(\mathrm{P},t) = \pi(h)$ almost surely for all $t$ and $\mathrm{P}$,
and if $\mathbb{E}[|H|^{1+\delta}]$ for some $\delta>0$ and if $\{t_\mathrm{P}\}$ is a sequence such that $\lim_{\mathrm{P}\to\infty} t_\mathrm{P}= \infty$,
then $\tilde{H}(\mathrm{P},t_\mathrm{P})$ converges to $\pi(h)$ in probability as $\mathrm{P}\to\infty$.
Next, we can construct confidence intervals for $\pi(h)$ based on $\tilde{H}(\mathrm{P},t)$,
following the end of Section 3 in \citet{Glynn1991}. Indeed, define
\begin{align*}
    \hat{\sigma}_1^2(\mathrm{P},t) &= \frac{1}{\mathrm{P}-1}\sum_{\mathrm{p}=1}^\mathrm{P} \left(\tilde{H}_\mathrm{p}(t) - \tilde{H}(\mathrm{P},t) \right)^2,\quad
    \tilde{\tau}(\mathrm{P},t) =  \frac{1}{\mathrm{P}}\sum_{\mathrm{p}=1}^\mathrm{P} \frac{1}{\tilde{N}_\mathrm{p}(t)} \sum_{n=1}^{\tilde{N}_\mathrm{p}(t)} C_{\mathrm{p}n}, \\
    \hat{\sigma}_2^2(\mathrm{P},t) &= \tilde{\tau}(\mathrm{P},t) \left[\left(\frac{1}{\mathrm{P}}\sum_{\mathrm{p}=1}^\mathrm{P} \frac{1}{\tilde{N}_\mathrm{p}(t)} \sum_{n=1}^{\tilde{N}_\mathrm{p}(t)} H_{\mathrm{p}n}^2 \right)- \tilde{H}(\mathrm{P},t)^2 \right],
    \label{eq:parallelunbiasedestimators:variance}
\end{align*}
where $\tilde{N}_\mathrm{p}(t) = \max(1,N_\mathrm{p}(t))$. Then we have the three following central limit theorems, 
\begin{eqnarray}
    \text{for fixed $t$ and $\mathrm{P}\to\infty$},\quad \frac{\sqrt{\mathrm{P}}}{\hat{\sigma}_1(\mathrm{P},t)}\left(\tilde{H}(\mathrm{P},t) - \pi(h)\right) &\to& \mathcal{N}(0,1)\label{eq:parallelclt:fixedt},\\
    \text{for fixed $\mathrm{P}$ and $t\to\infty$},\quad \frac{\sqrt{\mathrm{P}t}}{\hat{\sigma}_2(\mathrm{P},t)}\left(\tilde{H}(\mathrm{P},t) - \pi(h)\right) &\to& \mathcal{N}(0,1),\label{eq:parallelclt:fixedp}\\
    \text{if $t_\mathrm{P} \to \infty$ as $\mathrm{P}\to\infty$},\quad \frac{\sqrt{\mathrm{P}t_\mathrm{P}}}{\hat{\sigma}_2(\mathrm{P},t_\mathrm{P})}\left(\tilde{H}(\mathrm{P},t_\mathrm{P}) - \pi(h)\right) &\to& \mathcal{N}(0,1).
    \label{eq:parallelclt}
\end{eqnarray}
These results require moment conditions such as $\mathbb{E}[\tilde{H}_\mathrm{p}(t)^2]<\infty$.
The central limit theorem in \eqref{eq:parallelclt:fixedt} will be used to construct confidence intervals in Sections \ref{sec:ising} and \ref{subsec:variableselection}.

We conclude this section with a remark on the setting where
$t$ is fixed and the number of processors $\mathrm{P}$ goes to infinity.  There, the time to obtain
$\tilde{H}(\mathrm{P},t)$ would typically increase with $\mathrm{P}$. Indeed at least one
estimator needs to be completed on each processor for $\tilde{H}(\mathrm{P},t)$ to be
available.  The completion time behaves as the maximum of
independent copies of the cost $C$.  Under Assumption
\ref{assumption:meetingtime}, the completion time for $\tilde{H}(\mathrm{P},t)$ has expectation behaving as $\log \mathrm{P}$
when $\mathrm{P}\to \infty$, for fixed $t$.
Other tail assumptions \citep{middleton2018unbiased}
would lead to different behavior for the completion time associated with $\tilde{H}(\mathrm{P},t)$.

\section{Couplings of MCMC algorithms \label{sec:Practical-couplings}}

We consider couplings of various MCMC algorithms that 
satisfy Assumptions \ref{assumption:meetingtime}-\ref{assumption:sticktogether}. 
These couplings are widely applicable and do not require extensive analytical
knowledge of the target distribution. We stress that they are not 
optimal in general, and we expect that other constructions would yield
more efficient estimators.  We begin in Section \ref{subsec:maximalcoupling} by reviewing maximal couplings.

\subsection{Sampling from maximal couplings \label{subsec:maximalcoupling}}

A maximal coupling between two distributions $p$ and $q$ on a
space $\mathcal{X}$ is a distribution of a pair of random variables $(X,Y)$ that maximizes
$\mathbb{P}(X=Y)$,
subject to the marginal constraints $X\sim p$
and $Y\sim q$. 
We write $p$ and $q$ both for these distributions and for their probability density functions
with respect to a common dominating measure, and refer to the uniform distribution on the interval
$[a,b]$ by $\mathcal{U}([a,b])$. A procedure to sample from a maximal
coupling is described in Algorithm \ref{alg:maximalcoupling};
see e.g. Section 4.5 of Chapter 1 of \citet{thorisson2000coupling},
and \citet{johnson1998coupling} where it is termed $\gamma$-coupling.

We justify Algorithm \ref{alg:maximalcoupling} and compute its cost.
Denote by $(X,Y)$ the output of the algorithm. First, $X$ follows $p$ from step 1.
To prove that $Y$ follows $q$, introduce a measurable
set $A$. We write
$\mathbb{P}(Y\in A)=\mathbb{P}(Y\in A,\text{step 1})+\mathbb{P}(Y\in A,\text{step 2})$,
where the events $\{\text{step 1}\}$ and $\{\text{step 2}\}$ refer
to the algorithm terminating at step 1 or 2. We compute
\begin{align*}
\mathbb{P}\left(Y\in A,\text{step 1}\right) & =\int_{A}\int_{0}^{+\infty}\mathds{1}\left(w\leq q(x)\right)\frac{\mathds{1}\left(0\leq w\leq p(x)\right)}{p(x)}p(x)dwdx
  =\int_{A}\min(p(x),q(x))dx.
\end{align*}
We can deduce from this that $\mathbb{P}\left(\text{step 1}\right)=\int_{\mathcal{X}}\min(p(x),q(x))dx$.
For $\mathbb{P}\left(Y\in A,\text{step 2}\right)$
to be equal to $\int_{A}(q(x)-\min(p(x),q(x)))dx$, we need 
\begin{align*}
\int_{A}(q(x)-\min(p(x),q(x)))dx
 &=\mathbb{P}\left(Y\in A|\text{step 2}\right)\left(1-\int_{\mathcal{X}}\min(p(x),q(x))dx\right),
\end{align*}
and we conclude that 
the distribution of $Y$ given $\{\text{step 2}\}$
should for all $x$ have a density $\tilde{q}(x)$ equal to ${(q(x)-\min(p(x),q(x)))/(1-\int\min(p(x^{\prime}),q(x^{\prime}))dx^{\prime})}$. Step 2 is a standard rejection
sampler using $q$ as a proposal distribution to target $\tilde{q}$,
which concludes the proof that $Y\sim q$.
We also confirm that Algorithm \ref{alg:maximalcoupling} maximizes the probability of $\{X = Y\}$. Under the algorithm,
\[
	\mathbb{P}(X=Y)= \mathbb{P}(\text{step 1}) = \int_{\mathcal{X}}\min(p(x),q(x))dx=1-d_{\text{TV}}(p,q),
\]
where $d_{\text{TV}}(p,q)=\nicefrac{1}{2}\int_{\mathcal{X}}|p(x)-q(x)|dx$
is the total variation distance. 
By the coupling inequality
\citep{lindvall2002lectures},
this proves that the algorithm implements a maximal coupling.

To assess the cost of Algorithm \ref{alg:maximalcoupling}, 
note that step $1$ costs one draw from
$p$, one evaluation from $p$ and one from $q$. 
Each attempt in the rejection sampler of step 2 costs one draw from $q$, one evaluation
from $p$ and one from $q$. Hereafter we refer to the cost of one draw and
two evaluations by ``one unit''. 
Observe that the probability of acceptance in step 2 is given by
$ \mathbb{P}(W^{\star}\geq p(Y^{\star})) = 1-\int_{\mathcal{X}}\min\left(p(y),q(y)\right)dy$.
Then, the number of attempts in step 2 has a Geometric distribution with mean $(1-\int_{\mathcal{X}}\min\left(p(y),q(y)\right)dy)^{-1}$,
and step 2 itself occurs with probability $1-\int_{\mathcal{X}}\min\left(p(y),q(y)\right)dy$.
Therefore the overall expected cost is two units.
The expectation of the cost is the same for all distributions $p$ and $q$,
while the variance of the cost depends on $d_\text{TV}(p,q)$, and
in fact goes to infinity as this distance goes to zero.

\begin{algorithm}
\begin{enumerate}
\item Sample $X\sim p$ and $W|X\sim\mathcal{U}([0,p(X)])$. If $W\leq q(X)$,
output $(X,X)$.
\item Otherwise, sample $Y^{\star}\sim q$ and $W^{\star}|Y^{\star}\sim\mathcal{U}([0,q(Y^{\star})])$
until $W^{\star}>p(Y^{\star})$, and output $(X,Y^{\star})$. 
\end{enumerate}
\caption{Sampling from a maximal coupling of $p$ and $q$. \label{alg:maximalcoupling}}
\end{algorithm}

In Algorithm \ref{alg:maximalcoupling}, the value of $X$ is not used in the
generation of $Y^\star$ within step 2.  In other words, conditional on $\{X\neq
Y\}$, the two output variables are independent.  We might
prefer to correlate the outputs in the event $\{X \neq Y\}$,
e.g. in random walk MH as in the next
section.  We describe a maximal coupling presented in 
\citet{bou2018coupling}. It applies to distributions $p$ and
$q$ on $\mathbb{R}^d$ such that $X\sim p$ can be represented as $X =
\mu_1 + \Sigma^{1/2} \dot{X}$, and $Y \sim q$ as $Y = \mu_2 +
\Sigma^{1/2} \dot{Y}$, where the pair $(\dot{X},\dot{Y})$ follows a coupling of
some distribution $s$ with itself. The construction requires that $s$ is spherically symmetrical:
$s(x) = s(y)$ for all $x,y \in \mathbb{R}^d$ such that $\|x\| = \|y\|$,
where $\|\cdot\|$ denotes the Euclidean norm. For instance, if $s$ is a standard multivariate Normal
distribution, then $X \sim \mathcal{N}(\mu_1,\Sigma)$ and $Y \sim
\mathcal{N}(\mu_2,\Sigma)$.

Let $z = \Sigma^{-1/2}(\mu_1 - \mu_2)$ and $e = z / \|z\|$.
We independently draw $\dot{X} \sim s$ and $U \sim \mathcal{U}([0,1])$ and let
\[
    \dot{Y} = \begin{cases}
        \dot{X} + z & \text{if } U \leq \min\left(1, \frac{s(\dot{X} + z)}{s(\dot{X})}\right), \\
        \dot{X} - 2 (e^\prime \dot{X}) e & \text{otherwise}.
\end{cases}
\]
The above procedure outputs a pair $(\dot{X}, \dot{Y})$ that follows a coupling of $s$ with itself.
We then define $(X,Y) = (\mu_1 + \Sigma^{1/2} \dot{X}, \mu_2 + \Sigma^{1/2} \dot{Y})$. 
On the event $\{\dot{Y} = \dot{X} + z\}$, we have $X = Y$. On the event $\{\dot{Y} \neq \dot{X} + z\}$, the vector $\dot{X} - 2 (e^\prime \dot{X}) e $
is the reflection of $\dot{X}$ through the hyperplane orthogonal to $e$ that passes through the origin.
We show that the output $(X,Y)$ follows a maximal coupling of $p$ and $q$, which we
refer to as a maximal coupling with reflection on the residuals, or a ``reflection-maximal coupling''.
First we show that $\dot{Y}$ follows $s$, closely following the argument in \citet{bou2018coupling}.
For a measurable set $B$, we compute
\[
\begin{aligned}
    \mathbb{P}(\dot{Y} \in B) 
    & = \int \mathds{1}_B(x + z) \min\left(1, \frac{s(x + z)}{s(x)}\right) s(x) dx \\
    & \quad + \int \mathds{1}_B(x - 2 (e^\prime x) e) \max\left(0, 1- \frac{s(x + z)}{s(x)}\right) s(x) dx.
\end{aligned}
\]
The first integral above becomes
$\int \mathds{1}_B(w) \min\left(s(w - z), s(w) \right) dw$,
after a change of variables $w := x + z$.
To simplify the second integral we make the change of variables $w := x - 2(e^\prime x) e$. 
Since this corresponds to a reflection with respect to a plane orthogonal to $e$, we have $dw = dx$, and $x = w - 2(e^\prime w) e$, thus
\[
\begin{aligned}
    & \int \mathds{1}_B(x - 2 (e^\prime x) e) \cdot \max\left(0, s(x) - s(x + z)\right) d x \\
    & = \int \mathds{1}_B(w) \cdot \max\left(0, s(w) - s(w - z)\right) d w,
\end{aligned}
\]
where we have used $s(w - 2(e^\prime w) e) = s(w)$  and $s(w - 2(e^\prime w) e + z) = s(w-z)$,
respectively because 
$\|w - 2(e^\prime w) e\| = \|w\|$ and $\|w - 2(e^\prime w) e + z\|=\|w-z\|$.
Summing the two integrals we obtain ${\mathbb{P}(\dot{Y}\in B) = \int_B s(w)dw}$,
so that $\dot{Y}\sim s$.

To verify that the procedure corresponds to a maximal coupling of $p$ and $q$, we observe that
\[
\begin{aligned}
    \mathbb{P}(X \neq Y) & =
    \mathbb{P}(\dot{Y} \neq \dot{X} + z) 
 = 1 - \int \min \left(s(x), s(x+z) \right) d x \\
& = 1 - \int \min \left(s(\Sigma^{-1/2}(\tilde{x} - \mu_1)), s(\Sigma^{-1/2}(\tilde{x} - \mu_2)) \right) |\Sigma^{-1/2}| d \tilde{x},
\end{aligned}
\]
with the change of variable $\tilde{x}:= \mu_1 + \Sigma^{1/2} x$. The above is precisely the total variation distance
between $p$ and $q$, upon writing their densities in terms of the density of $s$.
Note that the computational cost associated with the above sampling technique is deterministic,
in contrast with the cost of Algorithm \ref{alg:maximalcoupling}.

Finally, for discrete distributions with common finite support, a procedure for
sampling from a maximal coupling is described in Section
\ref{subsec:variableselection}, with a cost that is also deterministic. 

\subsection{Metropolis--Hastings \label{subsec:Metropolis=002013Hastings}}

In Section \ref{subsec:exampleMH} we described a coupling of MH chains due to \citet{johnson1998coupling}; we summarize
the coupled kernel $\bar{P}((X_t,Y_{t-1}),\cdot)$ in the following procedure.
\begin{enumerate}
    \item Sample $(X^\star,Y^\star)| (X_t,Y_{t-1})$ from a maximal coupling of $q(X_t,\cdot)$ and $q(Y_{t-1},\cdot)$.
    \item Sample $U\sim \mathcal{U}([0,1])$.
    \item If $U\leq \min(1,\pi(X^\star)q(X^\star,X_t)/\pi(X_t)q(X_t,X^\star))$, then $X_{t+1}=X^\star$, otherwise $X_{t+1}=X_t$.
    \item If $U\leq \min(1,\pi(Y^\star)q(Y^\star,Y_{t-1})/\pi(Y_{t-1})q(Y_{t-1},Y^\star))$, then $Y_{t}=Y^\star$, otherwise $Y_{t}=Y_{t-1}$.
\end{enumerate}
Here we address the
verification of Assumptions \ref{assumption:marginaldistributions}-\ref{assumption:sticktogether} for this algorithm.
Assumption \ref{assumption:marginaldistributions} can be verified for
MH chains under conditions on the target
and the proposal \citep{nummelin:mcmc:2002,roberts2004general}.
In some settings the explicit drift function given in Theorem 3.2 of \citet{roberts1996geometric}
may be used to verify Assumption \ref{assumption:meetingtime} as in Section \ref{subsec:Bounding-coupling-probabilities}.
The probability of coupling at the next step given that the chains are in $X_t$ and $Y_{t-1}$
can be controlled as follows. First, the probability of proposing the same value $X^\star$ 
depends on the total variation distance between $q(X_{t},\cdot)$  and $q(Y_{t-1},\cdot)$,
which is typically strictly positive if $X_t$ and $Y_{t-1}$ are in bounded subsets of $\mathcal{X}$.
Furthermore, the probability of accepting $X^\star$
is often strictly positive on bounded subsets of $\mathcal{X}$, for instance when $\pi(x)>0$ for all $x\in \mathcal{X}$.
Assumption \ref{assumption:sticktogether} is satisfied by design thanks to the use of maximal couplings
and common uniform variable $U$ in the above procedure.

Different considerations drive the choice of proposal distribution in standard MCMC and in our proposed estimators. In the case of
random walk proposals with variance $\Sigma$, larger variances lead to
smaller total variation distances between $q(X_t,\cdot)$ and $q(Y_{t-1},\cdot)$
and thus larger probabilities of proposing identical values. However meeting
events only occur if proposals are accepted, which is unlikely if $\Sigma$ is
too large. This trade-off could lead to a different choice of $\Sigma$ than the 
optima known for the marginal chains \citep{roberts1997weak}, and deserves further investigation.

\begin{figure}
\begin{centering}
\subfloat[\label{fig:scaling:rwmh:maxcoupling} Maximum coupling.]{\begin{centering}
    \includegraphics[width=0.45\textwidth]{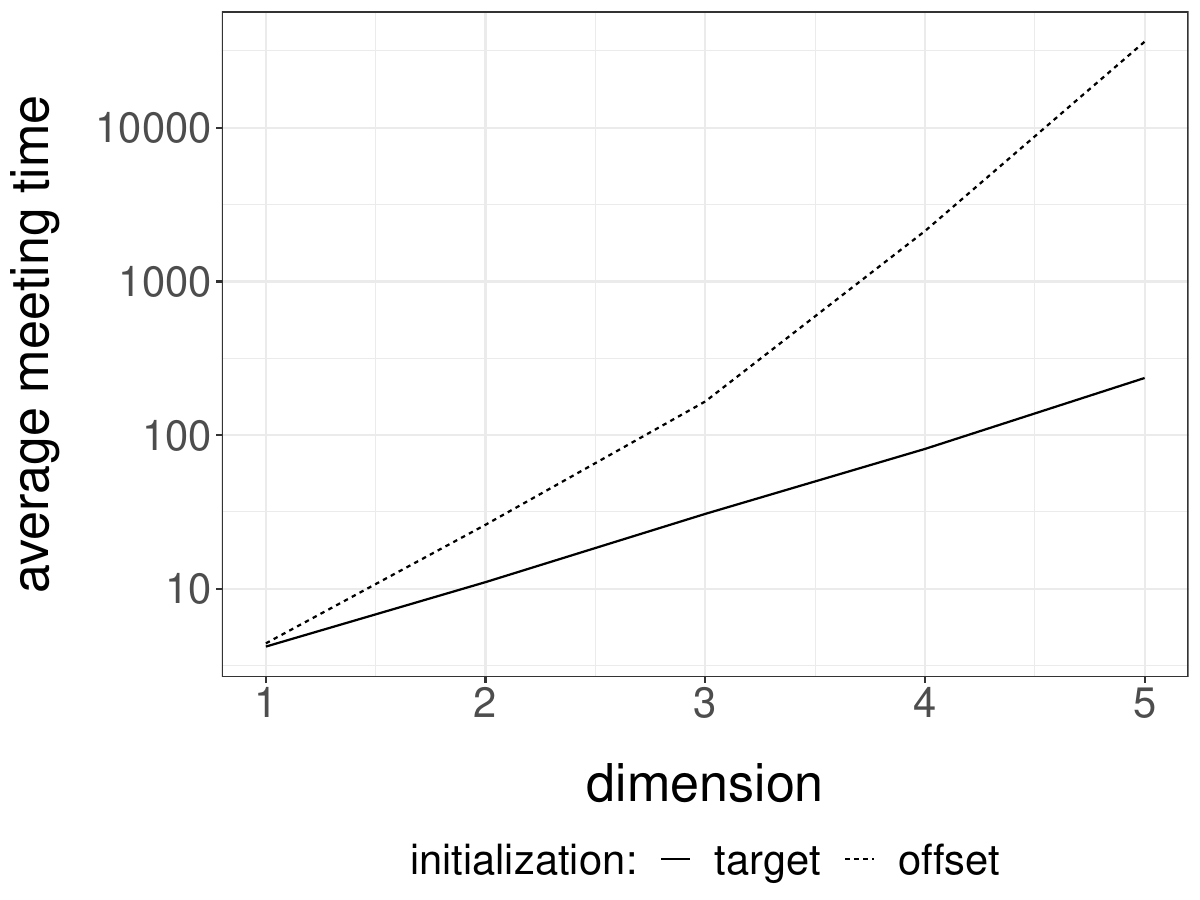}
\par\end{centering}
}\hspace*{1cm}
\subfloat[\label{fig:scaling:rwmh:reflmaxcoupling} Reflection-maximum coupling.]{\begin{centering}
    \includegraphics[width=0.45\textwidth]{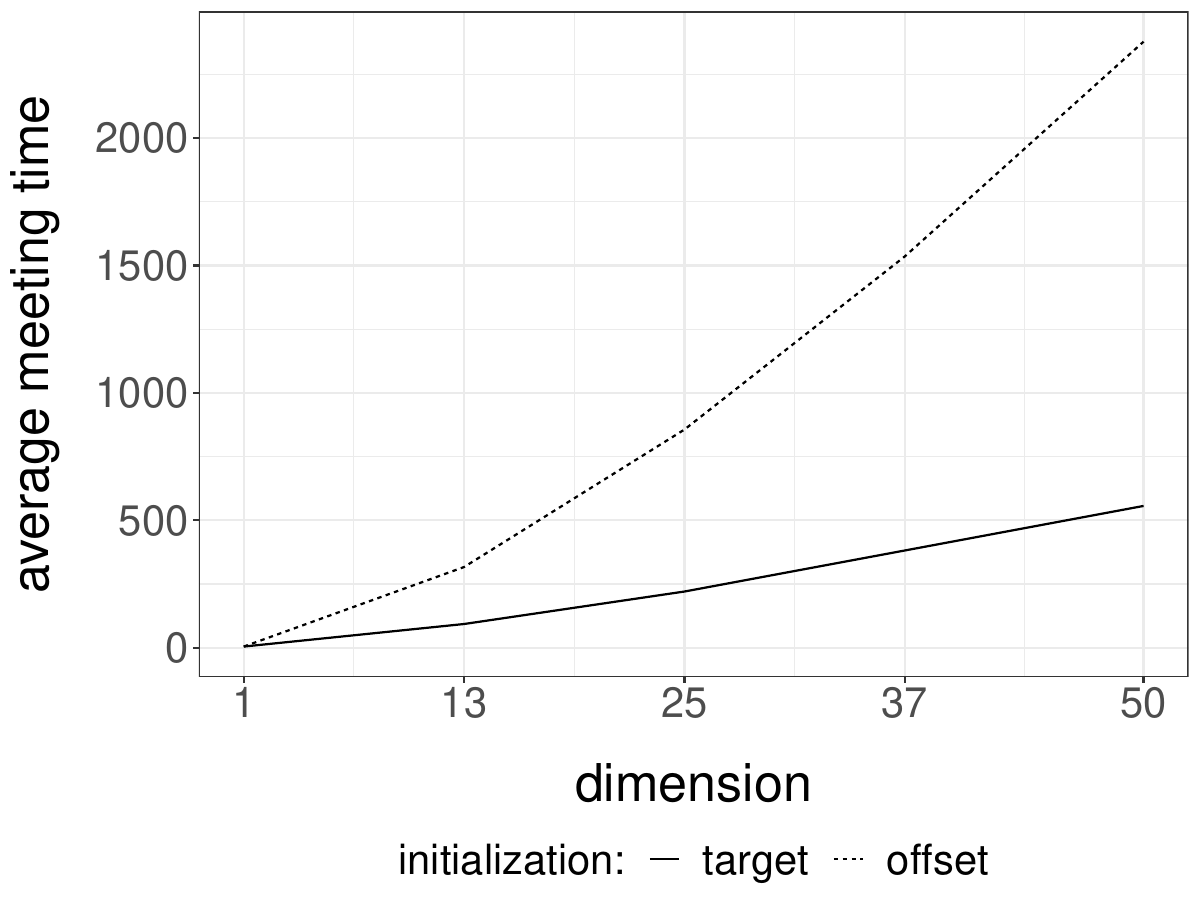}
\par\end{centering}
}
\par\end{centering}
\caption{\label{fig:scaling:rwmh}
Scaling of the average meeting time of a coupled MH algorithm with the dimension of the target $\mathcal{N}(0,V)$,
where $V$ is the inverse of a Wishart draw, as described in Section \ref{subsec:Metropolis=002013Hastings}.
The chains are either initialized from the target, or from a Normal $\mathcal{N}(1_d, I_d)$, where $1_d$ is a vector of ones (``offset'' in the legend).
Left: using maximal coupling of Algorithm \ref{alg:maximalcoupling}.
Right: using reflection-maximal coupling described in Section \ref{subsec:maximalcoupling}.
}
\end{figure}

We perform experiments with a $d$-dimensional Normal target distribution 
$\mathcal{N}(0, V)$, where $V$ is the inverse
of a matrix drawn from a Wishart distribution
with identity scale matrix and $d$ degrees of freedom. This setting, borrowed 
from \citet{hoffman2014no}, yields Normal targets with strong correlations and a dense precision matrix.
Below, each independent run is performed with an independent draw of $V$.
We consider Normal random walk proposals with variance $\Sigma$ set to $V/d$.
The division by $d$ heuristically follows from the scaling results of \citet{roberts1997weak}.
We initialize the chains either from the target distribution, 
or from a Normal centered at $(1,\ldots,1)$ with identity covariance matrix.
We first couple the proposals with a maximal coupling
given by Algorithm \ref{alg:maximalcoupling}. The resulting average meeting times, based on $1,000$ independent runs,
are given in Figure \ref{fig:scaling:rwmh:maxcoupling}. The plot indicates
an exponential increase of the average meeting times with the dimension, under both initialization strategies.
In passing, this illustrates that meeting times can be large even if the chains marginally start
at stationarity, i.e. in a setting where there is no burn-in bias.

Next we perform the same experiments with the reflection-maximum coupling
described in the previous section. The results are shown in Figure
\ref{fig:scaling:rwmh:reflmaxcoupling}.  The average meeting times now increase
at a rate that appears closer to linear in the dimension. This is to be compared
with established theoretical results on the linear performance of standard MH
estimators with respect to the dimension \citep{roberts1997weak}. 
A formal justification of the scaling observed in Figure \ref{fig:scaling:rwmh:reflmaxcoupling} is an open question,
and so is the design of more effective coupling strategies.

\subsection{Gibbs sampling \label{subsec:Gibbs}}

Gibbs sampling is another popular class of MCMC algorithms, in which 
components of a Markov chain are updated alternately by sampling from the target conditional distributions 
\citep[Chapter 10 of][]{robert:casella:2004}, implemented e.g. in the software packages JAGS \citep{plummer2003jags}.
In Bayesian statistics, these conditional distributions
sometimes belong to a standard family
such as Normal, Gamma, or Inverse Gamma. 
Otherwise, the conditional updates might require MH steps. 
We can introduce couplings in each conditional
update, using either maximal couplings of the target conditionals, if these are standard distributions, or maximal couplings of the proposal
distributions in MH steps targeting the target conditionals.
Controlling the probability of meeting at the next step over a set,
as required for the application of Proposition \ref{prop:vgeometric}, can be done on a case-by-case basis.
Drift conditions for Gibbs samplers also tend to rely on case-by-case arguments \citep[see e.g.][]{rosenthal1996analysis}.

\begin{figure}
\begin{centering}
\subfloat[\label{fig:scaling:gibbs:dense} With strong correlations.]{\begin{centering}
    \includegraphics[width=0.45\textwidth]{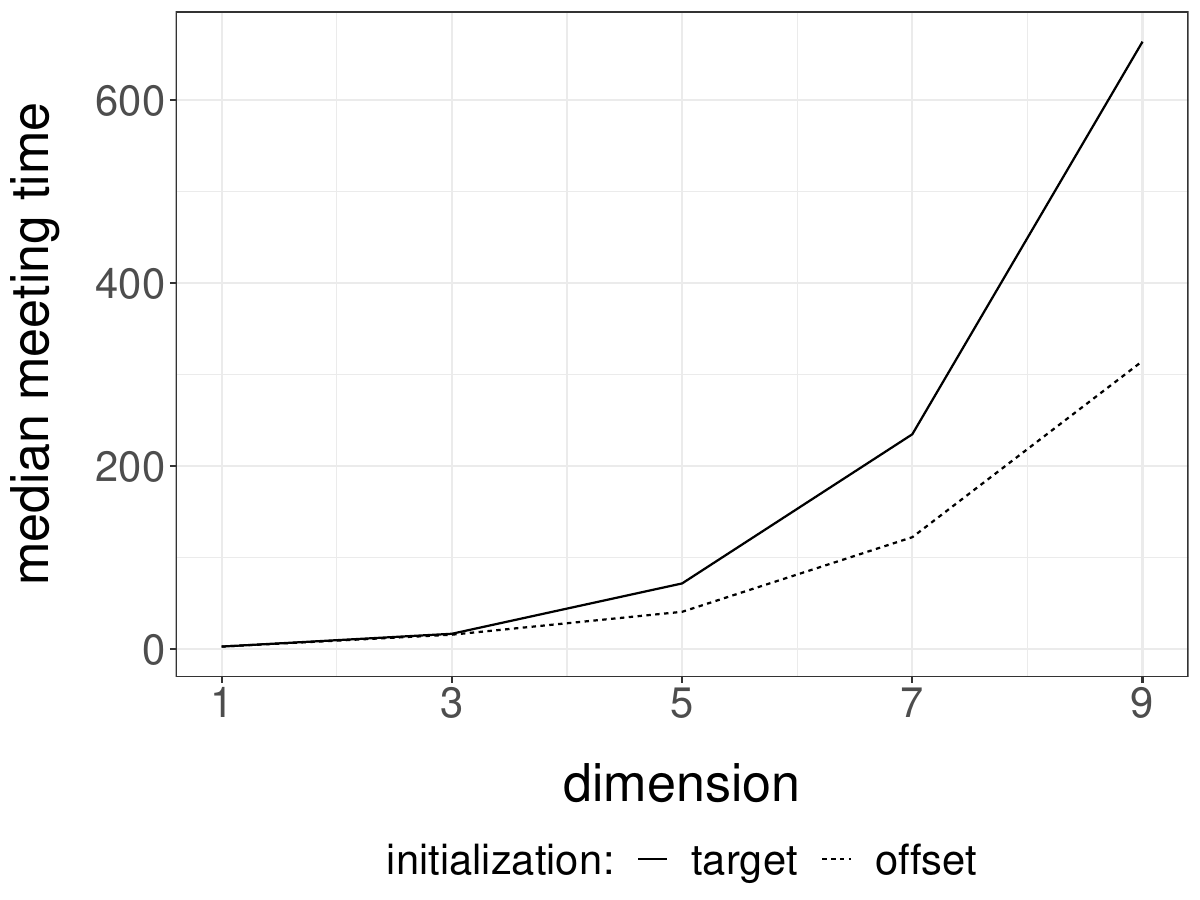}
\par\end{centering}
}\hspace*{1.0cm}
\subfloat[\label{fig:scaling:gibbs:sparse} With weak correlations.]{\begin{centering}
    \includegraphics[width=0.45\textwidth]{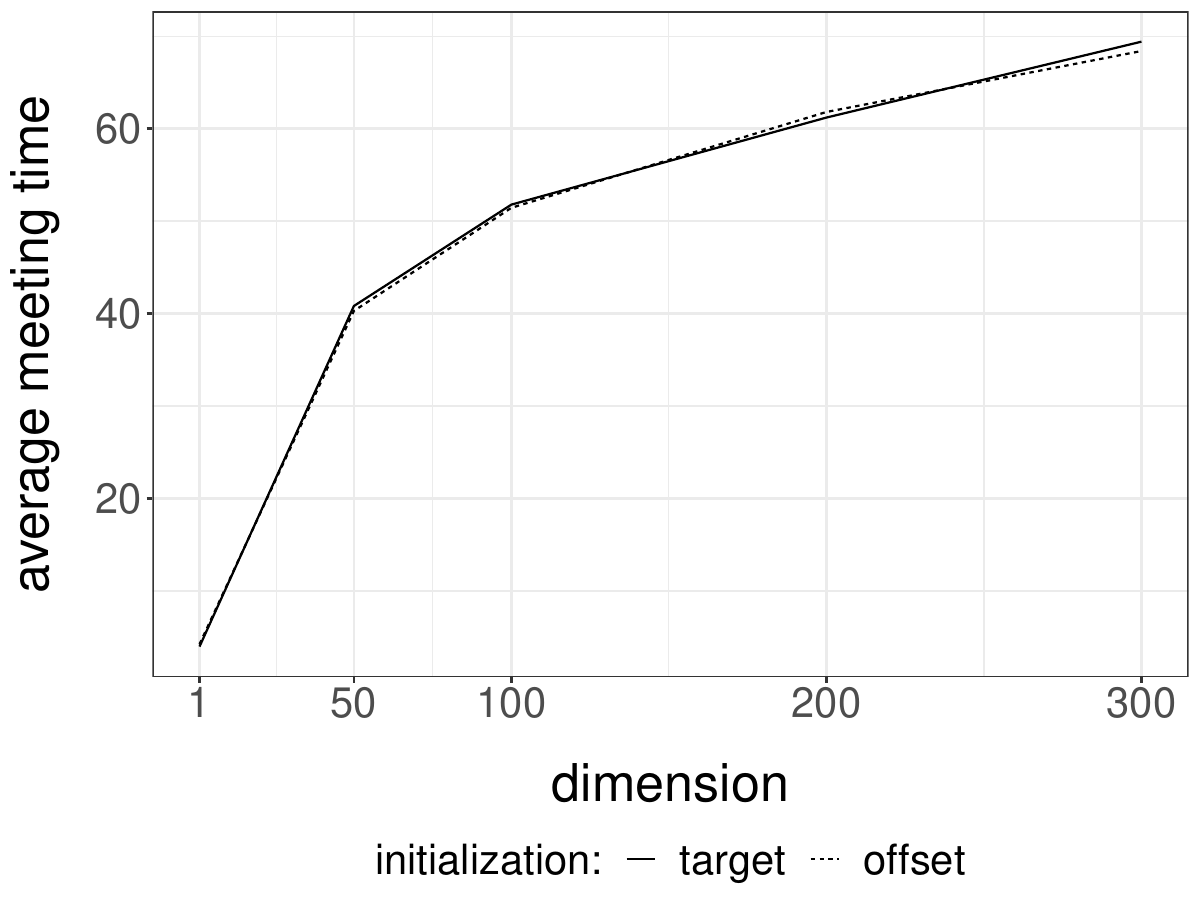}
\par\end{centering}
}\par\end{centering}
\caption{\label{fig:scaling:gibbs}
Scaling of the median (left) and average (right) meeting time of a coupled Gibbs algorithm with the dimension of the target $\mathcal{N}(0,V)$.
The chains are either initialized from the target, or from a Normal $\mathcal{N}(1_d, I_d)$, where $1_d$ is a vector of ones
(``offset'' in the legend).
Left: the covariance $V$ of the target is generated as the inverse of a Wishart sample, inducing strong correlations.
Right: the covariance $V$ is defined as $V_{ij} = 0.5^{-|i-j|}$, inducing weak correlations.
}
\end{figure}

Gibbs samplers tend to perform well for targets with weak correlations between the components being updated; otherwise Gibbs chains
are expected to mix poorly. We perform numerical experiments on Normal 
target distributions in varying dimensions to observe the effect of correlations on the meeting times of coupled Gibbs chains.
For each target $\mathcal{N}(0,V)$, we introduce an MH-within-Gibbs sampler, where
each univariate component $i$ is updated with a single Metropolis step, using Normal proposals 
with variance $V_{i,i}$. Here an iteration of the sampler refers to a complete scan of the components.
Figure \ref{fig:scaling:gibbs:dense} presents the median meeting times as a function of the dimension,
when $V$ is the inverse of a Wishart draw as in the previous section. 
In this highly correlated setting,
the meeting times scale poorly with the dimension. The plot presents the median instead of the average,
because we have stopped the runs after $500,000$ iterations; the median is robust to this truncation, but not the average.
We remark that shorter meeting times are obtained when initializing the chains away from the target distribution.

Next we consider a Normal target with covariance matrix $V$ defined by $V_{i,j} = 0.5^{-|i-j|}$, which induces weak correlations
among components; the inverse of $V$ is tridiagonal. In that case, the same Gibbs sampler performs much more favorably, as we can see from Figure \ref{fig:scaling:gibbs:sparse}.
The average meeting times seem to scale sub-linearly with the dimension, under both choices of initializations $\pi_0$.
Couplings of other Gibbs samplers will be encountered in the numerical experiments of Section \ref{sec:illustrations}.

\subsection{Coupling of other MCMC algorithms \label{subsec:otherMCMC}}

Among extensions of the MH algorithm, Metropolis-adjusted
Langevin algorithms \citep[e.g.][]{roberts1996exponential} are characterized
by the use of a proposal distribution given current state $X_t$ that is Normal with mean $X_t + h\nabla \log
\pi(X_t)/2$ and variance $h \Sigma$, with tuning parameter $h>0$ and
covariance matrix $\Sigma$. Maximal couplings or reflection-maximal couplings of the proposals
could be readily implemented to obtain faithful chains. Going further in the use of gradient information,
Hamiltonian or Hybrid Monte Carlo
\citep[HMC,][]{Duane:1987,Neal:1993,neal2011mcmc}
is a popular MCMC algorithm for large-dimensional targets. In  \citet{heng2017unbiased},
the framework of the present article is applied to pairs of Hamiltonian
Monte Carlo chains, with a focus of the verification of Assumptions \ref{assumption:marginaldistributions}-\ref{assumption:sticktogether} in that context. 
Such couplings are analyzed in detail in 
\citet{mangoubi2017rapid,bou2018coupling} to obtain convergence rates
for the underlying chains. We refer to \citet{heng2017unbiased}
for more details, and provide for completeness some experiments on the Normal target
described above in the supplementary materials.

The present article generalizes unbiased estimators obtained
by coupling conditional particle filters in \citet{jacob2017smoothing}. These 
algorithms, introduced in \citet{andrieu:doucet:holenstein:2010}, 
target the distribution of latent processes given observations and fixed parameters for
nonlinear state space models. The couplings
of conditional particle filters in \citet{jacob2017smoothing} involve a
 combination of common random numbers and maximal couplings. 
 Couplings of particle independent Metropolis--Hastings, which 
 is a particular case of Metropolis--Hastings with an independent proposal distribution,
 are simpler to design and considered in \citet{pmlr-v89-middleton19a}. 

The design of generic and efficient MCMC kernels is a topic of active ongoing
research \citep[see e.g.][and references
therein]{murray2010elliptical,goodman2010ensemble,pollock2016scalable,vanetti2017piecewise,titsias2017hamming}.
Any new kernel could lead to unbiased estimators with the proposed framework, as long as
appropriate couplings can be implemented. 

\section{Illustrations \label{sec:illustrations}}

Section \ref{subsec:Random-Walk-Metropolis-bimodal} illustrates
the impact of $k$, $m$, and the initial distribution $\pi_0$, identifying a
situation where some care is required.  Section
\ref{subsec:Gibbs-pump-failures} considers the removal of the bias from a Gibbs
sampler previously considered for perfect sampling and regeneration methods.
Section \ref{sec:ising} introduces an Ising model and a coupling of a replica
exchange algorithm, and we present experiments performed on parallel
processors.  Section \ref{subsec:variableselection} considers a
high-dimensional variable selection example, with an MH algorithm previously
shown to scale linearly with the number of variables.  Finally, Section
\ref{subsec:cut-distribution} focuses on the problem of approximating the cut
distribution arising in modular inference, which illustrates
the appeal of unbiased estimators beyond parallel computing.

\subsection{Bimodal target\label{subsec:Random-Walk-Metropolis-bimodal}}

We use a bimodal target distribution and a random walk MH 
algorithm to illustrate our method and highlight some of its limitations. In particular,
we consider a mixture of univariate Normal distributions with density 
$\pi(x)=0.5\cdot \mathcal{N}(x;-4,1)+0.5\cdot \mathcal{N}(x;+4,1)$,
which we sample from using random walk MH 
with Normal proposal distributions of variance $\sigma_q^2 = 9$.
This enables regular jumps between the modes of $\pi$.
We set the initial distribution $\pi_0$ to $\mathcal{N}(10,10^2)$, so that chains
are likely to start closer to the mode at $+4$ than the mode at $-4$.
Over $1,000$ independent runs, we find that the meeting time $\tau$ has an average of $20$ and 
a $99\%$ quantile of $105$. 

We consider the task of estimating $\int \mathds{1}(x > 3)\pi(dx) \approx 0.421$. 
First, we consider the choice of $k$ and $m$.
Over $1,000$ independent experiments, 
we approximate the expected cost $\mathbb{E}[2(\tau-1) + \max(1,m-\tau+1)]$, 
the variance $\mathbb{V}[H_{k:m}(X,Y)]$,
and compute the inefficiency as the product of the two (as in Section \ref{subsec:variance}).
We then divide the inefficiency by the asymptotic variance of the MCMC estimator, denoted by $V_\infty$,
which we obtain from $10^6$ iterations and a burn-in period of $10^4$ 
using the R package CODA \citep{plummer2006coda}. 

\begin{figure}
    \centering{}\subfloat[\label{fig:mixture:histogram:easy} $\sigma_q^2=3^2$ and $\pi_0 = \mathcal{N}(10,10^2)$.]{\includegraphics[width=0.3\textwidth]{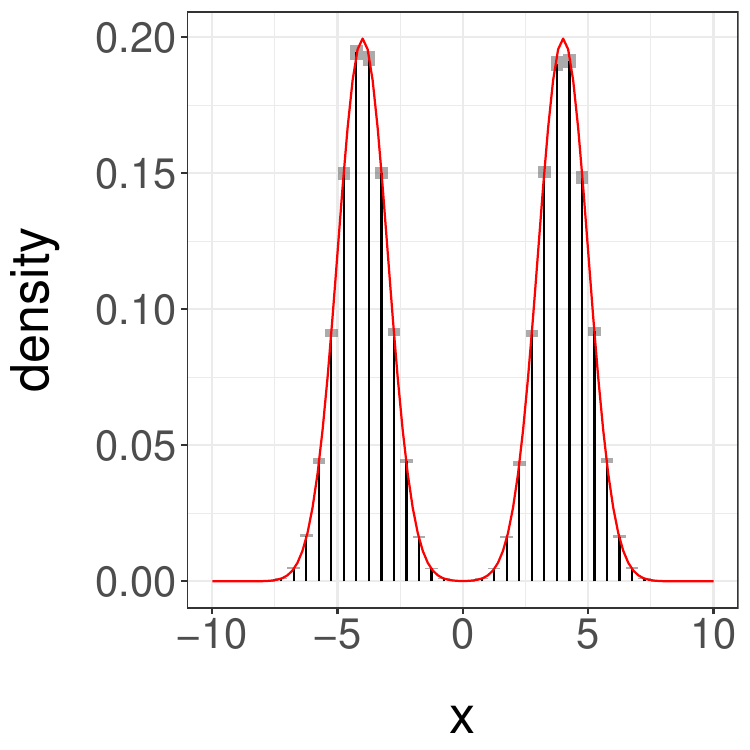}

}\subfloat[\label{fig:mixture:histogram:intermediate}$\sigma_q^2=1^2$ and $\pi_0 = \mathcal{N}(10,10^2)$.]{\includegraphics[width=0.3\textwidth]{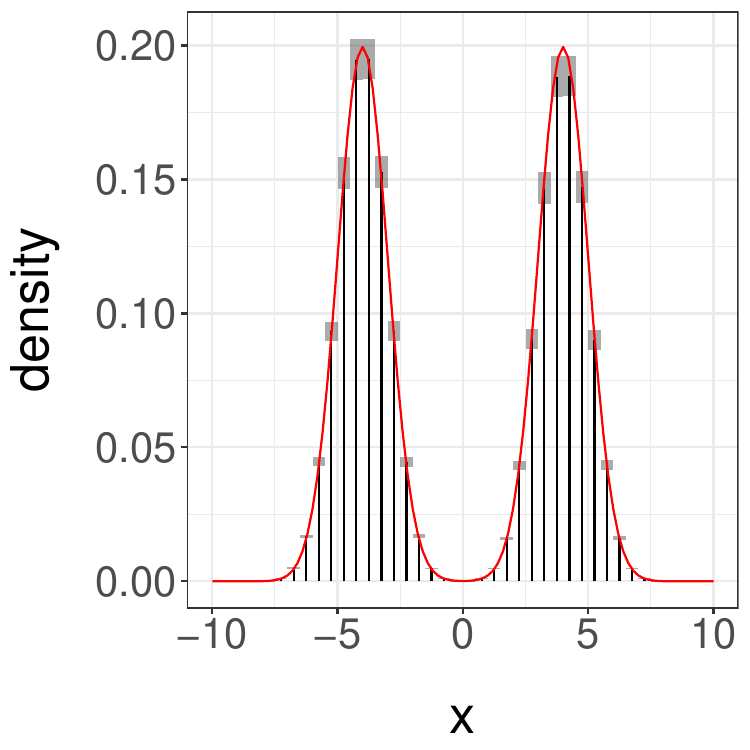}}
\subfloat[\label{fig:mixture:histogram:hard}$\sigma_q^2=1^2$ and $\pi_0 = \mathcal{N}(10,1^2)$.]{\includegraphics[width=0.3\textwidth]{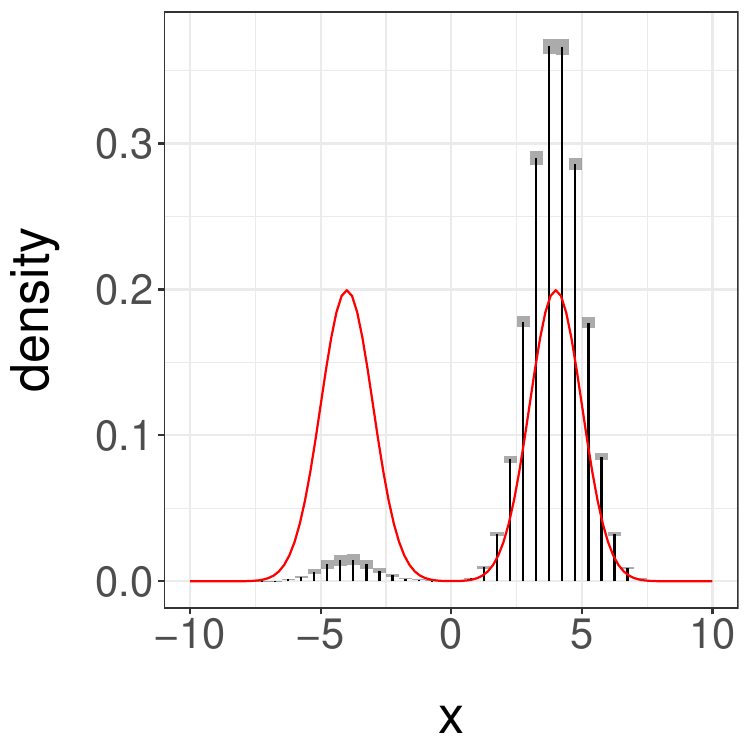}}
\caption{Histograms of the mixture target distribution of Section
    \ref{subsec:Random-Walk-Metropolis-bimodal}, obtained with the proposed unbiased estimators,
    based on a Normal random walk MH algorithm, with a proposal variance 
    $\sigma_q^2$ and an initial distribution $\pi_0$, 
    over $R=1,000$ experiments. The target density function is overlaid in solid lines.}
\end{figure}

We present the results in Table \ref{table:mixture:easy}. 
First, we see that the inefficiency is sensitive to the choice of $k$ and $m$.
Second, we see that when $k$ and $m$ are sufficiently large 
we can retrieve an inefficiency comparable to that of the underlying MCMC algorithm.
The ideal choice of $k$ and $m$ will depend on tradeoffs between inefficiency, the desired level of
parallelism, and the number of processors available.
We present a histogram of the target distribution, 
obtained using $k=200$, $m=2,000$, in Figure \ref{fig:mixture:histogram:easy}.
These histograms are produced by averaging unbiased estimators of expectations of indicator functions,
corresponding to consecutive intervals.
Confidence intervals at level $95\%$ 
are obtained from the central limit theorem and are represented as grey boxes,
with vertical bars showing the point estimates.

\begin{table}[ht]
\centering
\begin{tabular}{ccccc}
  \hline
$k$ & $m$ & Cost & Variance & Inefficiency / $V_\infty$ \\ 
  \hline
1 & $1 \times k$ &   37 & 4.1e+02 & 1878.4 \\ 
  1 & $10 \times k$ &   39 & 3.6e+02 & 1703.5 \\ 
  1 & $20 \times k$ &   45 & 3.0e+02 & 1624.8 \\ 
  100 & $1 \times k$ &  119 & 9.0e+00 &  130.6 \\ 
  100 & $10 \times k$ & 1019 & 2.3e-02 &    2.9 \\ 
  100 & $20 \times k$ & 2019 & 7.9e-03 &    1.9 \\ 
  200 & $1 \times k$ &  219 & 2.4e-01 &    6.5 \\ 
  200 & $10 \times k$ & 2019 & 5.3e-03 &    1.3 \\ 
  200 & $20 \times k$ & 4019 & 2.4e-03 &    1.2 \\ 
   \hline
\end{tabular}
\caption{Cost, variance and inefficiency divided by MCMC asymptotic variance $V_\infty$, for various choices of $k$ and $m$, for the test function $h:x\mapsto \mathds{1}(x > 3)$, in the bimodal target example of Section \ref{subsec:Random-Walk-Metropolis-bimodal}. \label{table:mixture:easy}} 
\end{table}

Next, we consider a more challenging case by setting $\sigma_q^2 = 1$, again with
 $\pi_0=\mathcal{N}(10,10^2)$. 
These values make it difficult for the chains to jump between the modes of $\pi$.
Over $R=1,000$ runs we find an average meeting time of $769$, with
a $99\%$ quantile of $9,186$. 
When the chains start in different modes, the meeting times are often dramatically
larger than when the chains start by the same mode.
One can still recover accurate estimates of the target
distribution, but $k$ and $m$ have to be set to larger values. 
With $k=20,000$ and $m=30,000$, we obtain the $95\%$ confidence interval
$[0.397, 0.430]$ for $\int \mathds{1}(x > 3)\pi(dx) \approx 0.421$.
We show a histogram of $\pi$ in Figure \ref{fig:mixture:histogram:intermediate}.

Finally we consider a third case, with $\sigma_q^2 = 1$ as before but now
with $\pi_0$ set to $\mathcal{N}(10,1)$.
This initialization makes it unlikely for a chain to start near the mode at $-4$. 
The pair of chains typically converge around the mode at $+4$ and meet in a small number of iterations.
Over $R=1,000$ replications, we find an average meeting time of $9$ and a $99\%$ quantile of $35$.
A $95\%$ confidence interval on ${\int \mathds{1}(x > 3)\pi(dx)}$ obtained from the estimators with
$k=50$, $m=500$ is $[0.799, 0.816]$, far from the true value of $0.421$.
The associated histogram of $\pi$ is shown in Figure \ref{fig:mixture:histogram:hard}.

Sampling $9,000$ additional estimators yields a $95\%$ confidence interval
$[-0.353, 1.595]$, again using $k=50$, $m=500$. Among these extra $9,000$ values, a few correspond
to cases where one chain jumped to the left-most mode before meeting the other. This resulted
in large meeting times and thus a large empirical variance for $H_{k:m}$. Upon noticing a large empirical variance
one can then decide to use larger values of $k$ and $m$.
We conclude that although our estimators are unbiased and are consistent in the limit as $R \to \infty$,
poor performance of the underlying Markov chains combined with ill-chosen initializations can still produce misleading results for any finite $R$,
such as $1,000$ in this example.

\subsection{Gibbs sampler for nuclear pump failure data\label{subsec:Gibbs-pump-failures}}

Next we consider a classic Gibbs sampler for a model of pump failure counts,
used e.g. in \citet{murdoch1998exact} to illustrate 
perfect samplers for continuous distributions, 
and in \citet{mykland1995regeneration} to illustrate their regeneration
approach.
Here we focus on a comparison with the regeneration
approach, which was motivated by similar practical concerns as this paper, in particular to avoid
an arbitrary choice of burn-in, construct confidence intervals on the expectations of interest, 
and make principled use of parallel processors.  In that paper the authors show how to construct regeneration times -- random
times between which the chain forms independent and identically distributed
``tours''. The authors define a consistent estimator for arbitrary test
functions, whose asymptotic variance takes a simple form.  The
estimator is then obtained by aggregating over these independent tours.

The data consist of operating times $(t_{n})_{n=1}^{K}$ and failure counts
$(s_{n})_{n=1}^{K}$ for $K=10$ pumps at the Farley-1 nuclear power station, as
first described in \citet{gaver1987robust}. The model specifies $s_{n}\sim
\text{Poisson}(\lambda_{n}t_{n})$ and $\lambda_{n}\sim
\text{Gamma}(\alpha,\beta)$, where
$\alpha=1.802$, $\beta \sim \text{Gamma}\left(\gamma,\delta\right)$,
$\gamma=0.01$, and $\delta=1$. The Gibbs sampler for this model consists of the following update steps:
\begin{align*}
    \lambda_n\ | \text{ rest} &\sim \text{Gamma}(\alpha+s_n, \beta + t_n) \quad \text{for }n = 1, \ldots, K, \\ 
    \beta\ | \text{ rest} &\sim \text{Gamma}(\gamma + 10\alpha, \delta + \sum_{n=1}^K \lambda_n).
\end{align*}
Here $\text{Gamma}(\alpha,\beta)$ refers
to the distribution with density $x\mapsto \Gamma(\alpha)^{-1} \beta^\alpha
x^{\alpha-1} \exp(-\beta x)$.
We initialize all parameter values to 1 (the initialization is not specified in \citet{mykland1995regeneration}).
To form our estimator we apply maximal
couplings at each conditional update of the Gibbs sampler, as described in
Section \ref{subsec:Gibbs}.

We begin by drawing $1,000$ meeting times independently. Following the
guidelines of Section \ref{subsec:variance}, we set $k=7$, corresponding to the
$99\%$ quantile of $\tau$ and $m = 10\cdot k = 70$. For the regeneration approach, 
\citet{mykland1995regeneration} gives a set of tuning parameters
which we adopt below.
Applying the regeneration approach to 1,000 Gibbs sampler runs of 5,000 iterations
each, we observe on average 1,996 complete tours per run with an average length of 2.50
iterations per tour. These values agree with the count of 1,967 tours of average
length 2.56 reported in \citet{mykland1995regeneration}. We observe 
a posterior mean estimate for $\beta$  of 2.47 with a variance of $1.89 \times 10^{-4}$ over the 1,000
independent runs, which implies an efficiency value of $(5,000 \cdot
1.89 \times 10^{-4})^{-1} = 1.06$. This exceeds the efficiency of $0.94$ achieved by our
estimator with the choice of $k=7$ and $m=70$. On the other hand, 
the regeneration approach often requires more extensive analytical work with the underlying Markov chain;
we refer to \citet{mykland1995regeneration} for a detailed description. For reference, the underlying Gibbs sampler achieves
an efficiency of $1.08$, based on a long run of $5\times 10^5$ iterations and a burn-in of $10^3$ iterations.
More extensive comparisons with other regeneration approaches such as that of \citet{brockwell:kadane:05}
would deserve investigation.

\subsection{Ising model\label{sec:ising}}

We consider an Ising model on a $32\times 32$ square lattice with periodic
boundaries.
This provides a setting where a
basic MCMC sampler can mix slowly depending on an inverse
temperature parameter $\theta$, and where a replica exchange strategy
as in \citet{geyer1991} can be helpful. We also use this example to illustrate the use
of our estimators on a large computing cluster, with the considerations reviewed in Section \ref{sec:glynnheidelberger}. 
For $i$ and $j$ in $\{1,\ldots,32\}^2$ we write
$i\sim j$ if $i$ and $j$ are neighbors in the square lattice with periodic
boundaries.  We write $x_i \in \{-1,+1\}$ for the spin at location $i$, and $x
= \{x_i\}$ for the full grid. We write $t(x)$ for the ``natural statistic''
${t(x) = 0.5 \sum_{i\in \{1,\ldots,32\}^2} \sum_{j\sim i} x_i x_j}$
summing the products of pairs of neighbors. 
The $0.5$ multiplier here results in each pair of neighboring sites only being counted once.
Under the model, the probability associated with a grid $x$ is
$\pi_\theta(x) \propto \exp(\theta t(x))$, where $\theta > 0$ denotes an
inverse temperature parameter that calibrates the degree of correlation between neighboring sites.

We consider a single-site Gibbs sampler, called a heat bath algorithm in this context,
to approximate the distribution $\pi_\theta$ given a value of $\theta$. 
One iteration of the algorithm consists of a sweep through all the locations $i\in\{1,\ldots,32\}^2$. For each $i$ we draw $x_i$ from its conditional distribution under $\pi_\theta$ given all the other spins.
It can be checked that the conditional probability of $\{x_i = +1\}$
given the other spins equals $\exp(\theta s_i)/(\exp(\theta s_i)+\exp(-\theta
s_i))$, where $s_i$ denotes the sum of spins over the four neighbors of $i$.
We initialize the chains by drawing spins uniformly in $\{-1,+1\}$ at each site,
independently across sites. 

\begin{figure}
\begin{centering}
\subfloat[\label{fig:ising:singlesitemeetings} Single-site Gibbs sampler.]{\begin{centering}
\includegraphics[width=0.45\textwidth]{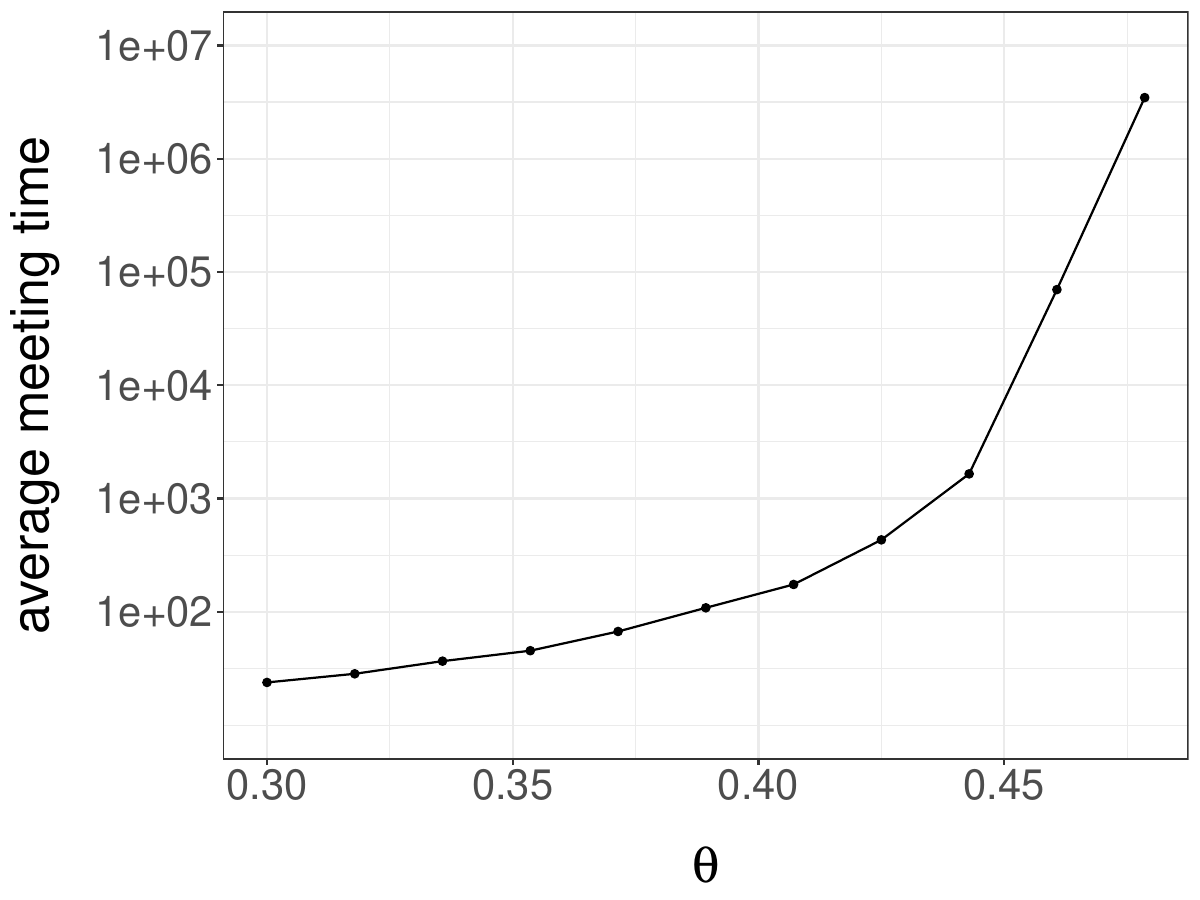}
\par\end{centering}
}\subfloat[\label{fig:ising:withswapmoves} Replica exchange, with $\theta^{(1)} =0.3$ and $\theta^{(N)} = 0.55$.]{\begin{centering}
    \includegraphics[width=0.45\textwidth]{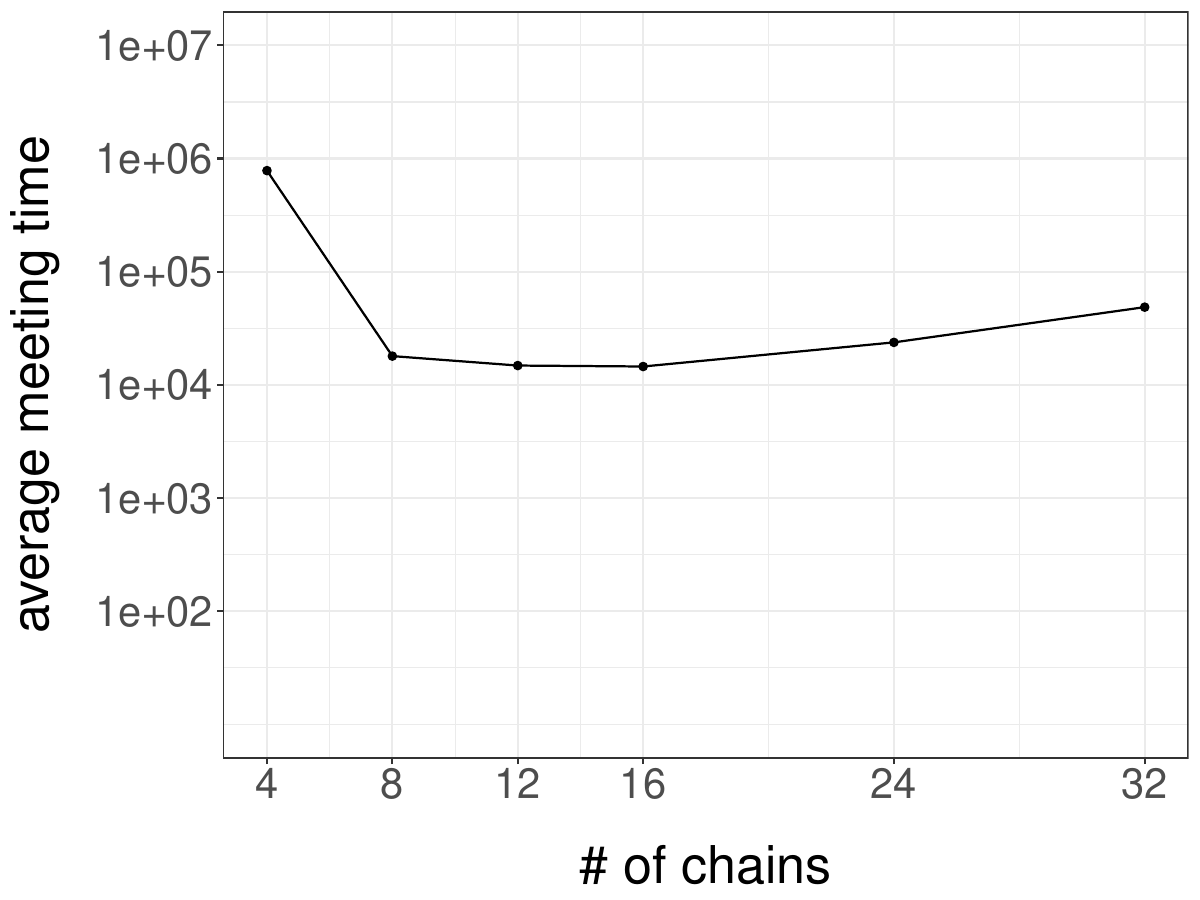}
\par\end{centering}
}
\par\end{centering}
\caption{\label{fig:ising:singlesite} For the Ising model of Section \ref{sec:ising}
on a $32\times 32$ grid, average meeting times corresponding to different coupled Markov chains.
Left: coupled single-site Gibbs sampler, for different inverse temperatures $\theta$. Right: coupled replica exchange algorithm
with $N$ chains, on a grid of values $\theta^{(1)}< \ldots<\theta^{(N)}$ with $\theta^{(1)} = 0.3$  and $\theta^{(N)} = 0.55$,
for different values of $N$.}
\end{figure}

A simple strategy to couple heat bath chains consists of sampling from the
maximal coupling of each conditional distribution. 
For a grid of $\theta$ values from $0.3$ and $0.48$,
we run 100 pairs of chains until they meet. 
We then plot the average meeting time as a function of $\theta$ in Figure
\ref{fig:ising:singlesitemeetings},
noting that the average meeting time increases
sharply to values above $10^6$ as $\theta$ 
approaches its critical value (see the related discussion in \citet{propp:wilson:1996}).
We conclude that it would be expensive to produce unbiased estimators based on the heat bath algorithm
for values of $\theta$ above $0.48$, for reasons related to the behavior of the underlying algorithm.

There are several ways to address the degeneracy of the heat bath algorithm
as $\theta$ increases.
Specialized algorithms have been proposed to jointly update groups of spins \citep{swendsen1987nonuniversal,wolff1989comparison}.
Here, we consider an approach based on an ensemble of $N$ chains that regularly exchange their states,
a technique often termed replica exchange or parallel tempering. 
Following e.g. \citet{geyer1991}, we introduce $N$ chains, $x^{(1)}$, \ldots, $x^{(N)}$, with each $x^{(n)}$ targeting $\pi_{\theta^{(n)}}$ with different values
of $\theta^{(n)}$ ordered as $\theta^{(1)} < \ldots < \theta^{(N)}$. Each iteration of
the algorithm proceeds as follows. With probability $p_{\text{swap}}\in (0,1)$, for $n\in
\{1,\ldots,N-1\}$ (sequentially), we propose exchanging the states $x^{(n)}$
and $x^{(n+1)}$ corresponding to $\theta^{(n)}$ and $\theta^{(n+1)}$. We accept
this swap with probability $\min(1,\pi_{\theta^{(n)}}(x^{(n+1)})\pi_{\theta^{(n+1)}}(x^{(n)})/(\pi_{\theta^{(n)}}(x^{(n)})\pi_{\theta^{(n+1)}}(x^{(n+1)})))$,
which simplifies to 
${\min(1,\exp((\theta^{(n)} - \theta^{(n+1)}) (t(x^{(n+1)}) - t(x^{(n)}))))}$. Otherwise
we perform a full sweep of single-site Gibbs updates, independently across chains.

A coupling of this algorithm involves a pair of ensembles with $N$ chains each; 
the two ensembles are identical if chain $n$ in the first ensemble equals chain $n$ in the second ensemble, for all $n\in\{1,\ldots,N\}$.
We use common random numbers to decide whether to perform swap moves or single-site Gibbs moves,
and whether to accept the proposed states in the event of a swap move.
In the event of a single-site Gibbs move, we maximally couple each conditional update.

Throughout the following experiments we use $p_{\text{swap}} = 0.01$, and introduce an equally spaced grid of $\theta$ values from 
 $\theta^{(1)} = 0.3$ to $\theta^{(N)} = 0.55$ 
 for several different choices of $N$.
We note that these grids includes $\theta$ values 
at which we have seen that the single-site Gibbs sampler mixes poorly.
Figure \ref{fig:ising:withswapmoves} shows the resulting average meeting times over 100 independent runs, as a function of the number of chains $N$.
The average meeting time first decreases with the number of chains, but then increases again. A possible explanation is
that the mixing of the chains first improves as $N$ increases, and then stabilizes; on the other hand it becomes
harder for the ensembles to meet when $N$ increases since all chains in the ensembles have to meet.
The minimum average meeting time is here attained for $N=16$ chains per ensemble.

\begin{figure}
\begin{centering}
\subfloat[\label{fig:ising:chronology} Chronology of the generation of unbiased estimators on 500 parallel processors over $30$ minutes.]{\begin{centering}
\includegraphics[width=0.45\textwidth]{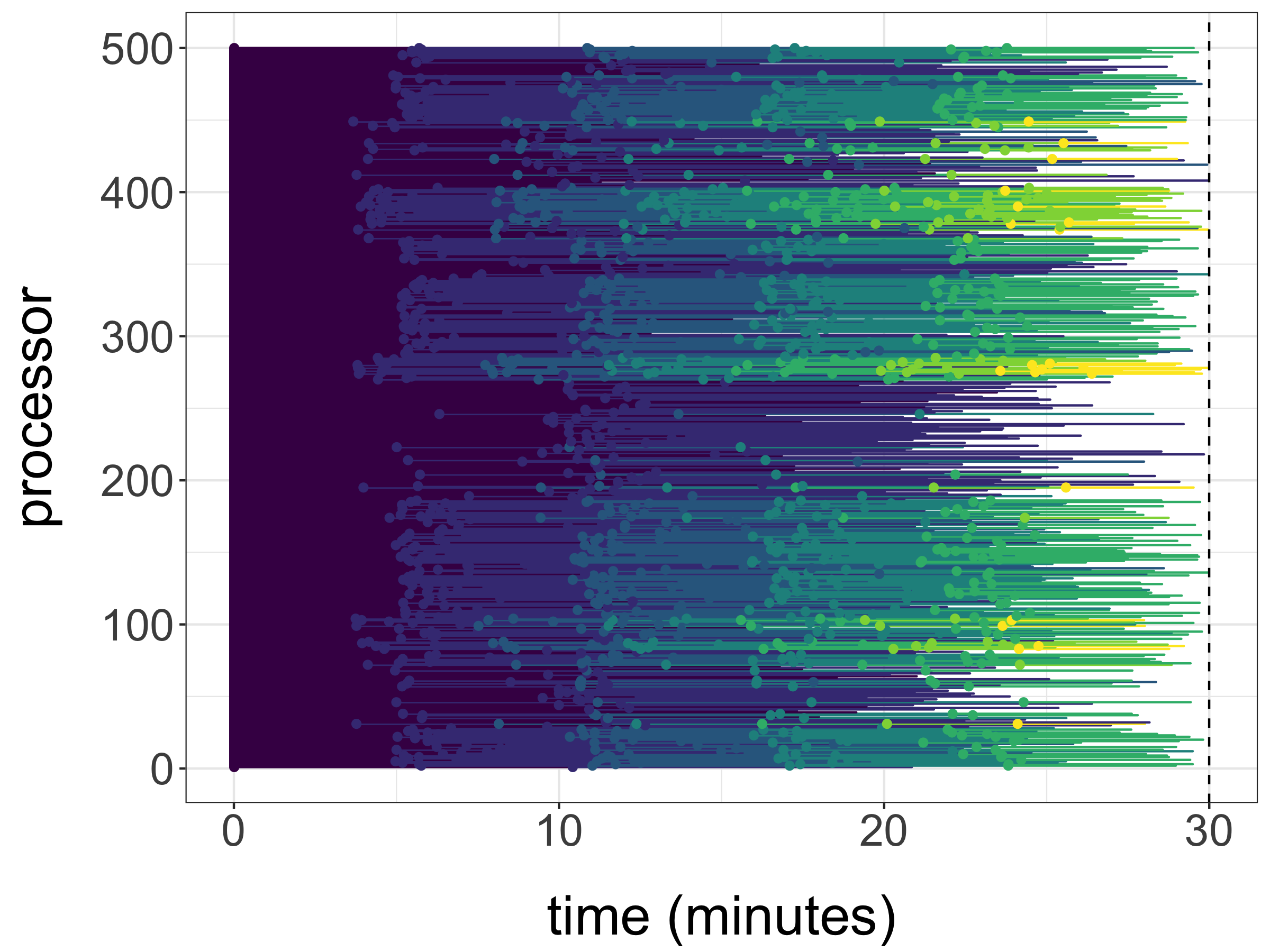}
\par\end{centering}
}\subfloat[\label{fig:ising:estimates} Estimates (top) and standard errors (bottom) for $\sum_{x} t(x) \pi_\theta(x)$ at different $\theta$.]{\begin{centering}
    \includegraphics[width=0.45\textwidth]{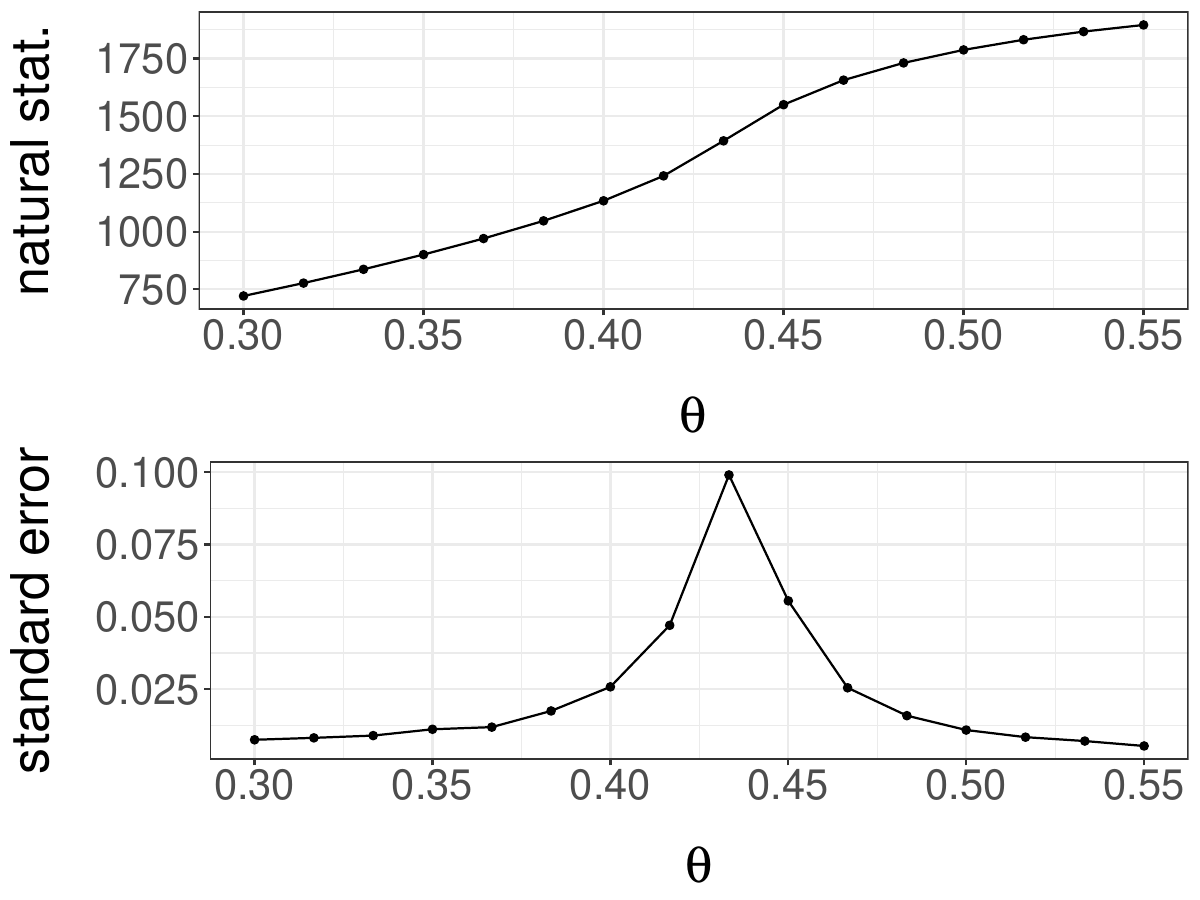}
\par\end{centering}
}
\par\end{centering}
\caption{\label{fig:ising:cluster} For the Ising model of Section \ref{sec:ising}
on a $32\times 32$ grid: on the left, chronology of the generation of unbiased estimators
on $500$ processors over $30$ minutes. Right: estimates (top) of the expected natural statistic,
$\sum_{x} t(x) \pi_\theta(x)$, and standard errors (bottom), for a grid of $16$ values $0.3=\theta^{(1)}< \ldots<\theta^{(N)}=0.55$.
Results obtained by coupling a replica exchange algorithm with $16$ chains.}
\end{figure}

Setting $N=16$, $k=10^5$ and $m=2\times 10^5$ we now illustrate the use of the proposed unbiased estimators on a cluster.
The test function is taken as $x\mapsto t(x)$ defined above, so that we estimate $\sum_{x} t(x) \pi_\theta(x)$ for different values of $\theta$.
We use $500$ processors to generate unbiased estimates with a time budget of $30$ minutes. Within that time, each processor generated between $1$ and $7$
estimators, with an average of $3.7$ estimators per processor and a total of $1858$ estimators. 
The chronology of the generation of these estimates
is illustrated in Figure \ref{fig:ising:chronology}. For each processor, horizontal segments of different colours indicate the duration associated with each estimator. The final estimates with standard errors are shown in Figure \ref{fig:ising:estimates}, where
we can see that the standard errors are very small compared to the values of the estimates, for each value of $\theta$. These 
standard errors were computed as $\hat\sigma_1(\mathrm{P},t)/\sqrt{\mathrm{P}}$ 
following \eqref{eq:parallelclt:fixedt}, the CLT corresponding to the large processor count limit.

\subsection{Variable selection \label{subsec:variableselection}}

For our next example we consider a variable selection problem following \citet{yang2016computational}
to illustrate the scaling of our proposed method on high-dimensional discrete state spaces.
For integers $p$ and $n$, let $Y\in\mathbb{R}^n$ represent a response variable
depending on covariates ${X_1,\ldots,X_p\in\mathbb{R}^n}$.
We consider the task of inferring a binary vector $\gamma \in \{0,1\}^p$ representing
which covariates to select as predictors of $Y$, 
with the convention that $X_i$ is selected if $\gamma_i = 1$.
For any $\gamma$, we write $|\gamma| = \sum_{i=1}^p \gamma_i$ for the number of selected covariates
and $X_\gamma$ for the $n \times |\gamma|$ matrix of covariates chosen by $\gamma$. 
Inference on $\gamma$
proceeds by way of a linear regression model relating $Y$ to $X_\gamma$, namely
$Y = X_\gamma \beta_\gamma + w$ with $w \sim \mathcal{N}(0,\sigma^2 I_n)$.

We assume a prior on $\gamma$ of $\pi(\gamma) \propto p^{-\kappa |\gamma|}
\mathds{1}(|\gamma| \leq s_0)$. This distribution puts mass only on vectors
$\gamma$ with fewer than $s_0$ ones, imposing a degree of sparsity.
Given $\gamma$ we assume a Normal prior for the regression
coefficient vector $\beta_\gamma\in \mathbb{R}^{|\gamma|}$ with zero mean and
variance $g \sigma^{2} (X_\gamma' X_\gamma)^{-1}$.  Finally, we give the precision
$\sigma^{-2}$ an improper prior $\pi(\sigma^{-2}) \propto 1/\sigma^{-2}$. This
leads to the marginal likelihood
\begin{align*}
    \pi(Y | X, \gamma) & \propto \frac{(1+g)^{-|\gamma|/2}}{(1 + g(1 - R^2_\gamma))^{n/2}}, \quad
    \text{where} \quad R_\gamma^2 = \frac{Y' X_\gamma (X'_\gamma X_\gamma)^{-1} X'_\gamma Y}{Y' Y}.
\end{align*}

To approximate the distribution $\pi(\gamma|X,Y)$,
\citet{yang2016computational} employ an MCMC algorithm whose kernel $P$ is a
mixture of two Metropolis kernels.  The first component $P_1(\gamma,\cdot)$
selects a coordinate $i\in \{1,\ldots,p\}$ uniformly at random and flips
$\gamma_i$ to $1 - \gamma_i$.  The resulting vector $\gamma^\star$ is then
accepted with probability $1 \wedge \pi(\gamma^\star|X,Y) / \pi(\gamma|X,Y)$,
where $a\wedge b$ denotes $\min(a,b)$ for $a,b\in\mathbb{R}$.
Sampling a vector $\gamma'$ from the second kernel $P_2(\gamma, \cdot)$
proceeds as follows. If $|\gamma|$ equals zero or $p$, then $\gamma'$ is set to
$\gamma$. Otherwise, coordinates $i_0$, $i_1$ are drawn uniformly among
$\{j:\gamma_j = 0\}$ and  $\{j:\gamma_j = 1\}$, respectively. 
The proposal $\gamma^\star$ has $\gamma^\star_{i_0} = \gamma_{i_1}$,
$\gamma^\star_{i_1} = \gamma_{i_0}$, and $\gamma^\star_j = \gamma_j$ for the
other components.  
Then $\gamma'$ is set to $\gamma^\star$ with probability $1
\wedge \pi(\gamma^\star|X,Y) / \pi(\gamma|X,Y)$, and to $\gamma$ otherwise.
The MCMC kernel $P(\gamma,\cdot)$ targets $\pi(\gamma|X,Y)$ by sampling
from $P_1(\gamma,\cdot)$ or from $P_2(\gamma,\cdot)$ with equal probability.
Note that each MCMC iteration can only benefit from parallel processors to a
limited extent, since $|\gamma|$ is always less than $s_0$, itself chosen to be
a small value; thus the calculation of $R_\gamma^2$ only involves linear algebra of small matrices.

We consider the following strategy to couple the above MCMC algorithm. To sample a pair of states $(\gamma',\tilde{\gamma}')$ given
$(\gamma,\tilde{\gamma})$, we first use a common uniform random variable
to decide whether to sample from a coupling $\bar{P}_1$ of $P_1$ to itself or
a coupling $\bar{P}_2$ of $P_2$ to itself.  The coupled kernel
$\bar{P}_1((\gamma,\tilde{\gamma}),\cdot)$ proposes flipping the same
coordinate for both vectors $\gamma$ and $\tilde{\gamma}$ and then uses a
common uniform random variable in the acceptance step.  For the coupled kernel
$\bar{P}_2((\gamma,\tilde{\gamma}),\cdot)$, we need to select two pairs of
indices, $(i_0,\tilde{i}_0)$ and $(i_1,\tilde{i}_1)$.  We obtain the first pair
by sampling from a maximal coupling of the discrete uniform distributions on
$\{j:\gamma_j = 0\}$ and $\{j: \tilde{\gamma}_j = 0\}$. This yields indices
$(i_0,\tilde{i}_0)$ with the greatest possible probability that
$i_0=\tilde{i}_0$.  We use the same approach to sample a pair
$(i_1,\tilde{i}_1)$ to maximize the probability that $i_1=\tilde{i}_1$.
Finally we use a common uniform variable to accept or reject the proposals.  If
either vector $\gamma$ or $\tilde{\gamma}$ has no zeros or no ones, then it is
kept unchanged.

We recall that one can sample from a maximal coupling of two discrete
probability distributions $q = (q_1,\ldots,q_N)$ and $\tilde{q} =
(\tilde{q}_1,\ldots, \tilde{q}_N)$ as follows.  First, let $c = (c_1, \dots,
c_N)$ be the distribution with probabilities $c_n = (q_n \wedge \tilde{q}_n) /
\alpha$ for $\alpha = \sum_{n=1}^N q_{n}\wedge \tilde{q}_{n}$ and define
residual distributions $q'$ and $\tilde{q}'$ with probabilities $q'_n = (q_n -
\alpha c_n) / (1-\alpha)$ and $\tilde{q}'_n = (\tilde{q}_n - \alpha
c_n)/(1-\alpha)$.  Then with probability $\alpha$, draw $i \sim c$ and output
$(i,i)$. Otherwise draw $i \sim q'$ and $\tilde{i} \sim \tilde{q}'$ and output
$(i,\tilde{i})$.  
The resulting pair follows a maximal coupling of $q$ and
$\tilde{q}$, since $\mathbb{P}(i=\tilde{i}) = \alpha = 1-d_{\text{TV}}(q, \tilde{q})$, and 
marginally $\mathbb{P}(i = n) = \alpha c_n + (1-\alpha) q'_n = q_n$, and likewise for $\mathbb{P}(\tilde{i} = n)$,
for all $n\in\{1,\ldots,N\}$.
The procedure involves $\mathcal{O}(N)$ operations for $N$ the
size of the state space.

We now consider an experiment like those of \citet{yang2016computational}.  We
define 
\[\beta^\star = \text{SNR}\sqrt{\sigma_0^2
\frac{\log(p)}{n}}(2,-3,2,2,-3,3,-2,3,-2,3,0,\ldots,0)' \in \mathbb{R}^p,\] 
and generate $Y$ given $X$ and $\beta^\star$ from the model
with $\sigma^2 = 1$, $\sigma_0^2 = 1$, 
 $n \in \{500,1000\}$,  $p \in \{1000, 5000\}$, and signal-to-noise parameter $\text{SNR} \in \{0.5, 1, 2\}$.
We also set $s_0 = 100$, $g = p^3$, and $\kappa =
2$ (exactly as in \citet{yang2016computational}; the value of $\kappa$ was obtained by personal communication) and generate the covariates $X$ using a multivariate normal distribution with covariance matrix $\Sigma$ either equal to a unit diagonal matrix or with entries $\Sigma_{ij} = \exp(-|i-j|)$. 
We refer to these two cases as the independent design and correlated design cases, respectively.
We draw from the initial distribution $\pi_0$ by creating a
vector of $p$ zeros, sampling $s_0$ coordinates uniformly from 
$\{1,\ldots,p\}$ without replacement, and setting the corresponding entries to
$1$ with probability $0.5$.

For different values of $n$, $p$ and SNR, and the two types of design, we run coupled chains 100 times independently until they meet. We report the average meeting times in Tables \ref{table:varselection:independent}
and  \ref{table:varselection:correlated}. The average meeting times are of the order of $10^4$ to $10^5$,
depending on the problem; the maximum is attained in the correlated design at $n=500,p=1000,\text{SNR}=2$.
In contrast with this, the experiments in \citet{yang2016computational} identify the scenario $n=500,p=5000,\text{SNR}=1$ as the most challenging one.
This discrepancy deserves further study; it could be due to variations from a synthetic data set to another,
or to differences in the criteria being reported. 

\begin{table}[ht]
\centering
\begin{tabular}{rrrrrr}
  \hline
 & n & p & SNR = 0.5 & SNR = 1 & SNR = 2 \\ 
  \hline
   & 500 & 1000 & 4937 & 7586 & 6031 \\ 
   & 500 & 5000 & 24634 & 25602 & 38083 \\ 
   & 1000 & 1000 & 4729 & 5893 & 4892 \\ 
   & 1000 & 5000 & 23407 & 46398 & 24712 \\ 
   \hline
\end{tabular}
\caption{Average meeting times in the independent design. \label{table:varselection:independent}} 
\end{table}

\begin{table}[ht]
\centering
\begin{tabular}{rrrrrr}
  \hline
 & n & p & SNR = 0.5 & SNR = 1 & SNR = 2 \\ 
  \hline
   & 500 & 1000 & 5536 & 5485 & 216996 \\ 
   & 500 & 5000 & 27535 & 28756 & 29083 \\ 
   & 1000 & 1000 & 4921 & 5451 & 5613 \\ 
   & 1000 & 5000 & 24101 & 29215 & 23043 \\ 
   \hline
\end{tabular}
\caption{Average meeting times in the correlated design. \label{table:varselection:correlated}} 
\end{table}

\begin{figure}
\begin{centering}
\subfloat[\label{fig:varselection:meetingtime} Meeting times divided by $p$ in violin plots.]{\begin{centering}
    \includegraphics[width=0.40\textwidth]{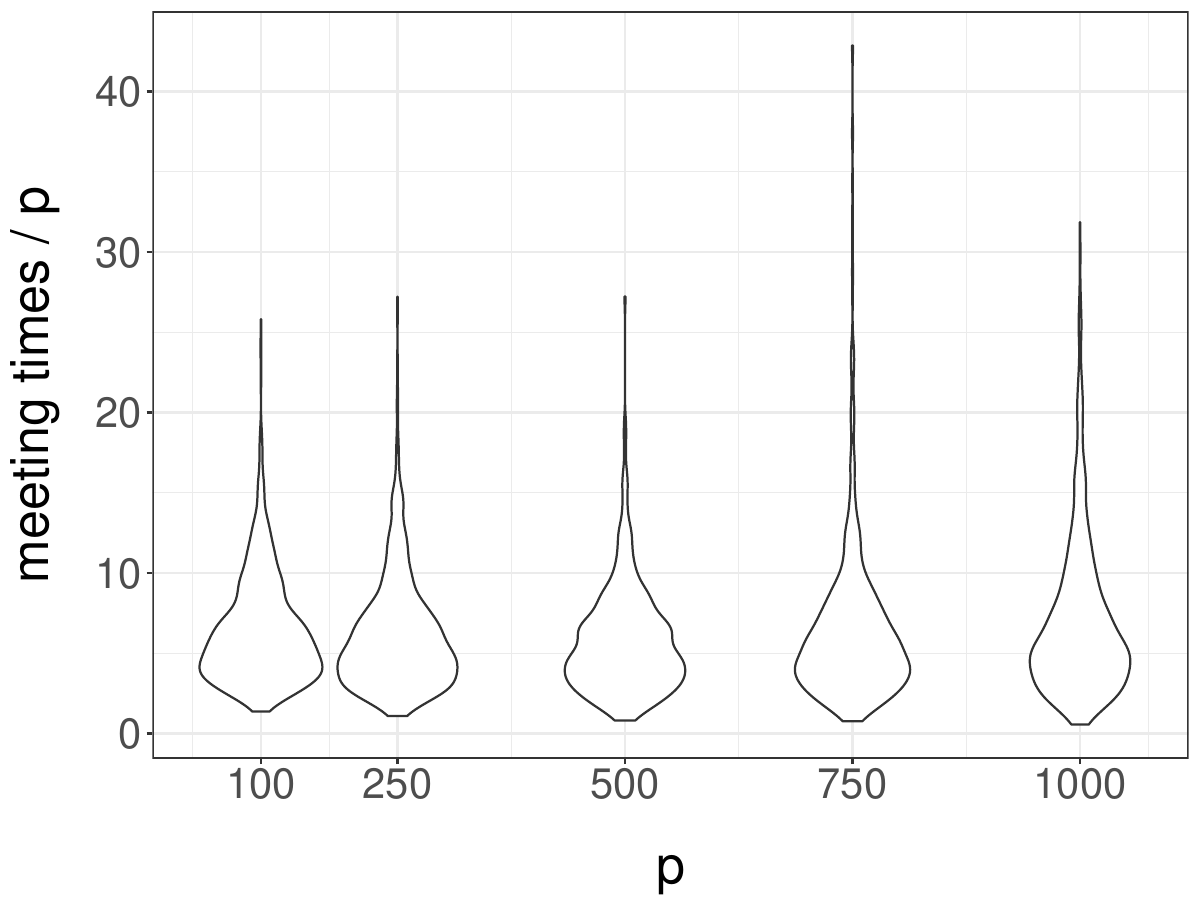}
\par\end{centering}
}
\hspace*{1cm}
\subfloat[\label{fig:varselection:estimates} Inclusion probabilities of the first $20$ variables.]{\begin{centering}
    \includegraphics[width=0.40\textwidth]{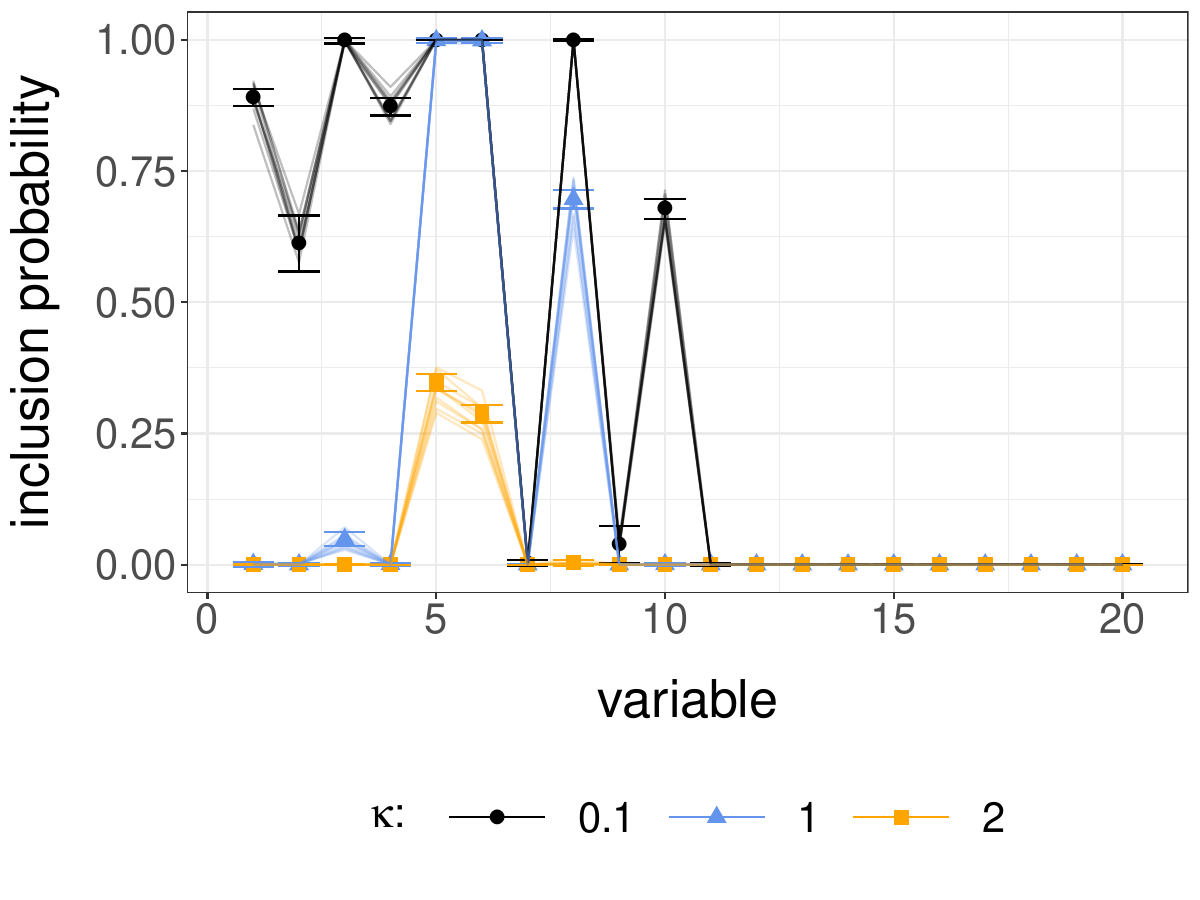}
\par\end{centering}
}
\par\end{centering}
\caption{\label{fig:varselection} Left: meeting times divided $p$ for $p\in\{100,250,500,750,1000\}$
and $n=500$, $\text{SNR}=1$, in the variable selection example of Section \ref{subsec:variableselection} with independent design, 
based on $R=1,000$ independent repeats. 
The violins represent the distributions of scaled meeting times for different $p$. 
Right: posterior probabilities of inclusion for the first $20$ variables, in the setting $n=500$, $p=1,000$, $\text{SNR}=1$,
for three different values of the prior hyperparameter $\kappa$.
The error bars representing $95\%$ confidence intervals were obtained after 120 minutes
of calculation on 600 processors, using $k=75,000$ and $m=150,000$. The solid lines represent standard
MCMC estimates based on 10 independent chains of length $10^6$.}
\end{figure}

To illustrate the impact of dimension, we focus on the independent design
setting with $n=500$ and $\text{SNR}=1$, and consider values of $p$ between
$100$ and $1000$. For each value of $p$, we run coupled chains $1,000$ times
independently until they meet.  We present violin plots representing the
distributions of meeting times divided by $p$ in Figure
\ref{fig:varselection:meetingtime}. The distribution of scaled meeting times
appears to be approximately constant as a function of $p$, suggesting that
meeting times increase linearly in $p$. This is consistent with the findings of
\citet{yang2016computational}, where mixing times are shown to increase
linearly in $p$.

Focusing now on the independent design case with $n=500$, $p=1{,}000$, and $\text{SNR}=1$
we consider different values of the prior hyperparameter $\kappa$ in $\{0.1,1,2\}$. 
We set $k=75{,}000$ and $m=150{,}000$ and generate
unbiased estimators on a cluster for $120$ minutes, using $200$ processors for each value of $\kappa$, and so $600$ processors in total. The test function is chosen 
so that the estimand $\pi(h)$ is the vector of inclusion probabilities
$\mathbb{P}(\gamma_i = 1 | X,Y)$ for $i\in \{1,\ldots,20\}$. 
Within the time budget, $39,282$ estimates were produced, with each processor producing
between 8 and 181 of these. The largest observed meeting time 
was $81,423$. The meeting times were similar across the three values of $\kappa$.

Figure \ref{fig:varselection:estimates} shows
the results in the form of $95\%$ confidence intervals shown as error
bars, using \eqref{eq:parallelclt:fixedt}, the CLT relevant when the time budget is fixed and the number of processors grows large.
We observe that $\kappa$ has a strong impact on the probability of 
including the first 10 variables in this setting, and that the 
most satisfactory results are obtained for $\kappa = 0.1$ rather than for $\kappa = 2$, recalling that $\beta^\star$ has non-zero entries in its first 10 components.
Note that the error bars are narrow but still noticeable, particularly for $\kappa = 0.1$.
On the same figure, the solid lines represent estimates obtained with 
10 independent MCMC runs with $10^6$ iterations each, discarding the first $10^5$ iterations as burn-in.
These MCMC estimates present noticeable variability in spite of the large number of iterations.
In a standard MCMC setting, we might run chains for more iterations until the estimates
agree across independent runs. In the proposed framework, we increase the precision by 
generating more independent unbiased estimators without necessarily modifying $k$ or $m$.

Figure \ref{fig:varselection:estimates} suggests that the variable selection procedure
considered here is sensitive to the prior hyperparameter $\kappa$; we refer to
\citet{yang2016computational}, and to
\citet{johnson2004,nikooienejad2016bayesian} for related discussions on Bayesian variable selection 
in high dimension and convergence of MCMC.

\subsection{Cut distribution \label{subsec:cut-distribution}}

Finally, our proposed estimator can be used to approximate the cut
distribution, which poses a significant challenge for existing MCMC methods
\citep{plummer2014cuts,jacob2017modularization}. This illustrates another 
appeal of the unbiasedness property, beyond the motivation for parallel computation.

Consider two models, one with parameters $\theta_{1}$ and data
$Y_{1}$ and another with parameters $\theta_{2}$ and data $Y_{2}$, where
the likelihood of $Y_2$ might depend on both $\theta_{1}$ and $\theta_{2}$. 
For instance the first model could be a regression 
with data $Y_1$ and coefficients $\theta_1$, 
and the second model could be another regression whose
covariates are the residuals, coefficients, or fitted values of the first regression \citep{pagan1984econometric,murphy2002estimation}.  
In principle one could introduce an encompassing model and conduct joint inference
on $\theta_{1}$ and $\theta_{2}$ via the posterior distribution.
In that case, misspecification of either model would lead to
misspecification of the ensemble and thus to a misleading quantification of
uncertainty, as noted in several studies 
\citep[e.g.][]{liu2009,plummer2014cuts,lunn2009combining,mccandless2010cutting,zigler2016central,blangiardo2011Bayesian}.

The cut distribution \citep{spiegelhalter2003winbugs,plummer2014cuts} allows the propagation
of uncertainty about $\theta_1$ to inference on $\theta_2$
while preventing misspecification in the second model from affecting estimation in the first. 
The cut distribution is defined as
\[
\pi_{\text{cut}}\left(\theta_{1},\theta_{2}\right)=\pi_{1}(\theta_{1})\pi_{2}(\theta_{2}|\theta_{1}).
\]
Here $\pi_{1}(\theta_{1})$ refers to the distribution of $\theta_1$ given $Y_1$ in the first
model alone, and $\pi_{2}(\theta_{2}|\theta_{1})$ refers to the distribution
of $\theta_{2}$ given $Y_{2}$ and $\theta_{1}$ in the second model.
Often, the density $\pi_{2}(\theta_{2}|\theta_{1})$ can only evaluated up to a constant
in $\theta_2$, which may vary with $\theta_1$. This makes the cut distribution difficult to approximate with MCMC algorithms \citep{plummer2014cuts}.

A naive approach consists of
first running an MCMC algorithm targeting $\pi_{1}(\theta_{1})$ to
obtain a sample $(\theta_{1}^{n})_{n=1}^{N_{1}}$, perhaps after
discarding a burn-in period and thinning the chain. Then for each
$\theta_{1}^{n}$, one can run an MCMC algorithm targeting $\pi_{2}(\theta_{2}|\theta_{1}^{n})$,
yielding $N_{2}$ samples. One might again discard some burn-in
and thin the chains, or just keep the final state of each chain. The resulting joint samples approximate the cut distribution.
However, the validity of this approach relies on a double limit 
in $N_{1}$ and $N_{2}$. Diagnosing convergence may also be
difficult given the number of chains in the second stage, each of which targets a different distribution $\pi_2(\theta_2|\theta_1^n)$.

If one could sample $\theta_1 \sim \pi_1$ and $\theta_2 |\theta_1 \sim \pi_2(\theta_2 |\theta_1)$,
then the pair $(\theta_1,\theta_2)$ would follow the cut distribution. The same two-stage rationale
can be applied in the proposed framework.
Consider a test function $(\theta_{1},\theta_{2})\mapsto h(\theta_{1},\theta_{2})$.
Writing $\mathbb{E}_\text{cut}$ for expectations with respect to $\pi_{\text{cut}}$, 
the law of iterated expectations yields
\begin{align*}
    \mathbb{E}_\text{cut}[h(\theta_1,\theta_2)] & =\int\left(\int h(\theta_{1},\theta_{2})\pi_{2}(d\theta_{2}|\theta_{1})\right)\pi_{1}(d\theta_{1})
    =\int\bar{h}(\theta_{1})\pi_{1}(d\theta_{1}).
\end{align*}
Here $\bar{h}(\theta_{1})=\int
h(\theta_{1},\theta_{2})\pi_{2}(d\theta_{2}|\theta_{1})$.  In the proposed
framework, one can make an unbiased estimator of $\bar{h}(\theta_1)$ for all
$\theta_1$, then plug these estimators into an unbiased estimator of the integral $\int
h(\theta_1)\pi_1(d\theta_1)$. This is perhaps clearer using the signed measure
representation of Section \ref{sec:signedmeasure}: one can obtain a signed
measure $\hat{\pi}_1 = \sum_{\ell=1}^N \omega_\ell
\delta_{\theta_{1,\ell}}$ approximating $\pi_1$, and then obtain an
unbiased estimator of $\bar{h}(\theta_{1,\ell})$ for all $\ell$, denoted by
$\bar{H}_\ell$.  Then the weighted average $\sum_{\ell=1}^N \omega_\ell
\bar{H}_\ell$ is an unbiased estimator of
$\mathbb{E}_{\text{cut}}[h(\theta_1,\theta_2)]$ by the law of iterated
expectations. Such estimators can be generated independently in parallel, and
their average provides a consistent approximation of an expectation with
respect to the cut distribution. 

We consider the example described in \citet{plummer2014cuts},
inspired by an investigation of the international correlation between
human papillomavirus (HPV) prevalence and cervical cancer incidence
\citep{maucort2008international}. The first module concerns HPV prevalence,
with data independently collected in $13$ countries. The parameter
$\theta_{1}=(\theta_{1,1},\ldots,\theta_{1,13})$ receives a Beta$(1,1)$ prior distribution
independently for each component.
The data $(Y_{1},\ldots,Y_{13})$ consist of $13$ pairs of integers. 
The first represents the number of women infected with high-risk HPV,
and the second represents population sizes. The likelihood specifies a Binomial model for $Y_{i}$,
independently for each component $i$. 
The posterior for this model is given by a product of Beta distributions.

The second module concerns the relationship between HPV prevalence
and cancer incidence, and posits a Poisson regression. The parameters
are $\theta_{2}=(\theta_{2,1},\theta_{2,2})\in\mathbb{R}^2$ and receive
a Normal prior with zero mean and variance $10^{3}$ per component. The likelihood in this module is 
given by
\[
Z_{1,i} \sim\text{Poisson}(\exp\left(\theta_{2,1}+\theta_{1,i}\theta_{2,2}+Z_{2,i}\right))
\quad \text{for }  i\in\left\{ 1,\ldots,13\right\},
\]
where the data $(Z_{1,i},Z_{2,i})_{i=1}^{13}$ are pairs of integers.  The first
component represents numbers of cancer cases, while the second is a number of
woman-years of follow-up.  The Poisson regression model might be misspecified,
motivating departures from inference based on the joint model \citep{plummer2014cuts}.

\begin{figure}
\begin{centering}
\subfloat[\label{fig:plummer:meetingtime}Histogram of meeting times.]{\begin{centering}
    \includegraphics[width=0.3\textwidth]{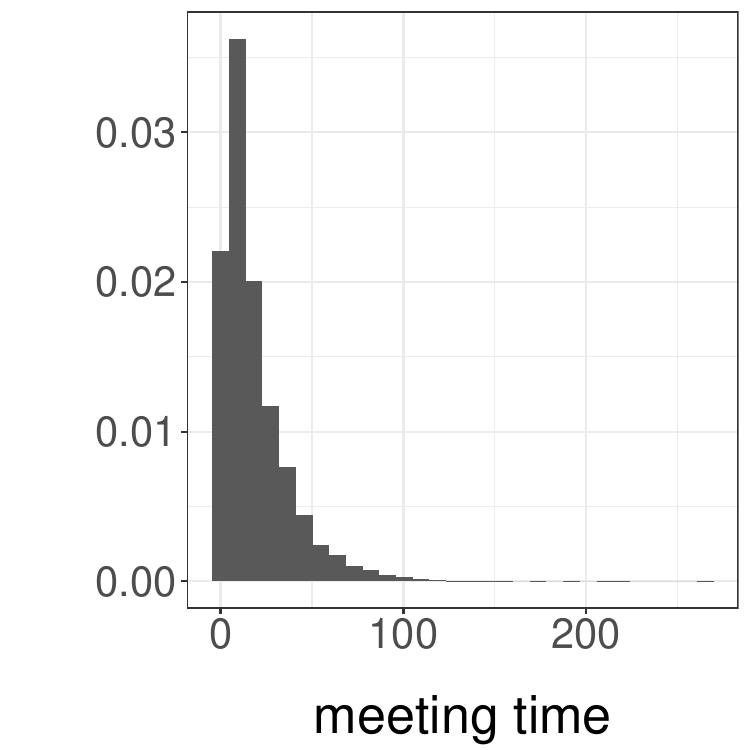}
\par\end{centering}
}\subfloat[\label{fig:plummer:histogramcomponent1}Histogram of $\theta_{2,1}$.]{\begin{centering}
    \includegraphics[width=0.3\textwidth]{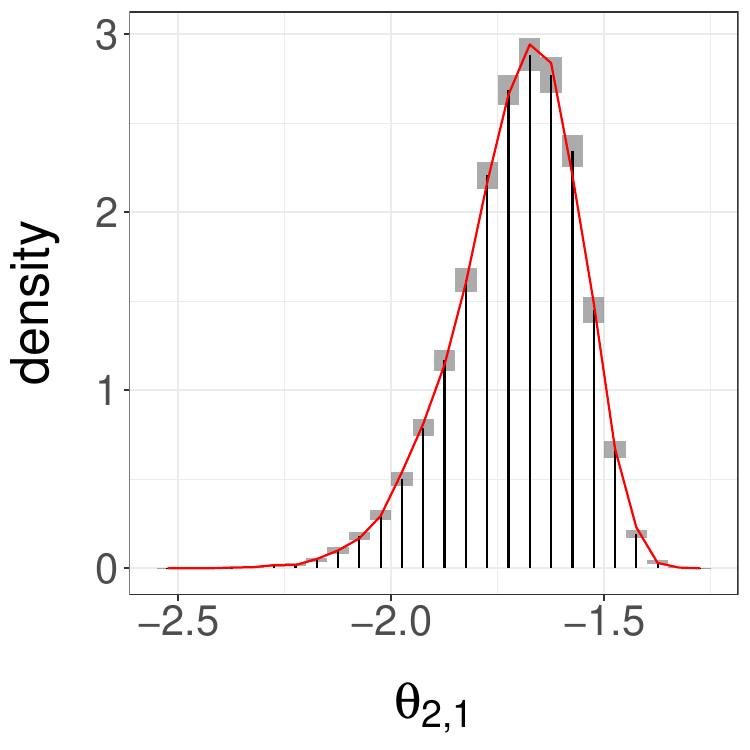}
\par\end{centering}
}\subfloat[\label{fig:plummer:histogramcomponent2}Histogram of $\theta_{2,2}$.]{\begin{centering}
    \includegraphics[width=0.3\textwidth]{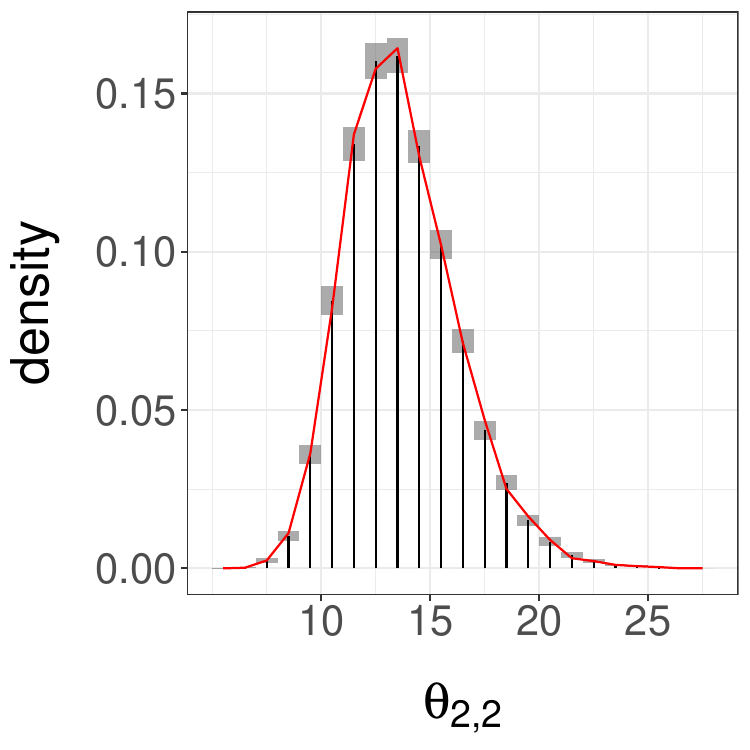}
\par\end{centering}
}
\par\end{centering}
\caption{\label{fig:cut:plummer}Meeting times (left) and histograms of $\theta_{2,1}$
and $\theta_{2,2}$ (middle and right), in the example of Section \ref{subsec:cut-distribution},
computed from $10,000$ unbiased estimators, with $k=100$ and $m = 1000$. Grey boxes indicate $95\%$ confidence intervals
on the histogram estimates. Overlaid solid lines indicate the marginal probability density functions of the cut distribution,
obtained by running $N=10,000$ MCMC chains for $1,000$ steps, each targeting $\pi_2(\theta_2|\theta_1^n)$ for $\theta_1^n$ drawn from $\pi_1(\theta_1)$, and retaining the last state only.
}
\end{figure}

Here we can draw directly from the first posterior, denoted by
$\pi_1(\theta_1)$, and obtain a sample $(\theta_{1}^{n})_{n=1}^{N}$. For each
$\theta_{1}^{n}$ we consider an MH algorithm targeting
$\pi_{2}(\theta_{2}|\theta_{1}^{n})$, using a Normal random walk proposal with
variance $\Sigma$. We couple this algorithm using reflection-maximal couplings
of the proposals as in Section \ref{subsec:maximalcoupling}.  In preliminary
runs, starting with a standard bivariate Normal as an initial distribution and
a proposal covariance matrix set to identity, we estimate the first two moments
of the cut distribution, and we use them to refine the initial distribution
$\pi_0$ and the proposal covariance matrix $\Sigma$.  With these settings we
obtain a distribution of meeting times shown in Figure
\ref{fig:plummer:meetingtime}.  We then set $k=100$, $m=10k$, and obtain
approximations of the cut distribution represented by histograms in Figures
\ref{fig:plummer:histogramcomponent1} and
\ref{fig:plummer:histogramcomponent2}, using $N=10,000$ unbiased estimators.  The overlaid curves correspond to a
kernel density estimate obtained by running $m=1,000$ steps of MCMC targeting
$\pi_2(\theta_2|\theta_1^n)$ with $\theta_1^n$ drawn from $\pi_1(\theta_1)$,
for $n\in\{1,\ldots,N\}$, and keeping the final $m$-th state of each chain. 
The proposed estimators can be refined by increasing the number  $N$ of independent replications,
whereas the MCMC estimators would converge only in the double limit of $N$ and $m$ going to infinity.

\section{Discussion \label{sec:discussion}}

By combining the powerful technique of \citet{glynn2014exact} with couplings of
MCMC algorithms, unbiased estimators of integrals with
respect to the target distribution can be constructed. Their efficiency can be controlled with
tuning parameters $k$ and $m$, for which we have proposed guidelines:
$k$ can be chosen as a large quantile of the meeting time $\tau$, and $m$ as a multiple of $k$.
Improving on these simple guidelines stands as a subject for future
research. In numerical experiments we have argued that the proposed estimators
yield a practical way of parallelizing MCMC computations in 
a range of settings.
We stress that coupling pairs of Markov chains does not improve their marginal
mixing properties, and that 
poor mixing of the underlying chains can lead to poor performance of the resulting estimator. 
The choice of initial distribution $\pi_0$ can have undesirable effects on the estimators, as in the multimodal example of Section \ref{subsec:Random-Walk-Metropolis-bimodal}.
Unreliable estimators would also result from stopping the
chains before their meeting time. 

Couplings of MCMC algorithms can be devised using maximal couplings, 
reflection couplings, and common random numbers. We have focused on couplings that can be implemented without
further analytical knowledge about the target distribution or about the MCMC kernels.
However, these couplings might result in prohibitively large meeting times,
either because the marginal chains mix slowly, as in \ref{subsec:Random-Walk-Metropolis-bimodal},
or because the coupling strategy is ineffective, as in Section \ref{subsec:Metropolis=002013Hastings}.

Regarding convergence diagnostics,
the proposed framework yields the following representation
for the total variation between $\pi_k$ and $\pi$, where 
$\pi_k$ denotes the marginal distribution of $X_k$:
\begin{eqnarray*}
     d_{\text{TV}}(\pi_k,\pi)  & = & \frac{1}{2} \sup_{h:\; |h|\leq 1} \left| \mathbb{E}[h(X_k)] - \mathbb{E}_{\pi}[h(X)]\right|\\
    & =&  \frac{1}{2}\sup_{h:\;|h|\leq 1} \left| \mathbb{E}\left[\sum_{t=k+1}^{\tau-1}\left(h(X_t)-h(Y_{t-1})\right) \right]\right|,
\end{eqnarray*}
Here the supremum ranges over all bounded measurable functions under the stated
assumptions.  The equality above has several consequences. For instance, the triangle inequality 
implies $d_{\text{TV}}(\pi_k,\pi)  \leq \min(1,
\mathbb{E}[\max(0, (\tau - k - 1))])$, and we can approximate
$\mathbb{E}[\max(0, (\tau - k - 1))]$ by Monte Carlo for a range of $k$ values.
This is pursued in \citet{biswas2019estimating}, where the proposed construction is 
extended to allow for arbitrary time lags between the coupled chains.

Thanks to its potential for parallelization, the proposed framework
can facilitate consideration of MCMC kernels that might be too expensive for
serial implementation.  For instance, one can improve MH-within-Gibbs samplers
by performing more MH steps per component update, HMC by using smaller step-sizes in the
numerical integrator \citep{heng2017unbiased}, and particle MCMC by using more particles
in the particle filters \citep{andrieu:doucet:holenstein:2010,jacob2017smoothing}.  We
expect the optimal tuning of MCMC kernels to be different in the proposed
framework than when used marginally. 

On top of enabling the application of the results of \citet{Glynn1991}
to accomodate budget constraints, the lack of bias of the proposed estimators
can be beneficial in combination with the law of total expectation,
to implement modular inference procedures as in Section \ref{subsec:cut-distribution}.
In \citet{rischard2018unbiased} the lack of bias 
is exploited in new estimators of Bayesian cross-validation criteria.
In \citet{chen2018blind} similar unbiased estimators 
are used in the expectation step of an expectation-maximization algorithm.
There may be other settings where the lack of bias is appealing,
for instance in gradient estimation for stochastic gradient descents \citep{doucet2017asymptotic}.

\subsubsection*{Acknowledgement.}

The authors are grateful to Jeremy Heng and Luc Vincent-Genod for useful
discussions. The authors gratefully acknowledge support by the National Science
Foundation through grants DMS-1712872 and DMS-1844695 (Pierre E. Jacob), and DMS-1513040 (Yves
F. Atchad\'e).

\bibliographystyle{abbrvnat}
\bibliography{Biblio}

\vspace*{2cm}

\appendix

The appendices contain the proofs of the results of the article, numerical
experiments with coupled Hamiltonian Monte Carlo on a Normal target of varying dimension,
Gibbs sampling on baseball batting average data, P\'olya-Gamma Gibbs sampling
on the German credit data, and Bayesian Lasso Gibbs sampling on the
diabetes data set.

\section{Proofs}

\subsection{Proof of Proposition \ref{prop:unbiased}}

We present the proof for $H_0(X,Y)$, a similar reasoning holds for $H_k(X,Y)$ and $H_{k:m}(X,Y)$.
We follow the same arguments as in \citet{glynn2014exact,vihola2015unbiased}.
To study $H_{0}(X,Y)=h(X_{0})+\sum_{t=1}^{\tau-1} (h(X_{t})-h(Y_{t-1}))$,
we introduce $\Delta_{0}=h\left(X_{0}\right)$ and $\Delta_{t}=h\left(X_{t}\right)-h\left(Y_{t-1}\right)$
for all $t\ge1$, and define $H_{0}^{n}(X,Y)=\sum_{t=0}^{n}\Delta_{t}$.
For simplicity we assume that $\Delta_{t}\in\mathbb{R}$, which corresponds
to studying the component-wise behavior of $H_{0}(X,Y)$.

We have $\mathbb{E}[\tau]<\infty$ from Assumption \ref{assumption:meetingtime},
so that the computing time of $H_0(X,Y)$ has a finite expectation.
Together with Assumption \ref{assumption:sticktogether}, this implies
that $H_{0}^{n}(X,Y)\to H_{0}(X,Y)$ almost surely as $n\to\infty$.
We now show that $H_{0}^{n}(X,Y)$ is a Cauchy sequence in $L_{2}$,
the complete space of random variables with finite two moments, 
that is, $\sup_{n^{\prime}\geq n}\mathbb{E}[(H_{0}^{n^{\prime}}(X,Y)-H_{0}^{n}(X,Y))^{2}]\to0$
as $n\to\infty$.  For $n^{\prime}\geq n$, we compute
\[
\mathbb{E}[(H_{0}^{n^{\prime}}(X,Y)-H_{0}^{n}(X,Y))^{2}]=\sum_{s=n+1}^{n^{\prime}}\sum_{t=n+1}^{n^{\prime}}\mathbb{E}[\Delta_{s}\Delta_{t}],
\]
and consider each term $\mathbb{E}[\Delta_{s}\Delta_{t}]$ for $(s,t)\in\{n+1,\ldots,n^{\prime}\}^{2}$.
The Cauchy-Schwarz inequality implies $\mathbb{E}[\Delta_{s}\Delta_{t}]\leq\left(\mathbb{E}[\Delta_{s}^{2}]\cdot\mathbb{E}[\Delta_{t}^{2}]\right)^{1/2}$,
and noting that $\mathbb{E}[\Delta_{t}^{2}]=\mathbb{E}[\Delta_{t}^{2}\cdot\mathds{1}(\tau>t)]$,
we can apply H\"older's inequality with $p=1+\eta/2$, $q=(2+\eta)/\eta$
for any $\eta>0$ to obtain

\[
\mathbb{E}[\Delta_{t}^{2}\cdot\mathds{1}(\tau>t)]\leq\mathbb{E}[|\Delta_{t}|^{2+\eta}]^{\nicefrac{1}{(1+\eta/2)}}\mathbb{\ \cdot\ E}[\mathds{1}(\tau>t)]^{\nicefrac{\eta}{(2+\eta)}}\leq\mathbb{E}[|\Delta_{t}|^{2+\eta}]^{\nicefrac{1}{(1+\eta/2)}}\cdot\left(C\delta^{t}\right)^{\nicefrac{\eta}{(2+\eta)}}.
\]
Here we have used Assumption \ref{assumption:meetingtime} to bound
$\mathbb{E}[\mathds{1}(\tau>t)]$. We can also use Assumption \ref{assumption:marginaldistributions}
together with Minkowski's inequality to bound $\mathbb{E}[|\Delta_{t}|^{2+\eta}]^{\nicefrac{1}{(1+\eta/2)}}$ by
a constant $\tilde{C}$, for all $t\geq 0$. Defining $\tilde{\delta}=\delta^{\nicefrac{\eta}{(2+\eta)}}\in(0,1)$
then gives the bound $\mathbb{E}[\Delta_{t}^{2}]\leq\tilde{C}\tilde{\delta}^{t}$
for all $t\geq 0$. This implies $\mathbb{E}[(H_{0}^{n^{\prime}}(X,Y)-H_{0}^{n}(X,Y))^{2}]\leq\bar{C}\tilde{\delta}^{n}$
for some other constant $\bar{C}$, and thus $(H_{0}^{n}(X,Y))$ is
Cauchy in $L_{2}$. This proves that its limit $H_{0}(X,Y)$ has finite
first and second moments. Assumption \ref{assumption:marginaldistributions}
implies that $\lim_{n\to\infty}\mathbb{E}[H_{0}^{n}(X,Y)]=\mathbb{E}_\pi[h(X)]$,
by a telescopic sum argument,
so we conclude that $\mathbb{E}[H_{0}(X,Y)]=\mathbb{E}_\pi[h(X)]$.
We can also obtain an
explicit representation of $\mathbb{E}[H_0(X,Y)^2]$
as the limit of $\mathbb{E}[H_{0}^{n}(X,Y)^{2}]$ when $n\to\infty$.

\subsection{Proof of Proposition \ref{prop:glivenko}}

We adopt a similar strategy to that of \citet{glynn2014exact}, Section 4. For
the $H_k(X,Y)$ case, the unbiased estimator of $F(s) = \mathbb{P}_\pi(X \leq
s)$ over $R$ samples is of the form
\[
\begin{aligned}
	\hat{F}^R(s) & = \frac{1}{R} \sum_{r=1}^R \left(
    \mathds{1}(X_k^{(r)} \leq s) + \sum_{\ell=k+1}^{\tau^{(r)}-1} (\mathds{1}(X_\ell^{(r)} \leq s) - \mathds{1}(Y_{\ell-1}^{(r)} \leq s) )
	\right).
\end{aligned}
\]
Here $\tau^{(r)}$ denotes the meeting time of the $r$-th independent run. We want to show that as $R \to \infty$, ${\sup_s |F(s) - \hat{F}^R(s)| \xrightarrow{a.s.} 0}$. Define
$G_{X,k}^R(s) = R^{-1} \sum_{r=1}^R 
    \mathds{1}(X_k^{(r)} \leq s)$.
For $\ell > k$, define
\[
	G_{X,\ell}^R(s) = \frac{1}{R} \sum_{r=1}^R \mathds{1}(X_\ell^{(r)} \leq s) \cdot \mathds{1}(\ell \leq \tau^{(r)}),
\qquad
	G_{Y,\ell}^R(s) = \frac{1}{R} \sum_{r=1}^R \mathds{1}(Y_{\ell-1}^{(r)} \leq s)  \cdot \mathds{1}(\ell \leq \tau^{(r)}).
\]
Then $\hat{F}^R(s) = G_{X,k}^R(s) + \sum_{\ell=k+1}^\infty (G_{X,\ell}^R(s) - G_{Y,\ell}^R(s))$. By the standard Glivenko-Cantelli theorem, as $\R \to \infty$ we have
\[
	\sup_s \left| G_{X,k}^R(s) - \mathbb{P}(X_k \leq s) \right| \xrightarrow{a.s.}  0,
\qquad
	\sup_s \left| G_{X,\ell}^R (s) - \mathbb{E}[\mathds{1}(X_\ell \leq s) \cdot \mathds{1}(\ell \leq \tau)] \right| \xrightarrow{a.s.}  0,
\]
\[
	\sup_s \left| G_{Y, \ell}^R (s) - \mathbb{E}[\mathds{1}(Y_{\ell-1} \leq s) \cdot \mathds{1}(\ell \leq \tau)] \right| \xrightarrow{a.s.}  0,
\]
for each $\ell > k$. Next, we observe that, for all $s,\ell$, 
\[
	\mathbb{E}[( \mathds{1}(X_\ell \leq s) -  \mathds{1}(Y_{\ell-1} \leq s)) \cdot \mathds{1}(\tau \geq \ell)] = \mathbb{P}(X_\ell \leq s) - \mathbb{P}(X_{\ell-1} \leq s).
\]
This holds for a simple reason in our setting. For any $h(\cdot)$ and any $\ell$,
\[
\begin{aligned}
	\mathbb{E}[h(X_\ell) - h(Y_{\ell-1})]  
	& =  \mathbb{E}[(h(X_\ell) - h(Y_{\ell-1}))\mathds{1}(\tau>\ell)] + \mathbb{E}[(h(X_\ell) - h(Y_{\ell-1}))\mathds{1}(\tau \leq \ell)] \\
	& = \mathbb{E}[(h(X_\ell) - h(Y_{\ell-1}))\mathds{1}(\tau>\ell)] 
\end{aligned}		
\]
since if $\tau \leq \ell$ then $X_\ell = Y_{\ell-1}$ by Assumption \ref{assumption:sticktogether}. Applying this result with $h(\cdot) = \mathds{1}(\cdot \leq s)$ yields the desired statement.

The above implies that for any finite $i \geq k$ we have
\[
\begin{aligned}
& \left| 
G_{X,k}^R(s) - \mathbb{P}(X_k \leq s) +
\sum_{\ell=k+1}^i \Big( G_{X,\ell}^R(s) - G_{Y,\ell}^R(s)
- \mathbb{E}[(\mathds{1}(X_\ell \leq s) - \mathds{1}(Y_{\ell-1} \leq s) ) \cdot \mathds{1}(\ell \leq \tau)] \Big) \right| \\
& = \left| G_{X,k}^R(s) - \mathbb{P}(X_k \leq s) + \sum_{\ell=k+1}^i \Big( G_{X, \ell}^R (s) - G_{Y,\ell}^R(s)
- (\mathbb{P}(X_\ell \leq s) - \mathbb{P}(X_{\ell-1} \leq s)) \Big) \right| \\
&= \left| 
\Big(G_{X,k}^R(s) + \sum_{\ell=k+1}^i (G_{X,\ell}^R(s) - G_{Y,\ell}^R(s))\Big) - \mathbb{P}(X_i \leq s) \right|.
\end{aligned}
\]
Hence
\[
	\sup_s \left| 
\Big(G_{X,k}^R(s) + \sum_{\ell=k+1}^i (G_{X, \ell}^R(s) - G_{Y, \ell}^R(s))\Big) - \mathbb{P}(X_i \leq s) \right| \to 0.
\]
We have assumed that $(X_t)_{t\geq0}$ converges to $\pi$ in total variation, which implies $\sup_s | \mathbb{P}(X_i \leq s) - F(s)| \to 0$ as $i\to\infty$. Also, we note that for all $s$,
\[
\left| \sum_{\ell > i} G_{X, \ell}^R(s) - G_{Y, \ell}^R(s) \right|
\leq
\sum_{\ell > i} \frac{1}{R} \sum_{r=1}^R \mathds{1}(\ell \leq \tau^{(r)}) \to 
\sum_{\ell > i} \mathbb{P}(\ell \leq \tau)
\]
almost surely by the strong law of large numbers. Assumption \ref{assumption:meetingtime} implies that this quantity goes to 0 as $i \to \infty$.

Combining these observations with the result obtained for finite $i$, we conclude that 
$\sup_s |\hat{F}^R(s) - F(s) | \to 0$ as $R \to \infty$, almost surely. 
The reasoning holds for the function $\hat{F}^{R}$ corresponding to $H_{k:m}(X,Y)$ instead of $H_k(X,Y)$; 
such a function is simply the average of a finite number of functions associated with $H_\ell(X,Y)$ for $\ell \in \{k,\ldots,m\}$.

\subsection{Proof of Proposition \ref{prop1}}
  
Throughout the proof $C$ denotes a generic finite constant whose actual value
may change from one appearance to the next. We will use the usual Markov chain
notation. In particular if $f:\;\mathcal{X}\times\mathcal{X}\to\mathbb{R}$ is a
measurable function then $\bar P [f](x,y) :=\int_{\mathcal{X}\times\mathcal{X}}
\bar P((x,y),dz) f(z)$. Note that from the construction of $\bar P$, if $f$ is
depends only on $x$ (resp. $y$), that is $f(x,y) = f(x)$ (resp. $f(x,y)=f(y)$),
then $\bar P [f](x,y) =  Pf(x)$ (resp. $\bar P[f](x,y) = Pf(y)$).

For $k\geq 1$, we consider the general problem of bounding $\mathbb{E}[S_{k}^2]$, where $S_{k}$ is of the form
\[S_{k} = \mathds{1}(\tau>k) \sum_{t=k}^{\tau-1} b_t\left(h(X_{t})-h(Y_{t-1})\right),\]
for some arbitrary bounded sequence $(b_t)_{t\geq 0}$. 
Fix an integer $N\geq k$, and set 
\[S_k^{(N)} = \mathds{1}(\tau>k) \sum_{t=k}^{N\wedge\tau-1} b_t\left(h(X_{t})-h(Y_{t-1})\right).\]
The same argument as in the proof of Proposition \ref{prop:unbiased} can be applied here and shows that $S_k^{(N)}$ is a Cauchy sequence in $L_2$ that converges to $S_k$, as $N\to\infty$, so that
\[\mathbb{E}[S_k^2] = \lim_{N\to\infty} \mathbb{E}\left[(S_k^{(N)})^2\right].\]
Since $|h|_{V^\beta}:=\sup_x |h(x)|/V^\beta(x)<\infty$, and under Assumption \ref{assumption:drift}, there exists a measurable function $g:\;\mathcal{X} \to\mathbb{R}$ such that $|g|_{V^\beta}<\infty$, and $g-Pg=h-\pi(h)$. To see this, first note that the drift condition (\ref{assumption:drift}) implies that for any $\beta\in(0,1/2)$, we have $PV^\beta(x) \leq \lambda^\beta V^\beta(x) + b^\beta\mathds{1}(x\in \mathcal{C})$, for all $x\in\mathcal{X}$. Indeed  by Jensen's inequality $PV^\beta(x) \leq (PV(x))^\beta \leq (\lambda V(x) + b\mathds{1}(x\in \mathcal{C}))^\beta \leq \lambda^\beta V^\beta(x) + b^\beta \mathds{1}(x\in \mathcal{C})$, using the fact that for all $x,y\geq 0$ and $\alpha\in[0,1]$, $(x+y)^\alpha \leq x^\alpha +y^\alpha$.  The drift condition in $V^\beta$ together with the fact that $P$ is $\phi$-irreducible and aperiodic implies that there exists $\rho_\beta\in (0,1)$, $C_\beta<\infty$ such that for all $x\in\mathcal{X}$, $n\geq 0$,
\begin{equation}\label{geo:rate}
\|P^n(x,\cdot)-\pi\|_{V^\beta} \leq C_\beta V^\beta(x) \rho_\beta^n,\end{equation}
where for a function $W:\;\Xset\to [1,\infty)$, the $W$-norm between two probability measures $\mu,\nu$ is defined as
\[\|\mu-\nu\|_W : = \sup_{f\textsf{ meas.}:\;|f|_W\leq 1} \;|\mu(f)-\nu(f)|,\]
and $|f|_W : =  \sup_x |f(x)|/W(x)$. This result can be found in Theorem 15.0.1 of \cite{meyn:tweedie:1993}. It follows from (\ref{geo:rate}) that $\sum_{j\geq 0} |P^j(h-\pi(h))(x)| \leq |h|_{V^\beta}\sum_{j\geq 0} \|P^j(x,\cdot) - \pi\|_{V^\beta} \leq \frac{C_\beta |h|_{V^\beta} V^\beta(x)}{1-\rho_\beta}<\infty$. Hence the function 
\[g(x) = \sum_{j\geq 0} P^j(h-\pi(h))(x),\;\;\;x\in\mathcal{X},\]
is well-defined and measurable (as a limit of a sequence of measurable functions) and satisfies $|g(x)| \leq C_\beta |h|_{V^\beta} V^\beta(x)/ (1-\rho_\beta)$. And since $PV^\beta$ is finite everywhere, by Lebesgue's dominated convergence we deduce that $Pg$ is finite everywhere as well and 
\[Pg(x) = \int g(y) P(x,dy) = \sum_{j\geq 0} \int (P^j(h-\pi(h))(y))P(x,dy) = \sum_{j\geq 1} P^j(h-\pi(h))(x).\]
Hence $g - Pg = h- \pi(h)$, as claimed.

Hence, with $Z_t := (X_{t},Y_{t-1})$, and $\bar g(x,y) = g(x)-g(y)$, we have
\[h(X_t)-h(Y_{t-1}) = \bar g(Z_t) - \bar P [\bar g](Z_t).\]
Using this and a telescoping sum argument, we write
\begin{eqnarray*}
S_k^{(N)} &=& \sum_{t=k}^{N-1} b_{t}\left(h(X_t)-h(Y_{t-1})\right)\mathds{1}(\tau> t)\\
& = & \sum_{t=k}^{N-1}b_t\left(\bar g(Z_t) - \bar P[\bar g](Z_t)\right)\mathds{1}(\tau> t)\\
& = & \sum_{t=k}^{N-1} b_{t}\left(\bar g(Z_{t+1}) - \bar P [\bar g](Z_t)\right)\mathds{1}(\tau> t) \\
&&+ \sum_{t=k}^{N-1} \left(b_{t}\bar g(Z_t)\mathds{1}(\tau> t) - b_{t+1}\bar g(Z_{t+1})\mathds{1}(\tau> t+1)\right)\\
&& + \sum_{t=k}^{N-1} \left(b_{t+1} \mathds{1}(\tau> t+1) - b_t\mathds{1}(\tau> t)\right) \bar g(Z_{t+1}).
\end{eqnarray*}
Since $\bar g(Z_{t+1}) = 0$ on $\{\tau=t+1\}$, the last term in the above
display reduces to $\sum_{t=k}^{N-1} (b_{t+1}-b_t) \bar
g(Z_{t+1})\mathds{1}(\tau> t+1)$, and we obtain
\begin{multline}\label{eq:prop31:S2}
S_k^{(N)} = \sum_{t=k}^{N-1} b_{t}\left(\bar g(Z_{t+1}) - \bar P [\bar g](Z_t)\right)\mathds{1}(\tau> t) \\
+ b_k\bar g(Z_k)\mathds{1}(\tau> k) - b_N\bar g(Z_N)\mathds{1}(\tau> N) + \sum_{t=k}^{N-1} (b_{t+1}-b_t) \bar g(Z_{t+1})\mathds{1}(\tau> t+1).
\end{multline}
Let $\mathcal{F}_t$ denote the sigma-algebra generated by the variables $X_0,(X_1,Y_0),\ldots,(X_t,Y_{t-1})$. Note that $\{\tau>t\}$ belongs to $\mathcal{F}_t$. Hence
\begin{eqnarray*}
\mathbb{E}\left[b_{t}\left(\bar g(Z_{t+1}) - \bar P [\bar g](Z_t)\right)\mathds{1}(\tau> t)\vert \F_{t}\right]
&=&  b_t \mathds{1}(\tau> t)\mathbb{E}\left[\bar g(Z_{t+1}) - \bar P[ \bar g](Z_t)\vert \F_{t}\right] \\
& = & b_t \mathds{1}(\tau> t)\left(\bar P[\bar g](Z_t) - \bar P [\bar g](Z_t) \right) 
 =  0.
\end{eqnarray*} 
In other words, $\left\{\left(\sum_{t=k}^{j} b_{t}\left(\bar g(Z_{t+1}) - \bar
P [\bar g](Z_t)\right)\mathds{1}(\tau> t),\mathcal{F}_j\right),\;k\leq j\leq
N-1\right\}$ is a martingale. The orthogonality of the martingale
increments gives
\[\mathbb{E}\left[\left(\sum_{t=k}^{N-1} b_{t}\left(\bar g(Z_{t+1}) - \bar P [\bar g](Z_t)\right)\mathds{1}(\tau> t)\right)^2\right] = \sum_{t=k}^{N-1} b_t^2\mathbb{E}\left[\left(\bar g(Z_{t+1}) - \bar P [\bar g](Z_t)\right)^2\mathds{1}(\tau> t)\right].\]
We use this together with (\ref{eq:prop31:S2}), the convexity of the squared norm, and Minkowski's inequality to conclude that
\begin{multline}\label{proof:propvar:eq2}
\mathbb{E}\left[(S_k^{(N)})^2\right] \leq 4\sum_{t=k}^{N-1}
b_t^2\mathbb{E}\left[\left(\bar g(Z_{t+1}) - \bar P [\bar
g](Z_t)\right)^2\mathds{1}(\tau> t)\right] + 4b_k^2\mathbb{E}\left[\bar
g^2(Z_k)\mathds{1}(\tau> k)\right]\\
  +4b_N^2\mathbb{E}\left[\bar g^2(Z_N)\mathds{1}(\tau> N)\right] +4\left[\sum_{t=k}^{N-1} |b_{t+1}-b_t|\mathbb{E}^{1/2}\left( \bar g^2(Z_{t+1})\mathds{1}(\tau> t+1)\right)\right]^2.
\end{multline}
Assumption \ref{assumption:drift} together with $\pi_0(V)<\infty$, implies that 
\begin{equation}\label{eq:control:moment}
    \sup_{n\geq 0}\mathbb{E}[V(X_n)]\leq C,\end{equation}
 for some finite constant $C$. Indeed, $\mathbb{E}[V(X_n)] = \int \pi_0( dx) P^n V(x)$, and a repeated application of the drift condition (\ref{assumption:drift}) implies that $P^n V(x) \leq \lambda^n V(x) + \frac{b}{1-\lambda}$, for all $x\in\mathcal{X}$. For any $t\geq 0$, and for $1<p=\frac{1}{2\beta}$, and $\frac{1}{p}+\frac{1}{q}=1$, we have 
\begin{eqnarray*}
\mathbb{E}\left[\left(V^\beta(X_t) + V^\beta(Y_{t-1})\right)^2\mathds{1}(\tau>t)\right] & \leq & \mathbb{E}^{1/p}\left[\left(V^\beta(X_t) + V^\beta(Y_{t-1})\right)^{2p}\right]\mathbb{P}^{1/q}(\tau>t)\;\;\mbox{(H\"older)}\\
& \leq & \left\{\mathbb{E}^{1/(2p)}\left[V^{2p\beta}(X_t)\right] + \mathbb{E}^{1/(2p)}\left[V^{2p\beta}(Y_{t-1})\right]\right\}^2 \\
&& \;\;\;\;\;\;\;\;\;\;\;\;\;\;\hspace{3cm}\times \mathbb{P}^{1/q}(\tau>t)\;\;\;\mbox{(Minkowski)}\\
& \leq & C\mathbb{P}^{1/q}(\tau>t)\hspace{4.8cm}(\mbox{by \eqref{eq:control:moment}})\\
& \leq & C\delta_\beta^{t} \hspace{6cm}(\mbox{by Assumption }\ref{assumption:meetingtime}),
\end{eqnarray*}
where $\delta_\beta = \delta^{1/q}$ with $\delta$ as in Assumption \ref{assumption:meetingtime}.
Note that all the expectations on the right hand side of (\ref{proof:propvar:eq2}) are of the form
$\mathbb{E}\left[ \bar g^2(Z_{t})\mathds{1}(\tau> t)\right]$, except for the term $\mathbb{E}\left[\left(\bar g(Z_{t+1}) - \bar P [\bar g](Z_t)\right)^2\mathds{1}(\tau> t)\right]$. However by the martingale difference property, $\mathbb{E}\left[\bar P [\bar g](Z_t)\left(\bar g(Z_{t+1}) - \bar P [\bar g](Z_t)\right)\mathds{1}(\tau> t)\right] = 0$, so that for all $t\geq k$,
\begin{eqnarray*}
\mathbb{E}\left[\left(\bar g(Z_{t+1}) - \bar P [\bar g](Z_t)\right)^2\mathds{1}(\tau> t)\right] &= &\mathbb{E}\left[\mathds{1}(\tau> t)\bar g(Z_{t+1})^2\right] - \mathbb{E}\left[\bar g(Z_{t+1})\bar P [\bar g](Z_t)\mathds{1}(\tau> t)\right]\\
& = & \mathbb{E}\left[\mathds{1}(\tau> t)\bar g(Z_{t+1})^2\right] - \mathbb{E}\left[\left(\bar P [\bar g](Z_t)\right)^2\mathds{1}(\tau> t)\right] \;( \mbox{by conditioning on } \F_t)\\
& \leq  & \mathbb{E}\left[\mathds{1}(\tau> t)\bar g(Z_{t+1})^2\right]\\
& \leq  & |g|_{V^\beta}^2\mathbb{E}\left[\mathds{1}(\tau> t)\left(V^\beta(X_{t+1}) + V^\beta(Y_t)\right)^2\right]\\
& \leq & C\delta_\beta^{t},
\end{eqnarray*}
using the same arguments as above. On the other hand, for all $t\geq k$, the term $\mathbb{E}\left[ \bar g^2(Z_{t})\mathds{1}_{\{\tau> t\}}\right]$ satisfies
\[\mathbb{E}\left[ \bar g^2(Z_{t})\mathds{1}(\tau> t)\right] \leq |g|_{V^\beta}^2\mathbb{E}\left[\left(V^\beta(X_t) + V^\beta(Y_{t-1})\right)^2\mathds{1}(\tau> t)\right]\leq C\delta_\beta^t,\]
as seen above. In conclusion, all the expectations appearing in \eqref{proof:propvar:eq2} are upper bounded by some constant times terms of the form $\delta_\beta^t$. 
We conclude that
\[\mathbb{E}\left[(S_k^{(N)})^2\right]  \leq C\left(b_k^2\delta_\beta^k + b_N^2\delta_\beta^N + \sum_{t=k}^{N-1}b_k^2\delta_\beta^t + \left[\sum_{t=k}^{N-1} |b_{t+1}-b_t|\delta_\beta^t\right]^2\right).\]
Letting $N\to\infty$, we conclude that
\[\mathbb{E}[S_k^2] \leq C \left(b_k^2\delta_\beta^k + \sum_{j\geq k}b_j^2\delta_\beta^j + \left[\sum_{j\geq k} |b_{j+1}-b_j|\delta_\beta^j\right]^2\right).\]
In the particular case of $\eta_{k:m}$, 
we have $\eta_{k:m} = \mathds{1}(\tau>k) \sum_{t=k}^{\tau-1}
\min\left(1,\frac{t+1-k}{m+1-k}\right)\left(h(X_{t+1})-h(Y_{t})\right)$. Hence
$b_k = 0$, 
$b_t = (t-k)/(m-k+1)$ if $k< t\leq m+1$, $b_t=1$ if $t> m+1$. We then
obtain the bound of Proposition \ref{prop1}.

\subsection{Proof of Proposition \ref{prop:vgeometric}}

Here $Z_n$ is defined as $(X_n,Y_{n-1})$ for all $n\geq 1$.
The assumption in \eqref{mino}, within the statement of Proposition \ref{prop:vgeometric}, implies that for $(x,y) \in \mathcal{C}\times
\mathcal{C}$, $\bar P$ can be written as a mixture \[\bar P\left((x,y),
dz\right) = \epsilon_{x,y}\nu_{x,y}( dz) + (1-\epsilon_{x,y}) R\left((x,y),
dz\right),\] where $\epsilon_{x,y}\geq \epsilon$, $\nu_{x,y}( dz)$ is a
restriction of $\bar P\left((x,y), dz\right)$ on $\mathcal{D}$ (that is for any measurable subset $A$ of $\mathcal{D}$, $\bar P\left((x,y), A\right) = P\left((x,y), A\cap\mathcal{D}\right)/P\left((x,y), \mathcal{D}\right)$), and
$R\left((x,y), dz\right)$ is the restriction of $\bar P\left((x,y), dz\right)$
on $(\mathcal{X}\times \mathcal{X})\setminus \mathcal{D}$. This means that whenever $(x,y)\in \mathcal{C}\times \mathcal{C}$ one can sample from $\bar P((x,y),\cdot)$ by drawing independently a Bernoulli random variable $J$, with probability of success $\epsilon_{x,y}$. Then if $J=1$, we draw from $\nu_{x,y}$, if $J=0$, we draw from $R\left((x,y), \cdot\right)$. From this
decomposition, the proof of the proposition follows the same lines as in
\cite{douc:moulines:rosenthal:04}, and we give the details only for completeness. We cannot directly invoke their result since their assumptions do not seem to apply to our setting.  

Set $\bar V(x,y) = \frac{1}{2}(V(x) + V(y))$. First we show that the bivariate
kernel satisfies a geometric drift towards $\mathcal{C}
\times \mathcal{C}$.  That is, there exists $\alpha\in (0,1)$ such that
\begin{equation}\label{biv:drift}
\bar P\bar V(x,y) \leq \alpha \bar V(x,y),\;\;\;(x,y)\notin\mathcal{C}\times\mathcal{C}.\end{equation}
 Indeed for $(x,y)\notin \mathcal{C}\times \mathcal{C}$, since $V\geq 1$, and $\mathcal{C} = \{V\leq L\}$, $\bar V(x,y) \geq (1+L)/2$. In other words, $\frac{1}{2} \leq \bar V(x,y)/(1+L)$. Therefore,
\[\bar P\bar V(x,y) = \frac{1}{2}\left(PV(x) + PV(y)\right) \leq \lambda\bar V(x,y) + \frac{b}{2}\leq \lambda\bar V(x,y) + \frac{b}{1+L} \bar V(x,y)\leq \alpha\bar V(x,y), \]
with $\alpha = \lambda +\frac{b}{1+L}<1$. We set
\[B = \max\left(1,\frac{1}{\alpha}\sup_{(x,y)\in\mathcal{C}\times\mathcal{C}}\frac{\bar P[\bar V\mathds{1}_{\D^c}](x,y)}{\bar V(x,y)}\right) \leq \frac{\lambda + b}{\alpha}.\]
In this section $\mathds{1}_\mathcal{S}(\cdot)$ refers to the indicator function on the set $\mathcal{S}$.
Let $N_n$ denote the number of visits to $\mathcal{C}\times\mathcal{C}$ by time $n$. Then
\[\mathbb{P}(\tau>n) = \mathbb{P}(\tau>n,\; N_{n-1}\geq j) + \mathbb{P}(\tau>n,\; N_{n-1}<j).\] 
The event $\{\tau>n,\; N_{n-1}\geq j\}$ implies that no success occurred within at least $j$ independent Bernoulli random variables each with probability of success at least $\epsilon$. Hence
\[\mathbb{P}(\tau>n,\; N_{n-1}\geq j)\leq (1-\epsilon)^j.\]
For the second term, we have (since $B\geq 1$, and the chains stay together after meeting via Assumption \ref{assumption:sticktogether}),
\[
\mathbb{P}(\tau>n,\; N_{n-1}\leq j-1) \leq \mathbb{P}\left(Z_n\notin \D, B^{-N_{n-1}} \geq B^{-(j-1)}\right)= \mathbb{P}\left(\mathds{1}_{\D^c}(Z_n) B^{-N_{n-1}}\geq B^{-(j-1)}\right).\]
Then use Markov's inequality to conclude that
\begin{multline*}
\mathbb{P}(\tau>n,\; N_{n-1}\leq j-1) \leq B^{j-1} \mathbb{E}\left[\mathds{1}_{\D^c}(Z_n) B^{-N_{n-1}}\right]\\
\leq B^{j-1} \mathbb{E}\left[\mathds{1}_{\D^c}(Z_n) B^{-N_{n-1}}\bar V(Z_n)\right] = \alpha ^n B^{j-1} \mathbb{E}[M_n],\end{multline*}
where $M_n = \mathds{1}_{\D^c}(Z_n) \alpha^{-n} B^{-N_{n-1}}\bar V(Z_n)$ (set $N_{0}=0$ so that $M_1$ is well-defined). The result follows by noting that $\{M_n,\F_n\}$ is a super-martingale, where $\F_n = \sigma(Z_1,\ldots,Z_n)$, so that $\mathbb{E}[M_n]\leq \mathbb{E}[M_1]\leq \pi_0(V)+\pi_0(PV)\leq (1+\lambda)\pi_0(V)+b$, which implies that
\[\mathbb{P}(\tau>n) \leq (1-\epsilon)^j + \alpha^n B^{j-1} \left((1+\lambda)\pi_0(V)+b\right).\]
Since $\alpha < 1$, there exists an integer $k_0\geq 1$ such that $\alpha B^{\frac{1}{k_0}}< 1$. In that case for $n\geq k_0$ one can take $j=\lceil n/k_0\rceil$, to get 
\[\mathbb{P}(\tau>n)\leq \left\{(1-\epsilon)^\frac{1}{k_0}\right\}^n + \left((1+\lambda)\pi_0(V)+b\right) \left\{\alpha B^{\frac{1}{k_0}}\right\}^n ,\]
as claimed.  

The argument that $\{M_n,\F_n\}$ is a super-martingale is as follows. We need to show that for all $n\geq 1$, $\mathbb{E}[M_{n+1}\vert \F_n]\leq M_n$. Note that $\mathbb{E}[M_{n+1}\vert \F_n] = 0\leq M_n$ on $Z_n\in \D$. So it is enough to assume that $Z_n \notin\D$. Now, suppose also that $Z_n\notin \mathcal{C}\times\mathcal{C}$. Then $N_n=N_{n-1}$, and
\begin{eqnarray*}
\mathbb{E}[M_{n+1}\vert \F_n] & = &  \alpha^{-n-1}\mathbb{E}\left[B^{-N_{n-1}} \mathds{1}_{\D^c}(Z_{n+1})\bar V(Z_{n+1})\vert \F_n\right],\\
& =& \alpha^{-n-1} B^{-N_{n-1}} \mathbb{E}\left[\mathds{1}_{\D^c}(Z_{n+1})\bar V(Z_{n+1})\vert Z_n\right],\\
& \leq & \alpha^{-n-1} B^{-N_{n-1}} \mathbb{E}\left[\bar V(Z_{n+1})\vert Z_n\right],\\
& \leq & \alpha^{-n} B^{-N_{n-1}} \bar V(Z_{n}),\\
& = & M_n.
\end{eqnarray*}
Suppose now that $Z_n\in \mathcal{C}\times\mathcal{C}$. Then $N_n = N_{n-1}+1$. Hence 
\begin{eqnarray*}
\mathbb{E}[M_{n+1}\vert \F_n] & = &  \alpha^{-n-1}B^{-N_{n-1}-1}\mathbb{E}\left[\mathds{1}_{\D^c}(Z_{n+1})\bar V(Z_{n+1})\vert \F_n\right],\\
& = & \alpha^{-n} B^{-N_{n-1}}\bar V(Z_n) \frac{1}{\alpha B}\frac{\bar P[\mathds{1}_{\D^c}\bar V](Z_n)}{\bar V(Z_n)} ,\\
& \leq & \alpha^{-n} B^{-N_{n-1}}\bar V(Z_n) = M_n.
\end{eqnarray*}

\section{Hamiltonian Monte Carlo on multivariate Normals\label{subsec:HMCscaling}}

As Section \ref{sec:Practical-couplings} in the main document,
we perform experiments with a $d$-dimensional Normal target distribution 
$\pi = \mathcal{N}(0, V)$, where $V$ is the inverse
of a matrix drawn from a Wishart distribution,
with identity scale matrix and $d$ degrees of freedom. We provide average meeting times
obtained in varying dimensions, when using the coupled HMC algorithm described in \citet{heng2017unbiased}.
The latter article presents similar experiments but for a different choice of matrix $V$, thus we provide
the present section for completeness.

Let us introduce a Markov kernel $P$ as a mixture of two kernels, an MH kernel $P_{\text{MH}}$
and an HMC kernel $P_{\text{HMC}}$. We first describe $P_{\text{MH}}$ and a coupling of it.
The kernel is an MH kernel with Normal random walk proposals, with a covariance matrix equal to $10^{-8}$
times the identity matrix. The coupled version of $P_{\text{MH}}$ uses a maximal coupling of the proposals
(as in Algorithm \ref{alg:maximalcoupling} of the main document).

The kernel $P_{\text{HMC}}$ corresponds to an HMC algorithm, with mass matrix given by the inverse of the target
variance $V$.
This preconditioning mechanism is motivated by considerations similar to those in \citet{girolami2011riemann}.
In the present case of Normal distributions, this is
particularly advantageous as it leads to a complete decoupling of the $d$ components of the target; see Proposition 3.1 in \citet{bou2018geometric}. 
We discretize Hamiltonian equations with a leap-frog integrator, using a stepsize of $\varepsilon = 0.1\times d^{-1/4}$,
and a number of steps of $L = 1 + \lfloor \varepsilon^{-1}\rfloor$, which corresponds to a trajectory length $\varepsilon L$ of approximately one.
The coupling of such Hamiltonian kernels is done by using common random numbers for the momentum variables, i.e. a synchronous 
coupling \citep{bou2018coupling}. The kernels $P_{\text{MH}}$ and $P_{\text{HMC}}$, and their coupled counterparts,
are combined into mixtures $P$ and $\bar{P}$, by assigning weights respectively of $0.05$ and $0.95$, i.e. an MH step is performed with probability $0.05$.

We consider two types of initialization $\pi_0$: either the target distribution $\pi$, 
or a Normal distribution $\mathcal{N}(1_d, I_d)$, with $1_d$ a vector of ones and $I_d$ the identity matrix.
With the latter initialization, we observed a very low acceptance rate when using the stepsize $\varepsilon$ given above.
We did not observe any issue when the chains were started from $\pi$. We also did not observe
such issues when $V$ was replaced by the identity matrix. 
In principle, smaller stepsizes could be
chosen, in order to increase the acceptance rate. However, this would result
in more expensive iterations as $L$ is defined as $1 + \lfloor \varepsilon^{-1}\rfloor$ above.
Instead, we resort to the following heuristic strategy, which appears to solve the issue in the present example. We draw an initial position 
from $\mathcal{N}(1_d,I_d)$,
then we perform $10$ steps of ``unadjusted HMC'', with $\varepsilon = 0.1\times d^{-1/4}$,
and $L = 1 + \lfloor \varepsilon^{-1}\rfloor$ as described above. By ``unadjusted HMC'', we refer to a scheme where the final point of the Hamiltonian
trajectory is accepted with probability one, i.e. no MH correction is applied. This initialization procedure is simply a redefinition of $\pi_0$,
and thus would not jeopardize the validity of the proposed estimators.

\begin{figure}
\centering
    \includegraphics[width=0.5\textwidth]{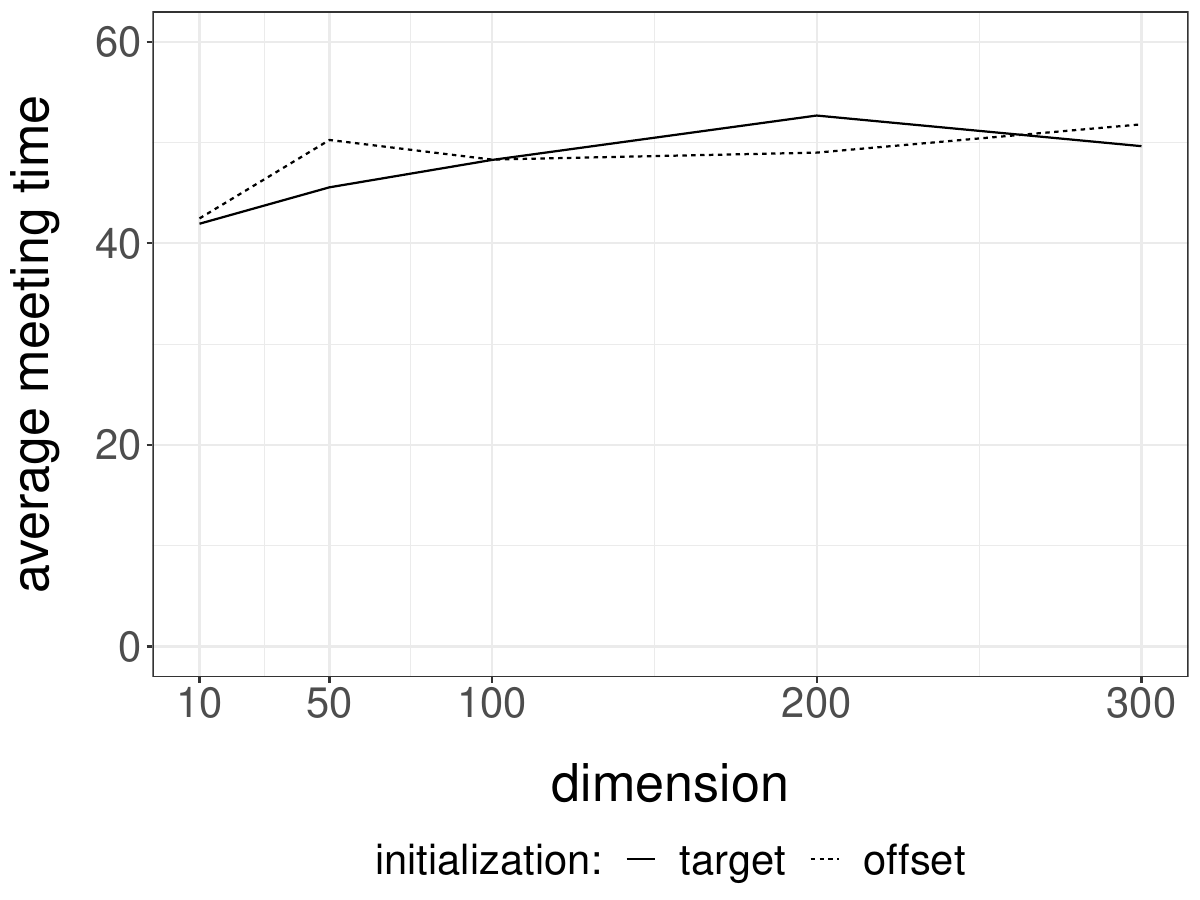}
\caption{\label{fig:scaling:hmc}
Scaling of the average meeting time of a coupled HMC algorithm with the dimension of the target $\mathcal{N}(0,V)$,
where $V$ is the inverse of a Wishart draw, as described in Section \ref{subsec:HMCscaling}.
The chains are either initialized from the target, or from a Normal $\mathcal{N}(1_d, I_d)$ followed by 10 steps 
of unadjusted HMC, as described in Section \ref{subsec:HMCscaling}; this is referred to as ``offset'' in the legend.
}
\end{figure}

The results under both initializations are shown in Figure \ref{fig:scaling:hmc},
where we observe average meeting times that increase very slowly with the dimension of the target distribution. Thanks to 
the initialization strategy described above, we obtain similar meeting times when
starting from the chains from $\pi$ or from $\mathcal{N}(1_d,I_d)$.

\section{Baseball batting averages \label{subsec:Gibbs-rosenthal2000}}

We consider a classic Gibbs sampler discussed in the context of parallel
computing in \citet{rosenthal2000parallel}.  In \citet{rosenthal1996analysis},
it was proved that the chain produced by this Gibbs sampler converges in total variation to within $1\%$
of its stationary distribution after at most $140$ iterations. This type of derivation 
is technically challenging and has been done only in specific cases.  Using this result,
\citet{rosenthal2000parallel} recommends to run parallel chains with a burn-in of $140$ iterations.
We will compare this value with choices of $k$ and $m$ for the proposed unbiased estimators.

The data $(Z_{n})_{n=1}^{K}$
are baseball players' batting averages taken from Table 1 of \citet{morris1983parametric},
where $K=18$. In the model, each $Z_{n}$ is assumed to follow $\mathcal{N}(\theta_{n},V)$
where $V$ is fixed to $0.00434$. Then, $\theta_{n}$ is assumed
to follow $\mathcal{N}(\mu,A)$, where $\mu$ is given a flat prior
and $A$ an Inverse Gamma $(a,b)$, with $a=-1$ and $b=2$. 
Here the Inverse Gamma $(\alpha,\beta)$ distribution 
has pdf $p(x; \alpha,\beta) = \Gamma(\alpha)^{-1} \beta^\alpha x^{-\alpha-1} \exp(-\beta / x)$.
The Gibbs
updates are as follows:
\begin{align*}
    A | \text{rest} &\sim \text{Inverse Gamma}(a+(K-1)/2, b+\sum_{n=1}^{K}(\theta_{n}-\bar{\theta})^{2}/2),
     \; \text{where} \; \bar{\theta}=K^{-1}\sum_{n=1}^{K}\theta_{n},\\
    \mu| \text{rest} &\sim  \mathcal{N}\left(\bar{\theta}, A/K\right),\\
    \theta_n | \text{rest} &\sim \mathcal{N}\left((V+A)^{-1}(\mu V+Z_{n}A), \left(V+A\right)^{-1}(AV)\right), \quad \text{for all }n \in \{1,\ldots,K\}.
\end{align*}
The initial values of the Gibbs sampler can be taken as $K^{-1}\sum_{n=1}^{K}Z_{n}$
for all $\theta_{n}$ \citep{rosenthal2000parallel}. 
We couple this Gibbs sampler using maximal couplings of the conditional updates.
The parameter space is $20$-dimensional, but the 
chains meet as soon as the components $(A,\mu)$ meet, by construction.
We consider the test function $h: (A,\mu,\theta_1,\ldots,\theta_K) \mapsto \theta_1$,
that is, we are interested in the posterior expectation of $\theta_1$,
which represents the mean of the batting average of the first player.

We first run the coupled chains $1,000$ times independently in parallel.
All observed meeting times were less or equal to $4$, so we consider 
$k$ equal to $1$, $3$ and $5$. Then, we consider $m$ equal to $k$, $5k$ and $10k$.
Over $10,000$ independent experiments, we approximate the expected cost
$\mathbb{E}[2(\tau-1) + \max(1,m-\tau+1)]$, the variance $\mathbb{V}[H_{k:m}(X,Y)]$,
and we compute the inefficiency as the product of expected cost and variance.
We then divide the inefficiency by the asymptotic variance of the MCMC estimator,
denoted by $V_\infty$,
obtained from $5\times 10^5$ iterations and a burn-in period of $10^3$, and the CODA
package \citep{plummer2006coda}. The results
are shown in Table \ref{table:baseball}. 
We see that when $k$ and $m$ are large enough we can retrieve
an inefficiency comparable to that of the underlying MCMC algorithm;
for instance $k = 3$ and $m = 30$ yields an efficiency close to $1$.
The value less than $1$ obtained for $k = 5$, $m = 50$ for the inefficiency ratio is likely due to Monte Carlo variability;
we expect a value greater or equal to $1$.

\begin{table}[ht]
\centering
\begin{tabular}{llrrr}
  \hline
$k$ & $m$ & Cost & Variance & Inefficiency / $V_\infty$\\ 
  \hline
  1 & $1 \times k $ & 5.1494 & 0.0070 & 5.2152 \\ 
  1 & $5 \times k $ & 8.0747 & 0.0013 & 1.5164 \\ 
  1 & $10 \times k$ & 13.0747 & 0.0006 & 1.1838 \\ 
  3 & $1 \times k $ & 6.0755 & 0.0061 & 5.3332 \\ 
  3 & $5 \times k $ & 18.0747 & 0.0005 & 1.1990 \\ 
  3 & $10 \times k$  & 33.0747 & 0.0002 & 1.0044 \\ 
  5 & $1 \times k $ & 8.0747 & 0.0059 & 6.8677 \\ 
  5 & $5 \times k $ & 28.0747 & 0.0003 & 1.1328 \\ 
  5 & $10 \times k$  & 53.0747 & 0.0001 & 0.9816 \\ 
   \hline
\end{tabular}
\caption{Cost, variance, and inefficiency divided by MCMC asymptotic variance $V_\infty$ for various choices of $k$ and $m$, 
    for the test function $h: (A,\mu,\theta_1,\ldots,\theta_K) \mapsto \theta_1$,  in the baseball batting averages example
    of Section \ref{subsec:Gibbs-rosenthal2000}. \label{table:baseball}} 
\end{table}

Since the efficiency of the underlying MCMC kernel is retrieved with $k=3$, $m = 30$, for an associated 
cost comparable to $33$ steps of MCMC, we see that we can perform in parallel what would be equivalent to 
MCMC runs of $33$ iterations. This is considerably less than the recommended burn-in of $140$ derived in \citet{rosenthal1996analysis},
which is a strong indication that this recommended burn-in is conservative.

\begin{figure}
\begin{centering}
\subfloat[\label{fig:baseball:histogramtheta1} Histogram of $\theta_1$.]{\begin{centering}
    \includegraphics[width=0.30\textwidth]{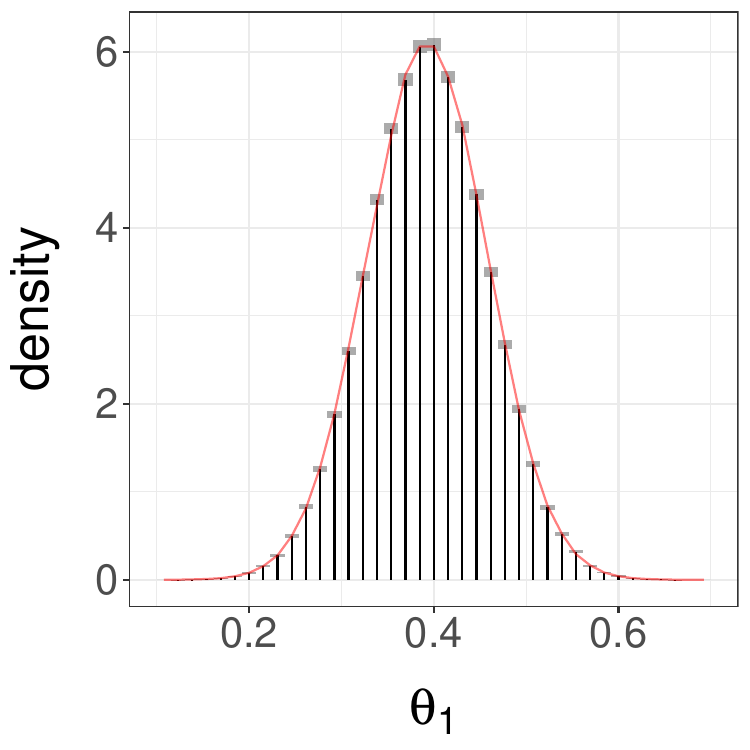}
\par\end{centering}
}
\subfloat[\label{fig:baseball:histogrammu} Histogram of $\mu$.]{\begin{centering}
    \includegraphics[width=0.30\textwidth]{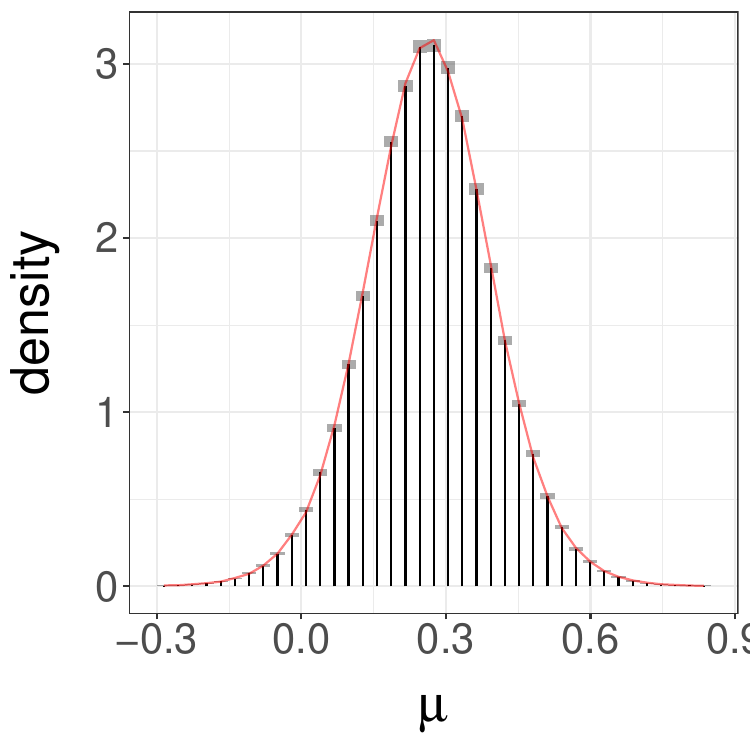}
\par\end{centering}
}
\subfloat[\label{fig:baseball:histogramA} Histogram of $A$.]{\begin{centering}
    \includegraphics[width=0.30\textwidth]{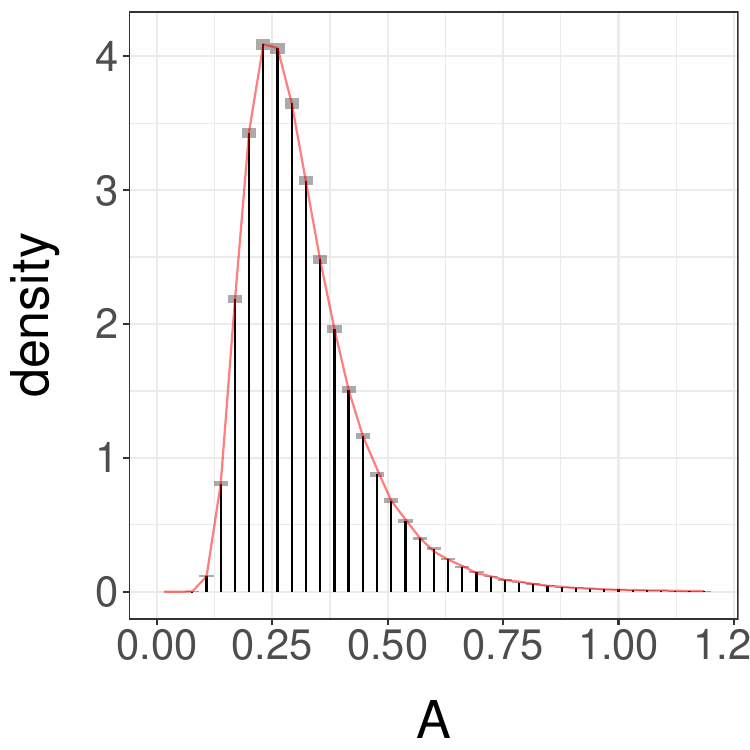}
\par\end{centering}
}
\par\end{centering}
\caption{\label{fig:baseball:histograms}
Gibbs sampling in the baseball batting averages example of Section \ref{subsec:Gibbs-rosenthal2000}.
Histograms of marginal posterior distributions of $\theta_1$, $A$ and $\mu$ in \ref{fig:baseball:histogramtheta1}, \ref{fig:baseball:histogrammu}, \ref{fig:baseball:histogramA},
produced using $R =10,000$ estimators, $k=3$, $m=30$.  }
\end{figure}

We plot histograms of $\theta_1$, $A$ and $\mu$ in Figures \ref{fig:baseball:histogramtheta1}, 
\ref{fig:baseball:histogrammu} and \ref{fig:baseball:histogramA}, obtained for $R=10,000$ estimators with $k=3$ and $m=30$.
The overlaid red curves indicate the marginal target densities estimated from an MCMC run
with $5\times 10^5$ iterations and a burn-in of $1,000$ (which is unnecessarily conservative).
These histograms confirm that accurate approximations of the posterior distribution are obtained with the proposed method.

\section{P{\'o}lya-Gamma Gibbs sampler for logistic regression\label{sub:PGG}}

Next, we turn to a more modern MCMC setting and demonstrate
that the proposed estimators can be constructed from 
the P{\'o}lya-Gamma Gibbs (PGG) sampler \citep{polson2013bayesian}.

We consider the logistic regression of an outcome $Y=(Y_{1},\ldots,Y_{n})\in\left\{ 0,1\right\} ^{n}$
on covariates $X=(x_{1},\ldots,x_{n})\in\mathbb{R}^{n\times p}$.
The model specifies $\mathbb{P}(Y_{i}=1|\beta)=\text{expit}\left(x_{i}^{T}\beta\right)$,
where $\text{expit}:z\mapsto1/\left(1+\exp(-z)\right)$, and the vector
$\beta\in\mathbb{R}^{p}$ is the regression parameter. Realizations
of $Y=(Y_{1},\ldots,Y_{n})$ are denoted by $y=(y_{1},\ldots,y_{n})$.
The prior distribution on $\beta$ is $\mathcal{N}\left(b,B\right)$,
where we set the mean $b$ to zero and the covariance $B$ as diagonal, with non-zero entries equal to $10$.
The P{\'o}lya-Gamma Gibbs (PGG) sampler \citep{polson2013bayesian} is
an MCMC algorithm targeting the posterior distribution,
extended with $n$ P{\'o}lya-Gamma variables $W=(W_{1},\ldots,W_{n})$.
The P{\'o}lya-Gamma distribution with parameters $(1,c)$, denoted by PG(1,c), has density defined for all $c\geq0$ as
\[
\forall x>0\quad\text{pg}(x;c)=\text{cosh}\left(\frac{c}{2}\right)\exp\left(-\frac{c^{2}x}{2}\right)\sum_{k=0}^{\infty}\left(-1\right)^{k}\frac{\left(2k+1\right)}{\sqrt{2\pi x^{3}}}\exp\left(-\frac{(2k+1)^{2}}{8x}\right).
\]
Under the extended target distribution the variables $W=(W_{1},\ldots,W_{n})$ 
are independent of each other given $\beta$, and have the property that $W_{i}$ follows
PG$(1,|x_{i}^{T}\beta|)$ for all $1\leq i\leq n$. 
The PGG sampler is a Gibbs sampler which alternates between the following updates:
\begin{align*}
    W_i | \text{rest} &\sim \text{PG}(1,|x_i^T \beta|) \quad \text{for all }i\in \{1,\ldots,n\},  \\
    \beta | \text{rest} &\sim \mathcal{N}(\Sigma(W)(X^{T}\tilde{y}+B^{-1}b), \Sigma(W)), \quad \text{with } \Sigma(W) = (X^{T}\text{diag}(W)X+B^{-1})^{-1},
\end{align*}
where $\tilde{y}=(y_{1}-\nicefrac{1}{2},\ldots,y_{n}-\nicefrac{1}{2})$.
The resulting chain targets the posterior
distribution and is uniformly ergodic \citep{choi2013polya}.
Here we initialize the algorithm with draws from the prior. 
We couple this chain using maximal couplings of the conditional updates.
For the P{\'o}lya-Gamma updates, 
the probability density functions are intractable but the ratio 
of two density evaluations can be calculated using the identity
\[
\forall x>0\quad 
\frac{\text{pg}\left(x;c_{2}\right)}{\text{pg}\left(x;c_{1}\right)}=\frac{\text{cosh}(c_{2}/2)}{\text{cosh}(c_{1}/2)}\exp\left(-\left(\frac{c_{2}^{2}}{2}-\frac{c_{1}^{2}}{2}\right)x\right),\]
which enables a fast implementation of the maximal coupling algorithm described in Algorithm \ref{alg:maximalcoupling} of the main document.

\begin{figure}
\begin{centering}
\subfloat[\label{fig:pgg:meetingtime} Histogram of meeting times.]{\begin{centering}
    \includegraphics[width=0.30\textwidth]{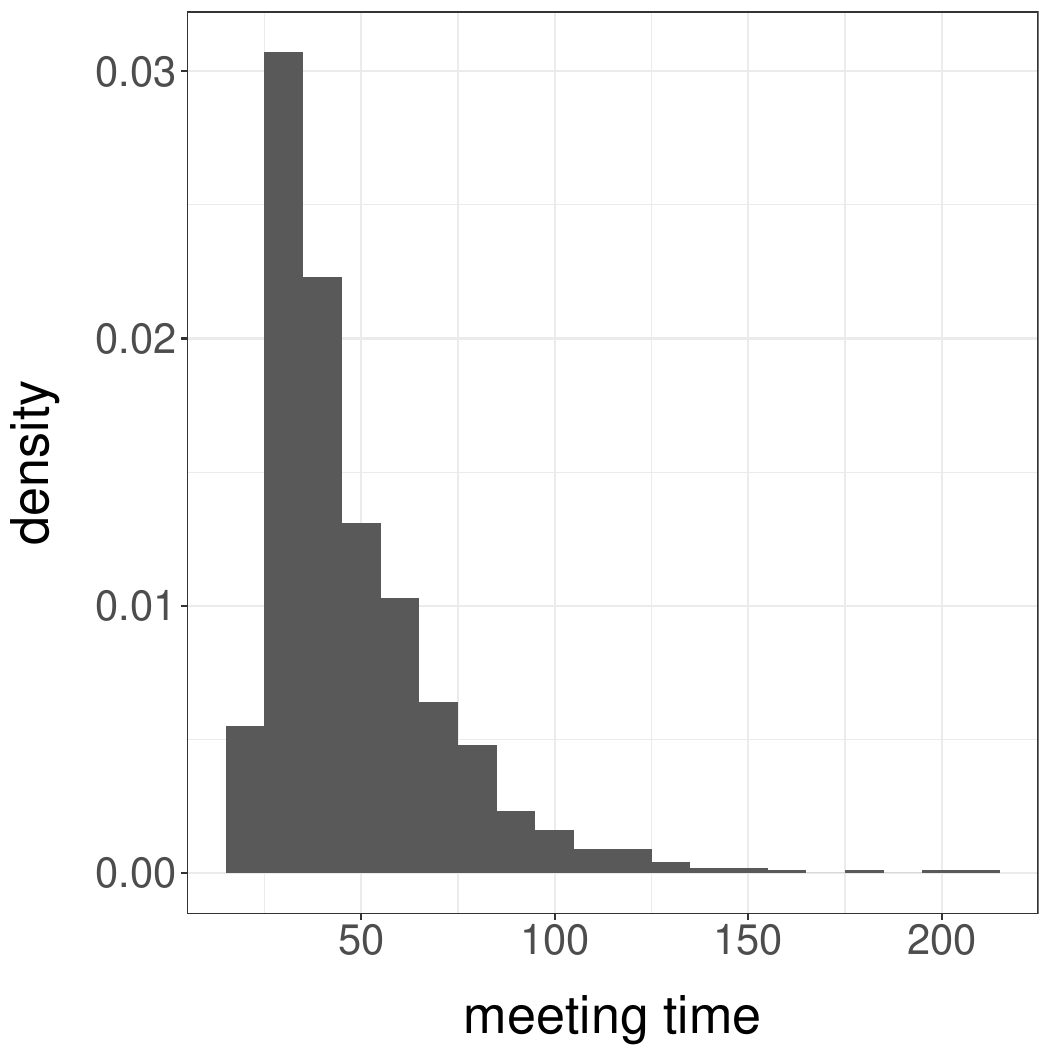}
\par\end{centering}
}
\subfloat[\label{fig:pgg:nmetw} Number of PG variables met.]{\begin{centering}
\includegraphics[width=0.3\textwidth]{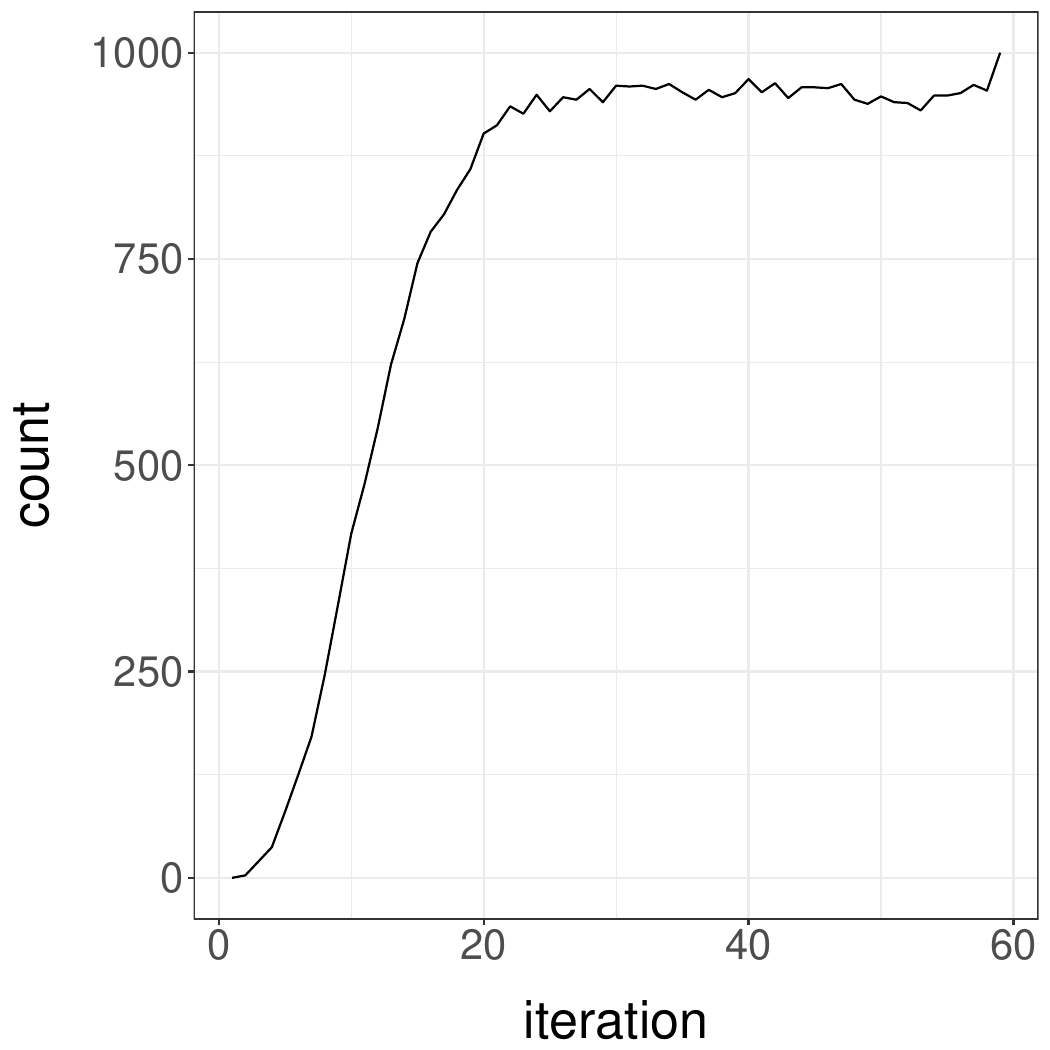}
\par\end{centering}
}
\subfloat[\label{fig:pgg:distw} PG distance $||W-\tilde{W}||_2$.]{\begin{centering}
\includegraphics[width=0.3\textwidth]{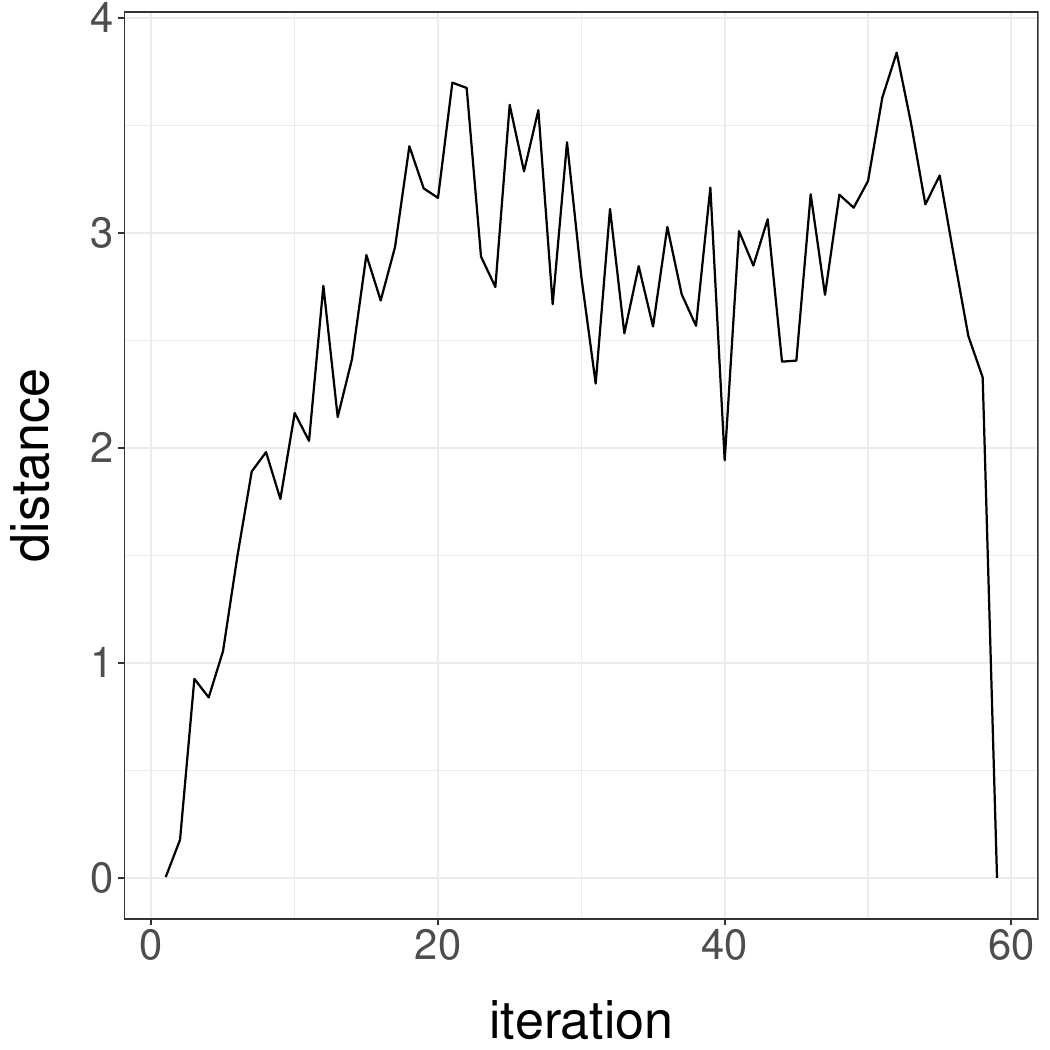}
\par\end{centering}
}

\subfloat[\label{fig:pgg:distbeta} Coefficient distance $||\beta-\tilde{\beta}||_2$.]{\begin{centering}
\includegraphics[width=0.3\textwidth]{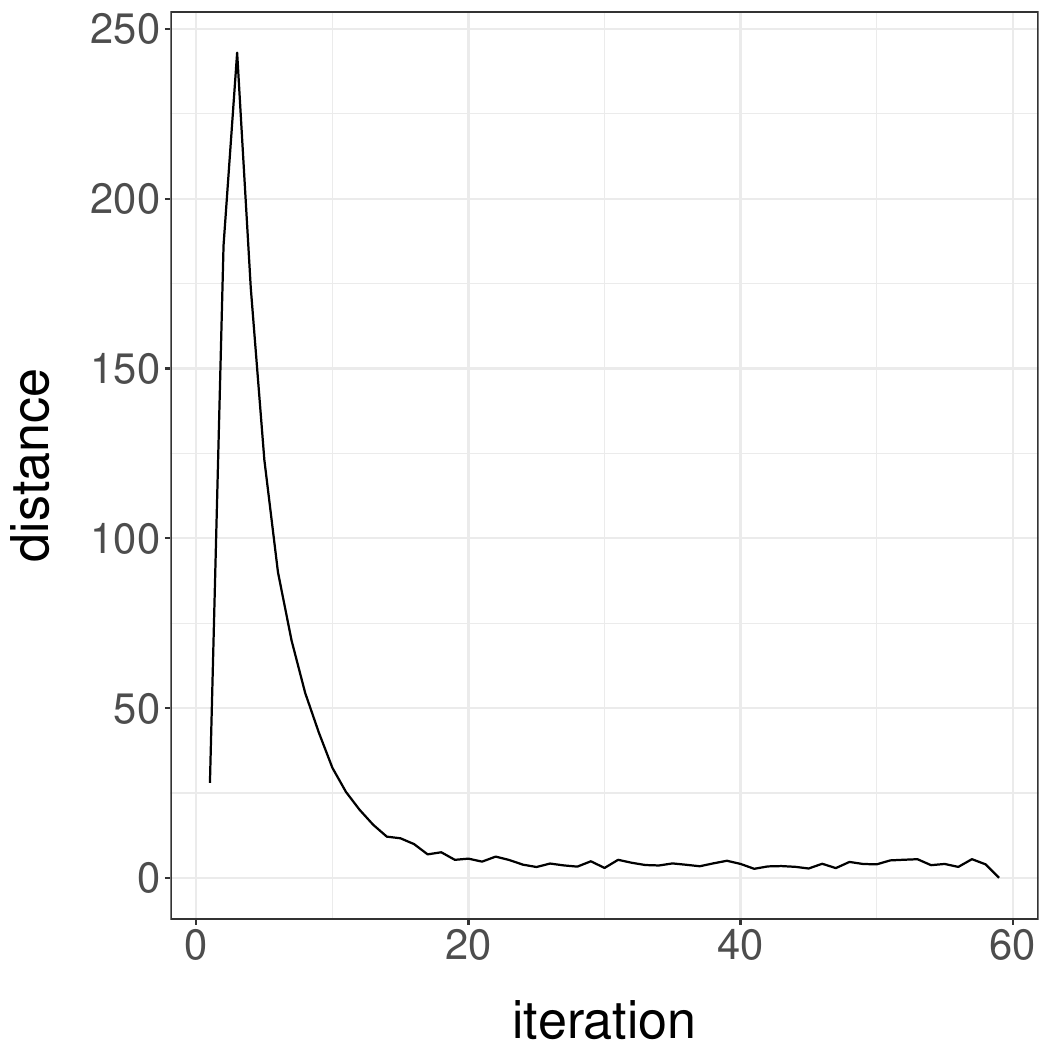}
\par\end{centering}
}
\subfloat[\label{fig:pgg:asympteff} Efficiency of $H_k(X,Y)$ vs. $k$.]{\begin{centering}
    \includegraphics[width=0.3\textwidth]{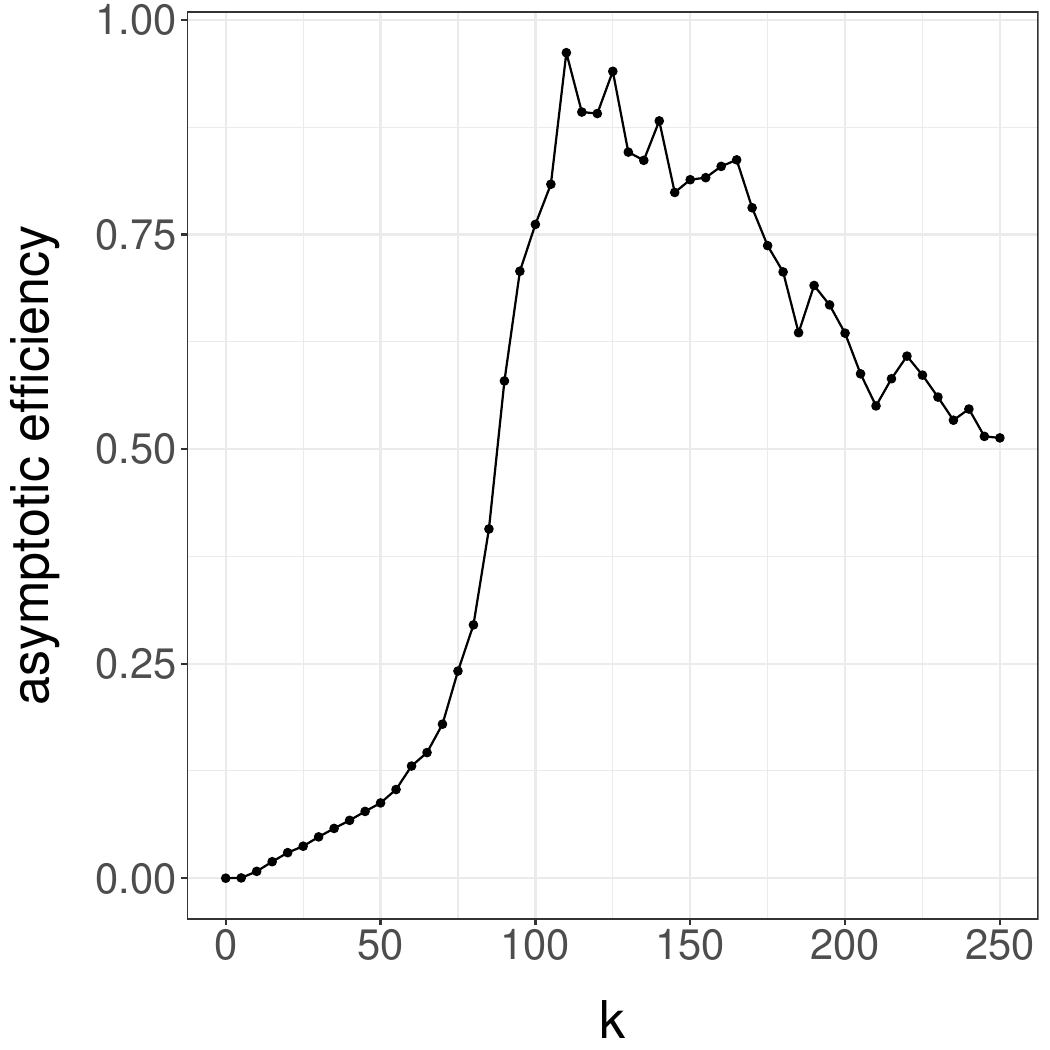}
\par\end{centering}
}
\subfloat[\label{fig:pgg:hist1} Histogram of $\beta$.]{\begin{centering}
    \includegraphics[width=0.3\textwidth]{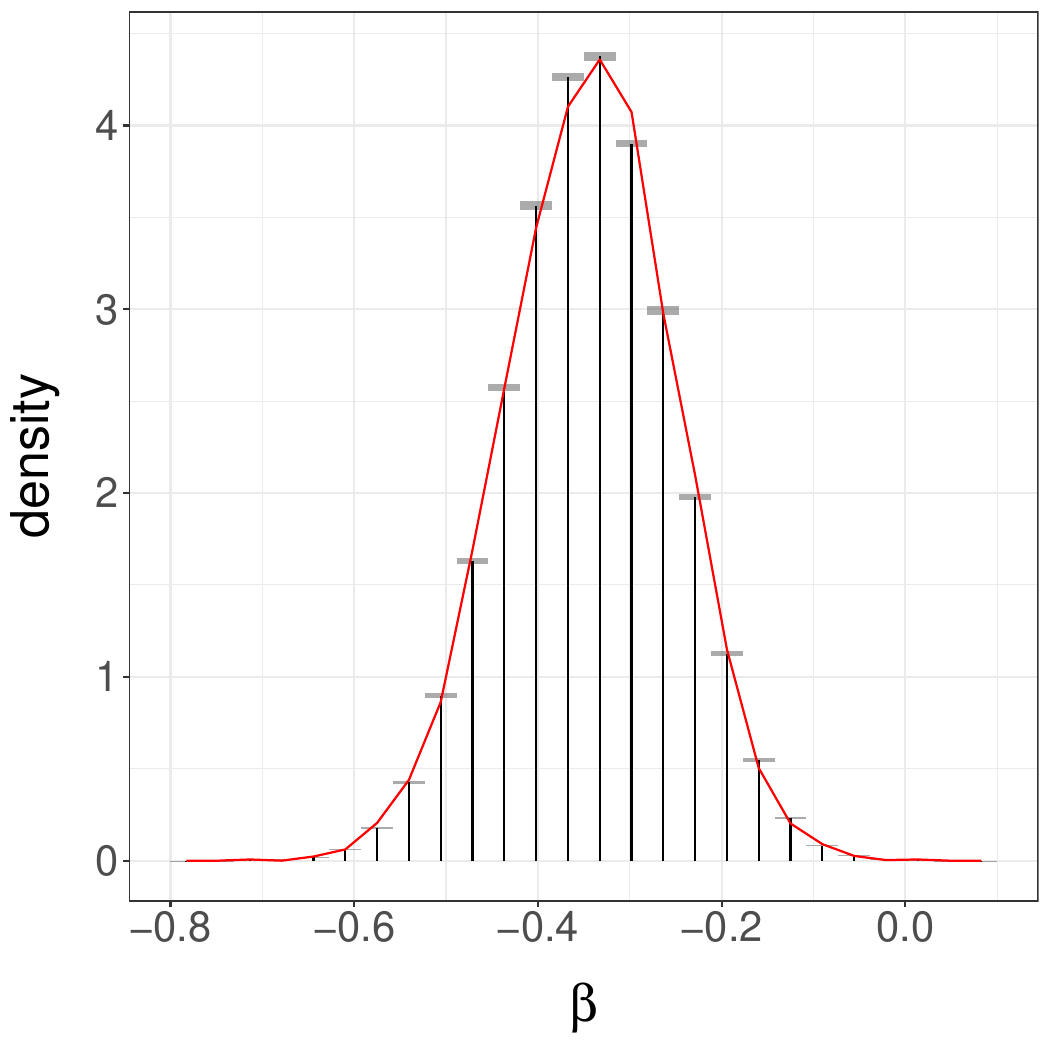}
\par\end{centering}
}
\par\end{centering}
\caption{\label{fig:pgg}
PGG sampling in the German credit example of Section \ref{sub:PGG}. 
Histogram of meeting times in \ref{fig:pgg:meetingtime}. 
Trace of the number of P{\'o}lya-Gamma variables met in \ref{fig:pgg:nmetw},
for one typical run of the coupled chains. 
Trace of the between-chain distance for P{\'o}lya-Gamma variables and regression coefficients in \ref{fig:pgg:distw}-\ref{fig:pgg:distbeta},
for one typical run of the coupled chains. 
Efficiency of $H_k(X,Y)$ as a function of $k$ in \ref{fig:pgg:asympteff}, computed over $R=1,000$ replicates.
Histogram of the marginal posterior for the `installment percent' regression coefficient in \ref{fig:pgg:hist1},
based on $R=1,000$ estimators, $k=110$ and $m=1,100$.
}
\end{figure}

We apply the proposed method to the German credit data of
\citep{lichman:2013}, a common example in binary regression and machine
learning studies such as \citet{polson2013bayesian,huang2007credit,west2000neural}. This dataset consists of $1,000$ loan application records,
700 of which were rated as creditworthy and 300 were rated as not creditworthy.
Each record includes 20 additional variables including loan purpose,
demographic information, bank account balances, marital, housing, 
employment status, and job type. Seven of these are quantitative
and the rest are categorical. After translating categorical variables
into indicators we obtain $p=49$ regressors on $n=1,000$ observations.
Histograms of the meeting times for $R=1,000$ coupled chains are shown in Figure \ref{fig:pgg:meetingtime}. 
These chains took between 18 and 164 steps to meet, with an average meeting
time of 48 iterations. 

The extended space of the Gibbs sampler is of dimension $n+p=1,049$. 
However the two chains meet as soon as either all the $n$ auxiliary PG variables or all the $p$ regression coefficients
meet. Since we use a maximal coupling of the update of the full vector of regression coefficients,
either all or none of these meet at each iteration; MH-within-Gibbs strategies could be employed instead. 
As we show in Figure \ref{fig:pgg:nmetw}, for one run of the coupled chains, the number of met PG variables rapidly increases to a plateau, 
at which point the chains are close enough to make coupling on the regression coefficients possible.
Figure \ref{fig:pgg:distw} shows the Euclidean distance between the PG variables of the two chains;
it starts at zero as an artefact of the initial values of the PG variables being set to zero.
Figure \ref{fig:pgg:distbeta} shows the Euclidean distance between the regression coefficients.
For this particular run, both PG variables and regression variables diverge at first,
before converging as an increasing number of PG variables meet.

Finally, we consider the choice of $k$ for the estimator $H_k(X,Y)$,
and of $m$ for the estimator $\bar{H}_{k:m}(X,Y)$. 
We consider the task of estimating the posterior mean of a particular 
regression coefficient corresponding to the installment payment as a percentage of disposable income.
The efficiency of $H_k(X,Y)$, defined as one over the product of the variance times the cost, 
is shown in Figure \ref{fig:pgg:asympteff} as a function of $k$. We see
that choosing a large quantile of the distribution of $\tau$, as shown in Figure 
\ref{fig:pgg:meetingtime}, would result in an efficiency close to its maximum.
We choose $k=110$, and,  in line with the heuristics suggested in the main document, we take $m$ to be $10k=1,100$,
that is, a large multiple of $k$.
For the estimation of the posterior mean, 
we find that the above values of $k$ and $m$ yield an inefficiency about $7$ times greater
to that of the underlying Gibbs sampler, based on a long run; larger values of $k$ and $m$ would further reduce this inefficiency ratio.

With these tuning parameters, we produce a histogram of the posterior distribution for 
that coefficient in Figure
\ref{fig:pgg:hist1}. We find 
agreement with a density estimated from a long MCMC run, depicted here by
the overlaid curve in red.
To summarize, in this example the proposed methodology
effectively allows to run PGG chains 
in parallel, in chunks of approximately $1,000$ iterations, while bypassing
the usual difficulties related to the choice of burn-in and the construction of confidence intervals.

In passing, in the particular case of the coupled PGG algorithm,
we can directly show that 
the meeting time has an $\varepsilon$ probability of
occurring at each step $t$, for large enough $t$ and some $\varepsilon>0$
that does not depend on the starting points of the chains. 
Denote the $\beta$-component of the two chains by 
$(\beta_{t})_{t\geq0}$ and $(\tilde{\beta}_{t})_{t\geq0}$.
From \citet{choi2013polya},
$(\beta_{t})_{t\geq0}$ and $(\tilde{\beta}_{t})_{t\geq0}$ are 
uniformly ergodic. Consider Fr\'echet's inequality $\mathbb{P}\left(A\cap B\right)\geq\mathbb{P}(A)+\mathbb{P}(B)-1$
and the events $A=\{\beta_{t}\in S\}$ and $B=\{\tilde{\beta}_{t-1}\in S\}$,
for some compact set $S$ of $\mathbb{R}^{p}$ and some $t\geq0$.
We can define $S$ such that $\{\beta_{t}\in S\}$ has probability
at least $0.5+\delta$, for some small $\delta>0$, provided $t$
is large enough, since
\[
    |\pi(S|y)-K^{t}(S|\beta)|\leq d_{\text{TV}}(\pi\left(\cdot|y\right), K^{t}(\cdot|\beta))<M\rho^{t},
\]
where $\beta$ is any starting point, $K^{t}(\cdot|\beta)$ is the
distribution of the $t$-th iterate of the chain starting from $\beta$,
$M>0$ and $\rho<1$ are constants independent of $\beta$, and
$d_{\text{TV}}$ stands for the total variation distance.
We take $S$ to be a compact set of $\mathbb{R}^{p}$ such that $\pi(S|y)>0.5+2\delta$.
There is some $t_{0}$ such that $t\ge t_{0}$ implies $K^{t}(S|\beta)>0.5+\delta$;
an identical reasoning can be done for the second chain. Using Fr\'echet's
inequality, $\mathbb{P}(\beta_{t}\in S,\tilde{\beta}_{t-1}\in S)>2\delta$.
On these events, $|x_{i}^{T}\beta_t|^{2}-|x_{i}^{T}\tilde{\beta}_{t-1}|^{2}$
are lower and upper-bounded for all $1\leq i\leq n$, which results
in a strictly non-zero probability of coupling each pair 
of P{\'o}lya-Gamma variables.
This, in turns, leads to an $\varepsilon>0$
probability of $\beta_{t+1}$ meeting with $\tilde{\beta}_{t}$.
Therefore, Assumption \ref{assumption:meetingtime} on the meeting time in the main document is satisfied.

\section{Bayesian Lasso \label{subsec:blasso}}

We consider the setting of Bayesian inference 
in regression models.  
The Bayesian Lasso \citep{park2008bayesian} assigns a hierarchical
prior on the parameters of a linear regression in such a way that
the posterior mode corresponds to the Lasso estimator \citep{tibshirani1996regression,efron2004least}.
The posterior distribution can be approximated by Gibbs sampling,
independently for a range of regularization parameters $\lambda>0$,
which can then be selected by cross-validation or as described in
\citet{park2008bayesian}; see also an alternative computational approach
in \citet{bornn2010efficient}. On
top of demonstrating the applicability of the proposed methodology
in this setting, we illustrate the use of confidence intervals to
guide the allocation of computational resources. 

The model and associated Gibbs sampler are as follows. Consider an
$n\times p$ matrix of standardized covariates $X$, and an $n$-vector
of outcomes $Y$. The centered outcomes are denoted by $\tilde{Y}$,
i.e. $\tilde{Y}=Y-\bar{Y}1_{n}$, where $\bar{Y}=n^{-1}\sum_{i=1}^{n}Y_{i}$
and $1_{n}$ is a $n$-vector of $1$'s. Introduce a regularization
parameter $\lambda\geq0$. The outcome $Y$ is assumed to follow $\mathcal{N}(\mu1_{n}+X\beta,\sigma^{2}I_{n})$,
where $\mu$ is the intercept, and $\beta$ is the $p$-vector of
regression coefficients; $I_{n}$ denotes a unit $n\times n$ diagonal
matrix. The hierarchical prior specifies $\beta\sim\mathcal{N}(0_{p},\sigma^{2}D_{\tau})$,
where $0_{p}$ is a $p$-vector of $0$'s, and $D_{\tau}=\text{diag}(\tau_{1}^{2},\ldots,\tau_{p}^{2})$.
The prior on $\tau_{j}^{2}$ is $\text{\text{Exponential}}(\lambda^{2}/2)$
for all $j\in\{1,\ldots,p\}$, and finally $\sigma^{2}\sim\text{Inverse Gamma}(a,\gamma)$.
A Gibbs sampler proposed in \citet{park2008bayesian} performs the following
updates, after having integrated $\mu$ out:
\begin{align*}
\beta|\text{rest} & \sim\mathcal{N}(A_{\tau}^{-1}X^{T}\tilde{Y},\sigma^{2}A_{\tau}^{-1}),\quad\text{where}\quad A_{\tau}=X^{T}X+D_{\tau}^{-1},\\
\sigma^{2}|\text{rest} & \sim\text{Inverse Gamma}(a+\frac{n-1}{2}+\frac{p}{2},\gamma+\left(\tilde{Y}-X\beta\right)^{T}\left(\tilde{Y}-X\beta\right)/2+\beta^{T}D_{\tau}^{-1}\beta/2),\\
\tau_{j}^{-2}|\text{rest} & \sim\text{Inverse Gaussian}\left(\left(\nicefrac{\lambda^{2}\sigma^{2}}{\beta_{j}^{2}}\right)^{\nicefrac{1}{2}},\lambda^{2}\right)\quad\text{for all }j\in\left\{ 1,\ldots,p\right\} .
\end{align*}
Here the Inverse Gaussian $(\mu,\lambda)$ distribution 
has pdf  $p(x;\mu,\lambda) = (\lambda / 2\pi x^3)^{\nicefrac{1}{2}} \exp(- \lambda (x-\mu)^2 / (2 \mu^2 x))$.
We initialize the sampler by setting $\beta_{j}=0,\tau_{j}^{2}=1$
for all $j\in\{1,\ldots,p\}$ and $\sigma^{2}=1$. We consider the
diabetes dataset used in \citet{efron2004least,park2008bayesian},
with $n=442$ individuals and $p=64$ covariates, as provided
in the \texttt{lars} package accompanying \citet{efron2004least}. The parameter space is
of dimension $2p+1=129$. We set $a=0$
and $\gamma=0$ throughout. To couple the Gibbs sampler we use a maximal
coupling of each conditional update, and focus on producing unbiased
estimators of the posterior means $\mathbb{E}_{\lambda}[\beta|Y,X]$
given $\lambda$. 

\begin{figure}
\begin{centering}
\subfloat[\label{fig:blasso:averagemeeting} Average meeting time  vs $\lambda$.]{\begin{centering}
\includegraphics[width=0.45\textwidth]{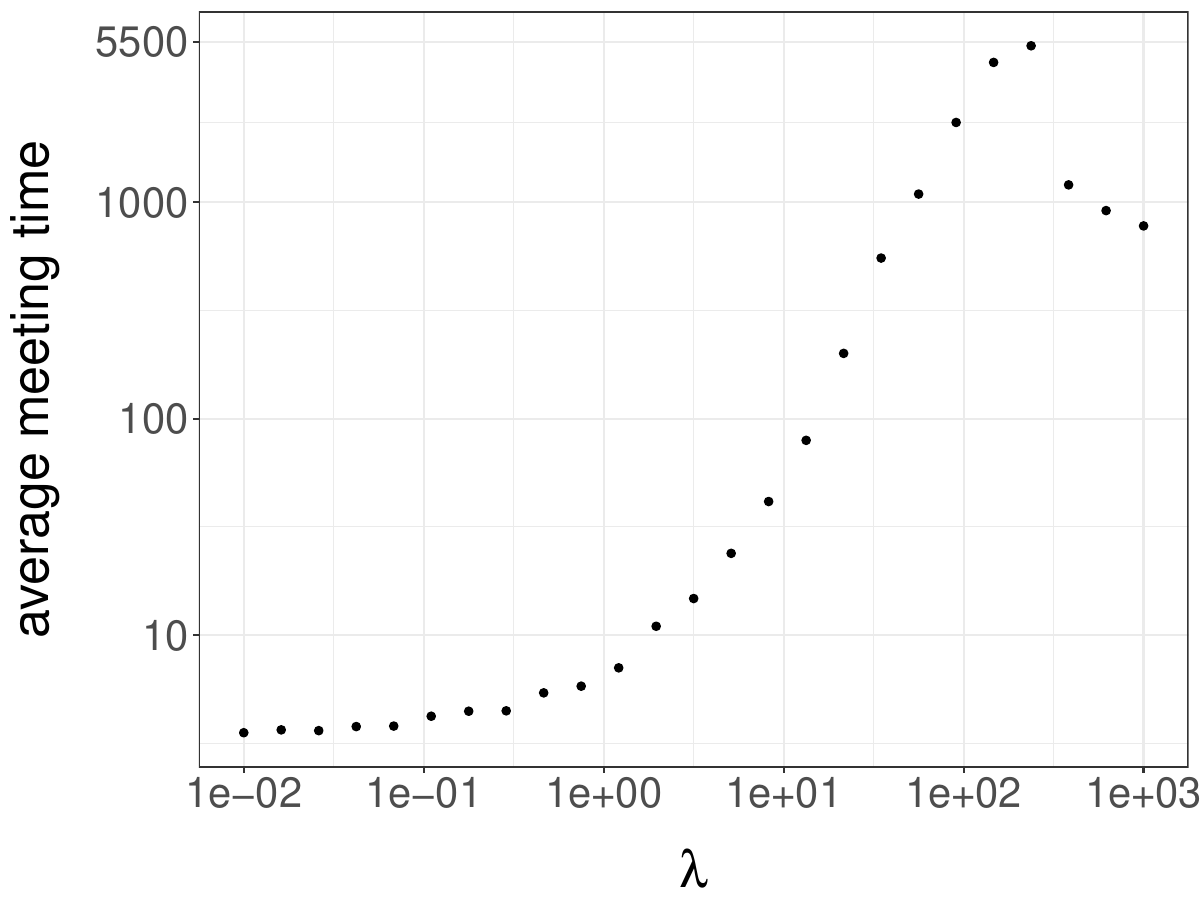}
\par\end{centering}
}
\subfloat[\label{fig:blasso:ess} Effective sample sizes vs
$\lambda$.]{\begin{centering}
\includegraphics[width=0.45\textwidth]{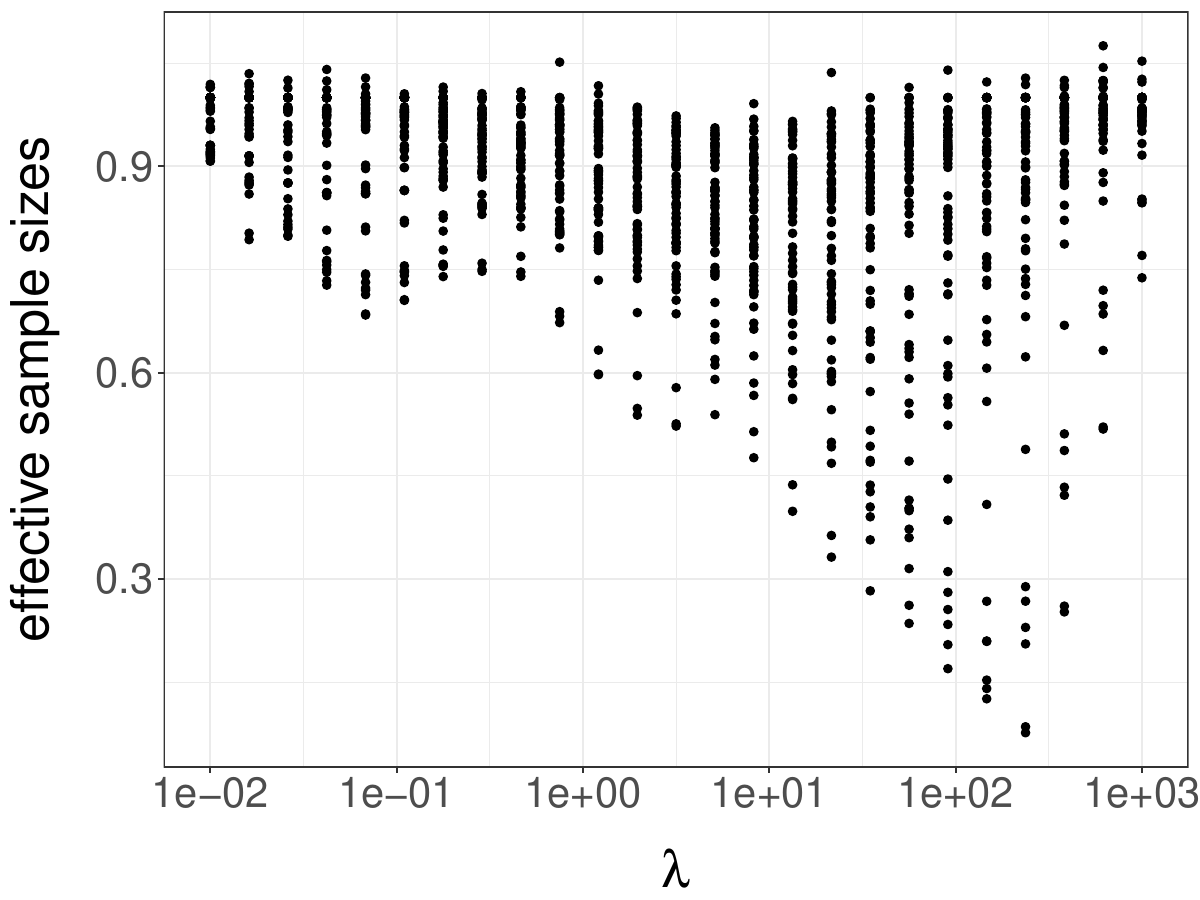}
\par\end{centering}
}
\par\end{centering}
\caption{\label{fig:bayesianlasso:mixing}Bayesian Lasso for the diabetes data as described
in Section \ref{subsec:blasso}. Figure \ref{fig:blasso:averagemeeting}
shows the average of $100$ independent meeting times
against the regularization parameter $\lambda$. Figure
\ref{fig:blasso:ess} shows the effective sample size
of each of the $64$ components of $\beta$, from MCMC runs of length $50,000$
and the CODA package, 
as a function
of $\lambda$. }
\end{figure}

For a range of values of $\lambda$ between $10^{-2}$ and $10^3$, we run $100$ coupled chains until they meet,
and plot the empirical average of the meeting times as a function of $\lambda$ in Figure \ref{fig:blasso:averagemeeting}.
First, we note that the average meeting times are very small for small values of
$\lambda$. Next we observe a peak located around $\lambda=10^{2}$,
which implies a similar peak for the computational cost of the proposed estimators.
This could be either a consequence of the mixing properties of the
underlying MCMC, or a defect of the coupling strategy.
To investigate this, we run the underlying Gibbs sampler for the same values of
$\lambda$, for $50,000$ iterations, discard the first $5,000$ iterations,
and compute the effective sample size using the CODA package \citep{plummer2006coda}.
We do so for each of the $64$ components of $\beta$, and plot the results
in Figure \ref{fig:blasso:ess}; each dot corresponds to the ESS of one of the components, for a particular value of $\lambda$. The effective sample size is divided by the number of iterations
post burn-in, and thus we expect a number between $0$ and $1$. We see a drastic loss of efficiency
around $\lambda = 10^2$, indicating that the Gibbs sampler of \citet{park2008bayesian}
mixes poorly for these values of $\lambda$.

\begin{figure}
\begin{centering}
\subfloat[\label{fig:blasso:p64:paths}Posterior mean of $\beta$ vs $\lambda$,
with $100$ estimators per $\lambda$.]{\begin{centering}
\includegraphics[width=0.45\textwidth]{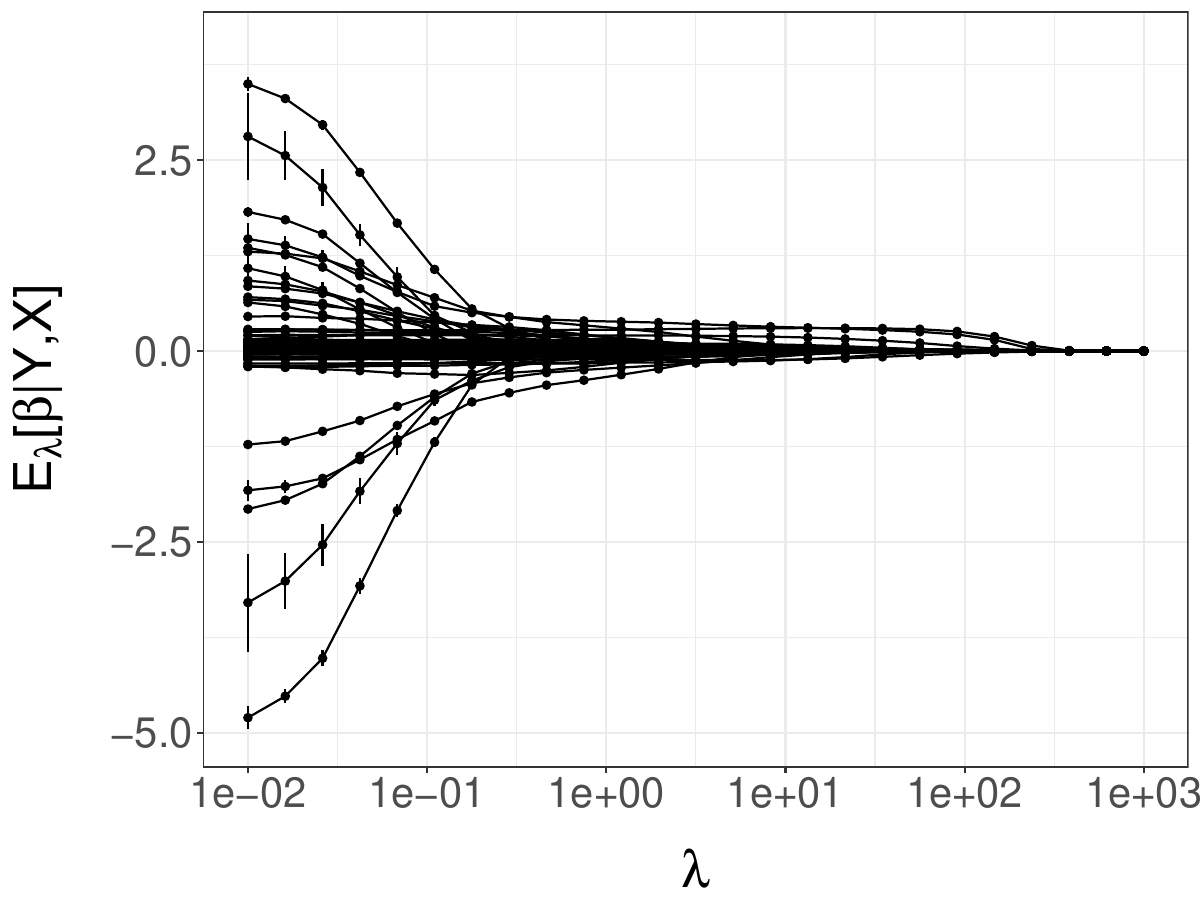}
\par\end{centering}
}\subfloat[\label{fig:blasso:p64:paths:refined}Posterior mean of $\beta$ vs
$\lambda$, with $1,000$ more estimators for the $10$ smallest values
of $\lambda$.]{\begin{centering}
    \includegraphics[width=0.45\textwidth]{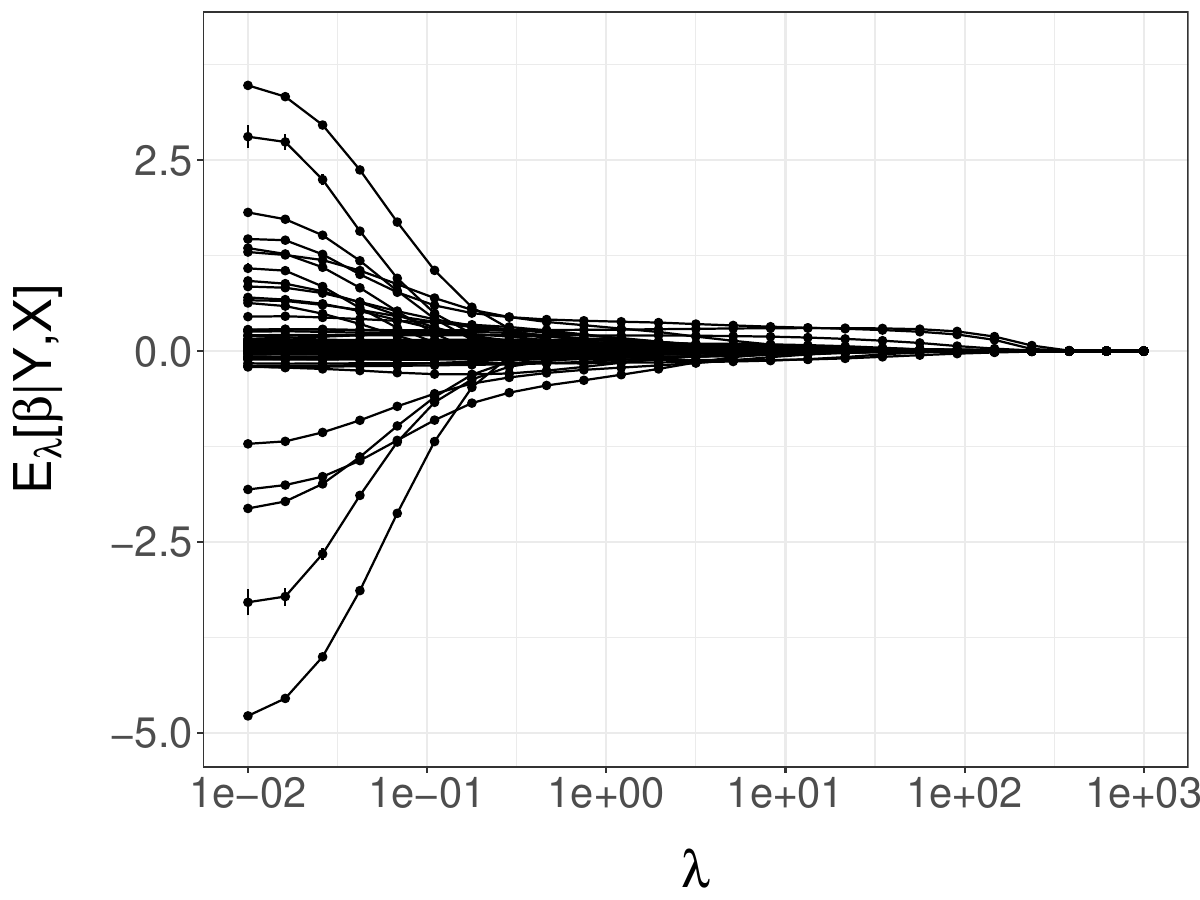}
\par\end{centering}
}
\par\end{centering}
\caption{\label{fig:blasso:p64}Bayesian Lasso for the diabetes data with $n=442$
and $p=64$, described in Section \ref{subsec:blasso}. Figure
\ref{fig:blasso:p64:paths} shows the paths of posterior means $\mathbb{E}_{\lambda}[\beta|Y,X]$
against the regularization parameter $\lambda$, obtained with $R=100$
estimators, $k$ chosen as one plus the $90\%$ quantile of the distribution
of meeting times, and $m=10k$. Figure \ref{fig:blasso:p64:paths:refined}
shows the estimates obtained after having run $1,000$ more estimators
for the first $10$ values of $\lambda$. }
\end{figure}

For each $\lambda$, we now choose $k$ as the $99\%$ quantile of the $100$ meeting times
previously obtained, and we choose $m = 10k$.
We produce $R=100$ unbiased estimators for each posterior
mean $\mathbb{E}_{\lambda}[\beta|Y,X]$, and plot them against $\lambda$
in Figure \ref{fig:blasso:p64:paths}, component-wise.
On top of each dot lies a $95\%$ confidence interval represented by a vertical segment:
these are too small to be noticeable except for the smallest values of $\lambda$. 

In order to refine the posterior mean estimates, we can allocate 
computational resources
based on the confidence intervals and the costs associated with each 
$\lambda$ (there are $25$ values of $\lambda$ in total). In order to tighten the most visible confidence intervals, 
we produce $1,000$ more estimators for the $10$ smallest values of
$\lambda$, and obtain the refined estimates shown in Figure \ref{fig:blasso:p64:paths:refined}.
That is, these estimates are obtained from $1,100$ unbiased estimators for the $10$ smallest
values of $\lambda$, and for $100$ for the other values of $\lambda$.
This refinement procedure could be automatized, adaptively producing more estimators
for values of $\lambda$ where confidence intervals are wider. 

\end{document}